\documentstyle[thmsa,a4,sw20lart]{article}

\input{tcilatex}
\begin{document}

\author{S. Manoff \\
{\it Bulgarian Academy of Sciences,}\\
{\it \ Institute for Nuclear Research and Nuclear Energy,}\\
{\it \ Blvd. Tzarigradsko Chaussee 72,}\\
{\it \ 1784 Sofia - Bulgaria}}
\title{{\sc Mechanics of Continuous Media in}{\bf \ }$(\overline{L}_n,g)${\sc %
-Spaces.}\\
{\bf IV. Stress (Tension) Tensor}}
\date{E-mail address: smanov@inrne.bas.bg}
\maketitle

\begin{abstract}
Basic notions of continuous media mechanics are introduced for spaces with
affine connections and metrics. Stress (tension) tensors are considered,
obtained by the use of the method of Lagrangians with covariant derivatives
(MLCD). On the basis of the covariant Noether's identities for the
energy-momentum tensors, Navier-Stokes' identities are found and generalized
Navier-Cauchy as well as Navier-Stokes' equations are considered over $(%
\overline{L}_n,g)$-spaces.

\end{abstract}

\tableofcontents

\section{Introduction}

The important relations between stresses (tensions) and deformations and
friction could also be established in continuous media mechanics in such
comprehensive spaces as spaces with affine connections and metrics [$(%
\overline{L}_n,g)$-spaces]. The main task of this paper is the consideration
of the structures connected to the relations between stresses and
deformations as a basis for the description of the dynamics of physical
system (continuous media and fluids) in $(\overline{L}_n,g)$-spaces.

In Section 2 energy-momentum tensors are considered, obtained by the use of
the method of Lagrangians with covariant derivatives (MLCD). In Section 3
the structure of the stress (tension) tensor is discussed. In Section 4
relations between the kinematic characteristics of the relative velocity, of
the friction velocity and the stress (tension) tensors are considered. In
Section 5, on the basis of the covariant Noether's identities for the
energy-momentum tensors, Navier-Stokes' identities are found and generalized
Navier-Cauchy and Navier-Stokes' equations are considered over $(\overline{L}%
_n,g)$-spaces. The last Section 6 comprises some concluding remarks. The
whole picture of a continuum media mechanics in $(\overline{L}_n,g)$-spaces
could be related to a classical (non-quantized) field theory.

All considerations are given in details (even in full details) for those
readers who are not familiar with the considered problems.

{\it Remark.} The present paper is the fourth part of a larger research
report on the subject with the title ''Contribution to continuous media
mechanics in $(\overline{L}_n,g)$-spaces'' with the following contents:

I. Introduction and mathematical tools.

II. Relative velocity and deformations.

III. Relative accelerations.

IV. Stress (tension) tensor.

The parts are logically self-dependent considerations of the main topics
considered in the report.

\subsection{Invariant projections of a mixed tensor field of second rank}

In the relativistic continuum media mechanics notions are introduced as
generalizations of the same notions of the classical continuum media
mechanics. This has been done by means of the projections of the (canonical,
symmetric of Belinfante or symmetric of Hilbert) energy-momentum tensors
along or orthogonal to a non-isotropic (non-null) contravariant vector field.

There are possibilities for using the projections for finding out the
physical interpretations of the determined energy-momentum tensors. In an
analogous way as in (pseudo) Riemannian spaces without torsion ($V_n$%
-spaces), the different relations between the quantities with well known
physical interpretations can be considered as well as their application for
physical systems described by means of mathematical models over
differentiable manifold with affine connections and metrics [$(\overline{L}%
_n,g)$-spaces].

The energy-momentum tensors are obtained as mixed tensor fields of second
rank of type 1 [by the use of the procedure on the grounds of the method of
Lagrangians with covariant derivatives (MLCD) \cite{Manoff-0}] 
\begin{equation}  \label{IX.1.-1}
G=G_\alpha \,^\beta \cdot e_\beta \otimes e^\alpha =G_i\,^j\cdot \partial
_j\otimes dx^i
\end{equation}

\noindent in contrast to the mixed tensor fields of second rank of type 2 
\begin{equation}  \label{IX.1.-2}
\overline{G}=\overline{G}\,^\beta \,_\alpha \cdot e^\alpha \otimes e_\beta =%
\overline{G}\,^j\,_i\cdot dx^i\otimes \partial _j\text{ .}
\end{equation}

The sets of vectors $\{e^\alpha \}$ and $\{e_\alpha \}$ are non-co-ordinate
(non-holonomic) covariant and contravariant basic vector fields
respectively, $(\alpha ,\beta =1,...,n)$.

The set of vectors $\{dx^i\}$ and $\{\partial _i\}$ are co-ordinate
(holonomic) covariant and contravariant basic vector fields respectively $%
(i,j=1,...,n)$, $\dim (\overline{L}_n,g)=n$.

To every covariant basic vector field a contravariant basic vector field can
be juxtaposed and vice versa by the use of the contravariant and covariant
metric tensor fields 
\begin{equation}  \label{IX.1.-3}
g(e_\gamma )=g_{\alpha \overline{\gamma }}\cdot e^\alpha \text{ ,\thinspace
\thinspace \thinspace \thinspace \thinspace \thinspace }g(\partial _j)=g_{i%
\overline{j}}\cdot dx^i\text{ ,\thinspace \thinspace \thinspace \thinspace }%
\overline{g}(e^\gamma )=g^{\alpha \overline{\gamma }}\cdot e_\alpha \text{
,\thinspace \thinspace \thinspace \thinspace }\overline{g}(dx^j)=g^{i%
\overline{j}}\cdot \partial _i\text{ .}
\end{equation}

On this basis a tensor field of type 2 can be related to a tensor field of
type 1 by the use of the covariant and contravariant tensor fields $%
g=g_{ij}\cdot dx^i.dx^j$ and $\overline{g}=g^{ij}\cdot \partial _i.\partial
_j$ [$dx^i.dx^j=(1/2)(dx^i\otimes dx^j+dx^j\otimes dx^i)$, $\partial
_i.\partial _j=(1/2)(\partial _i\otimes \partial _j+\partial _j\otimes
\partial _i)$] 
\begin{equation}
\overline{G}=g(G)\overline{g}=\overline{G}\,^\beta \,_\alpha \cdot e^\alpha
\otimes e_\beta =g_{\alpha \overline{\gamma }}\cdot G_\delta \,^\gamma \cdot
g^{\overline{\delta }\beta }\cdot e^\alpha \otimes e_\beta \text{ ,}
\label{IX.1.-5}
\end{equation}

\begin{equation}  \label{IX.1.-6}
\overline{G}\,^\beta \,_\alpha =g_{\alpha \overline{\gamma }}\cdot G_\delta
\,^\gamma \cdot g^{\overline{\delta }\beta }\text{ ,}
\end{equation}

\begin{equation}  \label{IX.1.-7}
G=\overline{g}(\overline{G})g=G_\alpha \,^\beta \cdot e_\beta \otimes
e^\alpha =g^{\beta \overline{\delta }}\cdot \overline{G}\,^\gamma \,_\delta
\cdot g_{\overline{\gamma }\alpha }\cdot e_\beta \otimes e^\alpha \text{ ,}
\end{equation}

\begin{equation}  \label{IX.1.-8}
G_\alpha \,^\beta =g^{\beta \overline{\delta }}\cdot \overline{G}\,^\gamma
\,_\delta \cdot g_{\overline{\gamma }\alpha }\text{ .}
\end{equation}

The Kronecker tensor field appears as a mixed tensor field of second rank of
type 1 
\[
Kr=g_\beta ^\alpha \cdot e_\alpha \otimes e^\beta =g_j^i\cdot \partial
_i\otimes dx^j\text{ } 
\]

\noindent and can be projected by means of the non-isotropic (non-null)
contravariant vector field $u$ and its projection metrics $h_u$ and $h^u$ [$%
h_u=g-\frac 1e\cdot g(u)\otimes g(u)$, $h^u=\overline{g}-\frac 1e\cdot
u\otimes u$, $e=g(u,u)\neq 0$] 
\[
Kr=\varepsilon _{Kr}\cdot u\otimes g(u)+u\otimes g(^{Kr}\pi
)+\,^{Kr}s\otimes g(u)+(^{Kr}S)g\text{ ,} 
\]

\noindent where 
\begin{equation}
\begin{array}{c}
\varepsilon _{Kr}=\frac 1{e^2}\cdot [g(u)](Kr)u=\frac 1{e^2}\cdot u_{%
\overline{\alpha }}\cdot u^{\overline{\alpha }}=\frac 1{e^2}\cdot g_{%
\overline{\alpha }\overline{\beta }}\cdot u^{\overline{\alpha }}\cdot
u^\beta = \\ 
=\frac 1{e^2}\cdot f^\rho \,_\beta \cdot f^\sigma \,\cdot f^\beta \,_\delta
\cdot g_{\rho \sigma }\cdot u^\gamma \cdot u^\delta =\frac 1{e^2}\cdot u^{%
\overline{i}}\cdot u_{\overline{i}}=\frac 1{e^2}\cdot g_{\overline{i}%
\overline{k}}\cdot u^{\overline{i}}\cdot u^k=\frac 1e\cdot k\text{ ,}
\end{array}
\label{IX.1.-42}
\end{equation}

\begin{equation}
k=\frac 1e\cdot [g(u)](Kr)u=\frac 1e\cdot u^{\overline{i}}\cdot u_{\overline{%
i}}\text{ ,\thinspace \thinspace \thinspace \thinspace \thinspace \thinspace
\thinspace \thinspace \thinspace \thinspace \thinspace \thinspace \thinspace
\thinspace \thinspace \thinspace \thinspace \thinspace \thinspace \thinspace
\thinspace \thinspace \thinspace }u_{\overline{i}}=g_{\overline{i}\overline{j%
}}\cdot u^j\text{ ,}  \label{IX.1.-43}
\end{equation}

\begin{equation}  \label{IX.1.-44}
\begin{array}{c}
^{Kr}\pi =\frac 1e\cdot [g(u)](Kr)h^u=\frac 1e\cdot g_{\overline{\beta }%
\overline{\gamma }}\cdot u^\gamma \cdot g_\delta ^\beta \cdot h^{\overline{%
\delta }\alpha }\cdot e_\alpha =\frac 1e\cdot g_{\overline{\beta }\overline{%
\gamma }}\cdot h^{\overline{\beta }\alpha }\cdot u^\gamma \cdot e_\alpha =
\\ 
=\frac 1e\cdot u_{\overline{\beta }}\cdot h^{\overline{\beta }\alpha }\cdot
e_\alpha =\,^{Kr}\pi ^\alpha \cdot e_\alpha \text{ ,}
\end{array}
\end{equation}

\begin{equation}  \label{IX.1.-45}
^{Kr}\pi ^\alpha =\frac 1e\cdot g_{\overline{\beta }\overline{\gamma }}\cdot
u^\gamma \cdot h^{\overline{\beta }\alpha }=\frac 1e\cdot u_{\overline{\beta 
}}\cdot h^{\overline{\beta }\alpha }\text{ ,\thinspace \thinspace \thinspace
\thinspace \thinspace \thinspace \thinspace }^{Kr}\pi ^i=\frac 1e\cdot g_{%
\overline{k}\overline{l}}\cdot u^l\cdot h^{\overline{k}i}=\frac 1e\cdot u_{%
\overline{k}}\cdot h^{\overline{k}i}\text{ ,\thinspace }
\end{equation}

\begin{equation}  \label{IX.1.-47}
^{Kr}s=\frac 1e\cdot h^u(g)(Kr)(u)=\,^{Kr}s^\alpha \cdot e_\alpha
=\,^{Kr}s^i\cdot \partial _i\text{ ,}
\end{equation}
\begin{equation}  \label{IX.1.-49}
^{Kr}s^\alpha =\frac 1e\cdot h^{\alpha \beta }\cdot g_{\overline{\beta }%
\overline{\delta }}\cdot u^{\overline{\delta }}\text{ ,\thinspace \thinspace
\thinspace \thinspace \thinspace \thinspace \thinspace \thinspace \thinspace 
}^{Kr}s^i=\frac 1e\cdot h^{ik}\cdot g_{\overline{k}\overline{l}}\cdot u^{%
\overline{l}}\text{ ,}
\end{equation}
\begin{equation}  \label{IX.1.-50}
^{Kr}S=h^u(g)(Kr)h^u=\,^{Kr}S^{\alpha \beta }\cdot e_\alpha \otimes e_\beta
=\,^{Kr}S^{ij}\cdot \partial _i\otimes \partial _j\text{ ,}
\end{equation}
\begin{equation}  \label{IX.1.-51}
^{Kr}S^{\alpha \beta }=h^{\alpha \gamma }\cdot g_{\overline{\gamma }%
\overline{\delta }}\cdot h^{\overline{\delta }\beta }\text{ ,\thinspace
\thinspace \thinspace \thinspace \thinspace \thinspace }^{Kr}S^{ij}=h^{ik}%
\cdot g_{\overline{k}\overline{l}}\cdot h^{\overline{l}j}\text{ ,}
\end{equation}
\[
h^u[g(u)]=[g(u)](h^u)=0\text{ ,\thinspace \thinspace \thinspace \thinspace
\thinspace \thinspace \thinspace }h^u(g)h^u=h^u\text{ ,} 
\]
\begin{equation}  \label{IX.1.-52}
\begin{array}{c}
(Kr) \overline{g}=\frac ke\cdot u\otimes u+u\otimes \,^{Kr}\pi
+\,^{Kr}s\otimes u+\,^{Kr}S\text{ ,} \\ 
(Kr)\overline{g}=g^{\overline{\alpha }\beta }\cdot e_\alpha \otimes e_\beta
=g^{\overline{i}j}\cdot \partial _i\otimes \partial _j\text{ .}
\end{array}
\end{equation}

The corresponding to the Kronecker tensor field mixed tensor field of type 2 
\begin{equation}
\overline{K}r=g(Kr)\overline{g}=g_{\alpha \overline{\gamma }}\cdot g^{%
\overline{\gamma }\beta }\cdot e^\alpha \otimes e_\beta =g_{i\overline{k}%
}\cdot g^{\overline{k}j}\cdot dx^i\otimes \partial _j\text{ }
\label{IX.1.-53}
\end{equation}

\noindent does not appear in general as a Kronecker tensor field.

{\it Special case}: $S=C:f^i\,_j=g_j^i$, \thinspace ($f^\alpha \,_\beta
=g_\beta ^\alpha $): 
\begin{equation}  \label{IX.1.54-59}
\begin{array}{c}
\overline{K}r=g_\beta ^\alpha \cdot e^\beta \otimes e_\alpha =g_j^i\cdot
dx^i\otimes \partial _j\text{ ,\thinspace \thinspace \thinspace } \\ 
k=1 \text{,\thinspace \thinspace \thinspace \thinspace \thinspace \thinspace 
}\varepsilon _{Kr}=\frac 1e\text{,\thinspace \thinspace \thinspace
\thinspace }^{Kr}\pi =0\text{ ,\thinspace \thinspace \thinspace \thinspace
\thinspace \thinspace }^{Kr}s=0\text{ ,\thinspace \thinspace \thinspace
\thinspace \thinspace }^{Kr}S=h^u\text{ ,} \\ 
Kr=\frac 1e\cdot u\otimes g(u)+(h^u)g=\overline{g}(g)\text{ ,\thinspace
\thinspace \thinspace \thinspace \thinspace \thinspace }(Kr)\overline{g}=%
\overline{g}\text{ .}
\end{array}
\end{equation}

The representation of the tensor fields of the type 1 by the use of the
non-isotropic (non-null) contravariant vector field $u$ and its projective
metrics $h^u$ and $h_u$ corresponds in its form to the representation of the
viscosity tensor and the energy-momentum tensors in the continuum media
mechanics in $V_3$- or $V_4$-spaces, where $\varepsilon _G$ is the inner
energy density, $^G\pi $ is the conductive momentum, $e\cdot \,^Gs$ is the
conductive energy flux density and $^GS$ is the stress tensor density. An
analogous interpretation can also be accepted for the projections of the
energy-momentum tensors found by means of the method of Lagrangians with
covariant derivatives (MLCD).

\section{Energy-momentum tensors and the rest mass density}

The covariant Noether identities (generalized covariant Bianchi identities)
can be considered as identities for the components of mixed tensor fields of
second rank of the first type. The second covariant Noether identity \cite
{Manoff-01} 
\[
\overline{\theta }_\alpha \,^\beta -\,_sT_\alpha \,^\beta \equiv \overline{Q}%
\,_\alpha \,^\beta 
\]

\noindent can be written in the form 
\begin{equation}  \label{IX.2.-1}
\theta -\,_sT\equiv Q\text{ ,}
\end{equation}

\noindent where 
\begin{equation}  \label{IX.2.-2}
\begin{array}{c}
\theta = \overline{\theta }_\alpha \,^\beta \cdot e_\beta \otimes e^\alpha =%
\overline{\theta }_i\,^j\cdot \partial _j\otimes dx^i\text{ ,} \\ 
_sT=\,_sT_\alpha \,^\beta \cdot e_\beta \otimes e^\alpha =\,_sT_i\,^j\cdot
\partial _j\otimes dx^i \text{ ,} \\ 
Q=\overline{Q}_\alpha \,^\beta \cdot e_\beta \otimes e^\alpha =\overline{Q}%
_i\,^j\cdot \partial _j\otimes dx^i\text{ ,}
\end{array}
\end{equation}

The tensor of second rank $\theta $ is the generalized canonical
energy-momentum tensor (GC-EMT) of the type 1; the tensor $_sT$ is the
symmetric energy-momentum tensor of Belinfante (S-EMT-B) of the type 1; the
tensor $Q$ is the variational energy-momentum tensor of Euler-Lagrange
(V-EMT-EL) of the type 1.

The second covariant Noether identity for the energy-momentum tensors of the
type 1 is called {\it second covariant Noether identity of type 1.}

By means of the non-isotropic contravariant vector field $u$ and its
corresponding projective metric the energy-momentum tensors can be
represented in an analogous way as the mixed tensor fields of the type 1.

The structure of the generalized canonical energy-momentum tensor and the
symmetric energy-momentum tensor of Belinfante for the metric and non-metric
tensor fields has similar elements and they can be written in the form 
\begin{equation}  \label{IX.2.-3}
\begin{array}{c}
G=\,_kG-L\cdot Kr \text{ ,} \\ 
\theta =\,_k\theta -L\cdot Kr\text{ ,\thinspace \thinspace \thinspace
\thinspace \thinspace \thinspace \thinspace \thinspace \thinspace \thinspace
\thinspace }_sT={\it T}-L\cdot Kr\text{ ,}
\end{array}
\end{equation}

\noindent where 
\begin{equation}  \label{IX.2.-6}
_k\theta =\,_k\overline{\theta }_\alpha \,^\beta \cdot e_\beta \otimes
e^\alpha =\,_k\overline{\theta }_i\,^j\cdot \partial _j\otimes dx^i\text{ .}
\end{equation}

On the analogy of the notions of the continuum media mechanics $_kG$ is
called {\it viscosity tensor field}.

$G$ and $_kG$ can be written by means of $u$, $h^u$ and $h_u$ in the form 
\begin{equation}  \label{IX.2.-9}
\begin{array}{c}
G=\varepsilon _G\cdot u\otimes g(u)+u\otimes g(^G\pi )+\,^Gs\otimes
g(u)+\,(^GS)g \text{ ,} \\ 
_kG=\varepsilon _k\cdot u\otimes g(u)+u\otimes g(^k\pi )+\,^ks\otimes
g(u)+\,(^kS)g\text{ .}
\end{array}
\end{equation}

From the relation (\ref{IX.2.-3}) the relations between the different
projections of $G$ and $_kG$ follow. If we introduce the abbreviation $%
\varepsilon _G=\rho _G$, then 
\begin{equation}
\begin{array}{c}
\varepsilon _k=\rho _G+\frac 1e\cdot L\cdot k\text{ , \thinspace \thinspace
\thinspace \thinspace \thinspace \thinspace \thinspace \thinspace }^k\pi
=\,^G\pi +L\cdot ^{Kr}\pi \text{ ,\thinspace \thinspace \thinspace
\thinspace \thinspace \thinspace \thinspace \thinspace \thinspace }%
^ks=\,^Gs+L\cdot ^{Kr}s\text{ ,} \\ 
^kS=\,^GS+L\cdot ^{Kr}S\text{ ,\thinspace \thinspace \thinspace \thinspace
\thinspace \thinspace \thinspace \thinspace \thinspace \thinspace \thinspace
\thinspace }^G\overline{S}:=\,^kS\text{ \thinspace \thinspace \thinspace
\thinspace ,}
\end{array}
\label{IX.2.-10}
\end{equation}

\noindent and $G$ can be written by means of (\ref{IX.2.-3}) in the form 
\begin{equation}  \label{IX.2.-14}
G=(\rho _G+\frac 1e\cdot L\cdot k)\cdot u\otimes g(u)-L\cdot Kr+u\otimes
g(^k\pi )+\,^ks\otimes g(u)+\,(^kS)g\text{ ,}
\end{equation}

\noindent where 
\[
\rho _G=\frac 1{e^2}\cdot [g(u)](G)(u) 
\]

\noindent is the {\it rest mass density }of the energy-momentum tensor $G$
of the type 1. This type of representation of a given energy-momentum tensor 
$G$ by means of the projective metrics of $u$ and $\rho _G$ is called {\it %
representation of }$G${\it \ by means of the projective metrics of the
contravariant non-isotropic (non-null) vector field }$u${\it \ and the rest
mass density} $\rho _G$.

If the viscosity tensor field $\,_kG=0$ then the energy-momentum tensor $%
G\sim (\theta $,\thinspace $_sT)$ will have the simplest form $G=\,-L\cdot
Kr $. At the same time, the relation $\varepsilon _k=\rho _G+\frac 1e\cdot
L\cdot k=0$ is valid. It leads to the relation $\rho _G=-\frac 1e\cdot
L\cdot k$. For $L=p$, it follows that $p=-\frac 1k\cdot \rho _G\cdot e$,
i.e. the pressure $p$ is proportional to the mass density $\rho _G$ and to
the rest mass energy $\rho _G\cdot e$ respectively. Therefore, if a flow has
no viscosity then the rest mass density (and rest mass energy respectively)
are acting in their motion as a pressure of the flow. Since $k=\frac 1e\cdot
[g(u)](Kr)u=\frac 1e\cdot u^{\overline{i}}\cdot u_{\overline{i}}$, we have 
\[
p=-\frac 1{u^{\overline{i}}\cdot u_{\overline{i}}}\cdot \rho _G\cdot e^2%
\text{ .} 
\]

{\it Special case:} $V_n$-spaces: $S=C$, $k=1$, $u^{\overline{i}}\cdot u_{%
\overline{i}}=u^i\cdot u_i=e$, $p=-\rho _G\cdot e$.

Therefore, the pressure in a viscous-free flow is caused by the existence of
a rest mass density of the flow. If we can measure a pressure in a
viscous-free flow we can conclude that its rest mass density is different
from zero.

There are other possibilities for representation of $G$ by means of $u$ and
its corresponding projective metrics.

If we introduce the abbreviation 
\begin{equation}  \label{IX.2.-15}
p_G=\rho _G\cdot u+\,^G\pi \text{ ,}
\end{equation}

\noindent where $p_G$ is the {\it momentum density} of the energy-momentum
tensor $G$ of the type 1, then $G$ can be written in the form 
\begin{equation}  \label{IX.2.-16}
\begin{array}{c}
G=u\otimes g(\rho _G\cdot u+\,^G\pi )+\,^Gs\otimes g(u)+(^GS)g \text{ ,} \\ 
G=u\otimes g(p_G)+\,^Gs\otimes g(u)+(^GS)g\text{ .}
\end{array}
\end{equation}

The representation of $G$ by means of the last relation is called {\it %
representation of }$G${\it \ by means of the projective metrics of the
contravariant non-isotropic contravariant vector field }$u${\it \ and the
momentum density} $p_G$.

By the use of the relations 
\begin{equation}  \label{IX.2.-17}
g(^G\pi ,u)=0\text{ ,\thinspace \thinspace \thinspace \thinspace \thinspace
\thinspace \thinspace \thinspace \thinspace }(^GS)[g(u)]=0\text{ ,}
\end{equation}

\noindent valid (because of their constructions) for every energy-momentum
tensor $G$ and the definitions 
\begin{equation}  \label{IX.2.-18}
e_G=G(u)=(G)(u)=e\cdot (\rho _G\cdot u+\,^Gs)\text{ ,\thinspace \thinspace
\thinspace \thinspace }g(u,u)=e\neq 0\text{ ,}
\end{equation}
where $e_G$ is the {\it energy flux density} of the energy-momentum tensor $%
G $ of the type 1, the tensor field $G$ can be written in the form 
\begin{equation}  \label{IX.2.-19}
\begin{array}{c}
G=(\rho _G\cdot u+\,^Gs)\otimes g(u)+u\otimes g(^G\pi )+(^GS)g \text{ ,} \\ 
G=\frac 1e\cdot e_G\otimes g(u)+u\otimes g(^G\pi )+(^GS)g\text{ .}
\end{array}
\end{equation}

The representation of $G$ by means of the last expression is called {\it %
representation of }$G${\it \ by means of the projective metrics of the
contravariant non-isotropic vector field }$u${\it \ and the energy flux
density} $e_G$.

The generalized canonical energy-momentum tensor $\theta $ can be
represented, in accordance with the above described procedure, by the use of
the projective metrics of $u$ and the rest mass density $\rho _\theta $%
\begin{equation}  \label{IX.2.-20}
\theta =\,_k\theta -L\cdot Kr\text{ ,\thinspace \thinspace \thinspace
\thinspace \thinspace \thinspace \thinspace }_k\theta =\theta +L\cdot Kr%
\text{ ,}
\end{equation}
\begin{equation}  \label{IX.2.-21}
\theta =(\rho _\theta +\frac 1e\cdot L\cdot k)\cdot u\otimes g(u)-L\cdot
Kr+u\otimes g(^\theta \overline{\pi })+\,^\theta \overline{s}\otimes
g(u)+(^\theta \overline{S})g\text{ ,}
\end{equation}

\noindent where 
\begin{equation}  \label{IX.2.-23}
_k\theta =\,_k\overline{\theta }_\alpha \,^\beta \cdot e_\beta \otimes
e^\alpha \text{ ,\thinspace \thinspace \thinspace \thinspace \thinspace }_k%
\overline{\theta }_\alpha \,^\beta =\overline{t}_\alpha \,^\beta -K_\alpha
\,^\beta -\overline{W}_\alpha \,^{\beta \gamma }\,_\gamma +L\cdot g_\alpha
^\beta \text{ ,}
\end{equation}
\begin{equation}  \label{IX.2.-24}
\rho _\theta =\frac 1{e^2}\cdot [g(u)](\theta )(u)\text{ , \thinspace
\thinspace \thinspace \thinspace }k=\frac 1e\cdot [g(u)](Kr)(u)\text{ ,}
\end{equation}
\begin{equation}  \label{IX.2.-25}
\begin{array}{c}
\rho _\theta =\frac 1{e^2}\cdot g_{\overline{\alpha }\overline{\beta }}\cdot
u^\beta \cdot \overline{\theta }_\gamma \,^\alpha \cdot u^{\overline{\gamma }%
}=\frac 1{e^2}\cdot \overline{\theta }_\gamma \,^\alpha \cdot u_{\overline{%
\alpha }}\cdot u^{\overline{\gamma }}= \\ 
=\frac 1{e^2}\cdot g_{\overline{i}\overline{j}}\cdot u^j\cdot \overline{%
\theta }_k\,^i\cdot u^{\overline{k}}=\frac 1{e^2}\cdot \overline{\theta }%
_k\,^i\cdot u_{\overline{i}}\cdot u^{\overline{k}}\text{ ,}
\end{array}
\end{equation}
\begin{equation}  \label{IX.2.-26}
\varepsilon _{_k\theta }=\rho _\theta +\frac 1e\cdot L\cdot k\text{ ,}
\end{equation}
\begin{equation}  \label{IX.2.-27}
^\theta \overline{\pi }=\frac 1e\cdot [g(u)](_k\theta )h^u=\,^\theta 
\overline{\pi }^\alpha \cdot e_\alpha =\,^\theta \overline{\pi }^i\cdot
\partial _i\text{ ,}
\end{equation}
\begin{equation}  \label{IX.2.-28}
^\theta \overline{\pi }^\alpha =\frac 1e\cdot \,_k\overline{\theta }_\gamma
\,^\beta \cdot u_{\overline{\beta }}\cdot h^{\overline{\gamma }\alpha }\text{
,\thinspace \thinspace \thinspace \thinspace }^\theta \overline{\pi }%
^i=\frac 1e\cdot \,_k\overline{\theta }_l\,^k\cdot u_{\overline{k}}\cdot h^{%
\overline{l}i}\text{ ,}
\end{equation}
\begin{equation}  \label{IX.2.-29}
^\theta \overline{s}=\frac 1e\cdot h^u(g)(_k\theta )(u)=\,^\theta \overline{s%
}^\alpha \cdot e_\alpha =\,^\theta \overline{s}^i\cdot \partial _i\text{ ,}
\end{equation}
\begin{equation}  \label{IX.2.-30}
^\theta \overline{s}^\alpha =\frac 1e\cdot h^{\alpha \beta }\cdot g_{%
\overline{\beta }\overline{\gamma }}\cdot \,_k\overline{\theta }_\delta
\,^\gamma \cdot u^{\overline{\delta }}\text{ ,\thinspace \thinspace
\thinspace \thinspace \thinspace }^\theta \overline{s}^i=\frac 1e\cdot
h^{ij}\cdot g_{\overline{j}\overline{k}}\cdot \,_k\overline{\theta }%
_l\,^k\cdot u^{\overline{l}}\text{ ,}
\end{equation}
\begin{equation}  \label{IX.2.-31}
^\theta \overline{S}=h^u(g)(_k\theta )h^u=\,^\theta \overline{S}\,^{\alpha
\beta }\cdot e_\alpha \otimes e_\beta =\,^\theta \overline{S}\,^{ij}\cdot
\partial _i\otimes \partial _j\text{ ,}
\end{equation}
\begin{equation}  \label{IX.2.-32}
^\theta \overline{S}\,^{\alpha \beta }=h^{\alpha \gamma }\cdot g_{\overline{%
\gamma }\overline{\delta }}\cdot \,_k\overline{\theta }_\kappa \,^\delta
\cdot h^{\overline{\kappa }\beta }\text{ ,\thinspace \thinspace \thinspace
\thinspace \thinspace \thinspace \thinspace \thinspace }^\theta \overline{S}%
\,^{ij}=h^{ik}\cdot g_{\overline{k}\overline{l}}\cdot \,_k\overline{\theta }%
_m\,^l\cdot h^{\overline{m}j}\text{ ,}
\end{equation}
\begin{equation}  \label{IX.2.-33}
\begin{array}{c}
(\theta ) \overline{g}=(\rho _\theta +\frac 1e\cdot L\cdot k)\cdot u\otimes
u-L\cdot Kr(\overline{g})+u\otimes \,^\theta \overline{\pi }+\,^\theta 
\overline{s}\otimes u+\,^\theta \overline{S}= \\ 
=\theta ^{\alpha \beta }\cdot e_\alpha \otimes e_\beta =\theta ^{ij}\cdot
\partial _i\otimes \partial _j\text{ ,}
\end{array}
\end{equation}
\begin{equation}  \label{IX.2.-34}
\theta ^{\alpha \beta }=g^{\beta \overline{\gamma }}\cdot \overline{\theta }%
_\gamma \,^\alpha =\overline{\theta }_\gamma \,^\alpha \cdot g^{\beta 
\overline{\gamma }}\text{ ,\thinspace \thinspace \thinspace \thinspace
\thinspace }\theta ^{ij}=\overline{\theta }_k\,^i\cdot g^{\overline{k}j}%
\text{ ,}
\end{equation}
\begin{equation}  \label{IX.2.-35}
\begin{array}{c}
g(\theta )=(\rho _\theta +\frac 1e\cdot L\cdot k)\cdot g(u)\otimes
g(u)-L\cdot g(Kr)+g(u)\otimes g(^\theta \overline{\pi })+ \\ 
+g(^\theta \overline{s})\otimes g(u)+g(^\theta \overline{S})g\text{ ,}
\end{array}
\end{equation}
\begin{equation}  \label{IX.2.-36}
\begin{array}{c}
g(\theta )=\theta _{\alpha \beta }\cdot e^\alpha \otimes e^\beta =\theta
_{ij}\cdot dx^i\otimes dx^j \text{ ,} \\ 
\theta _{\alpha \beta }=g_{\alpha \overline{\gamma }}\cdot \overline{\theta }%
_\beta \,^\gamma \text{ ,\thinspace \thinspace \thinspace \thinspace
\thinspace \thinspace \thinspace }\theta _{ij}=g_{i\overline{k}}\cdot 
\overline{\theta }_j\text{ }^k\text{ ,}
\end{array}
\end{equation}
\begin{equation}  \label{IX.2.-40}
\begin{array}{c}
g(Kr)=g_{\alpha \overline{\beta }}\cdot e^\alpha \otimes e^\beta =g_{i%
\overline{j}}\cdot dx^i\otimes dx^j\text{ ,\thinspace \thinspace \thinspace
\thinspace }g_{\alpha \overline{\beta }}=g_{\alpha \gamma }\cdot f^\gamma
\,_\beta =g_{\gamma \alpha }\cdot f^\gamma \,_\beta =g_{\overline{\beta }%
\alpha }\text{ ,} \\ 
g_{i\overline{j}}=g_{\overline{j}i}\text{ ,\thinspace \thinspace \thinspace
\thinspace }(Kr)\overline{g}=Kr(\overline{g})=g^{\overline{\alpha }\beta
}\cdot e_\alpha \otimes e_\beta =g^{\overline{i}j}\cdot \partial _i\otimes
\partial _j\text{ .}
\end{array}
\end{equation}

In a co-ordinate basis $\theta $, $(\theta )\overline{g}$ and $g(\theta )$
can be represented in the forms 
\begin{equation}  \label{IX.2.-46}
\overline{\theta }_i\,^j=(\rho _\theta +\frac 1e\cdot L\cdot k)\cdot
u_i\cdot u^j-L\cdot g_i^j+\,^\theta \overline{\pi }_i\cdot u^j+u_i\cdot
\,^\theta \overline{s}^j+g_{i\overline{k}}\cdot \,^\theta \overline{S}\,^{jk}%
\text{ ,}
\end{equation}
\begin{equation}  \label{IX.2.-47}
\theta ^{ij}=\overline{\theta }_k\,^i\cdot g^{\overline{k}j}=(\rho _\theta
+\frac 1e\cdot L\cdot k)\cdot u^i\cdot u^j-L\cdot g^{\overline{i}j}+u^i\cdot
\,^\theta \overline{\pi }^j+\,^\theta \overline{s}^i\cdot u^j+\,^\theta 
\overline{S}\,^{ij}\text{ ,}
\end{equation}
\begin{equation}  \label{IX.2.-48}
\theta _{ij}=g_{i\overline{k}}\cdot \overline{\theta }_j\,^k=(\rho _\theta
+\frac 1e\cdot L\cdot k)\cdot u_i\cdot u_j-L\cdot g_{i\overline{j}}+u_i\cdot
\,^\theta \overline{\pi }_j+\,^\theta \overline{s}_i\cdot u_j+\,g_{i%
\overline{k}}\cdot ^\theta \overline{S}\,^{kl}\cdot g_{\overline{l}j}\text{ ,%
}
\end{equation}

\noindent where 
\begin{equation}  \label{IX.2.-49}
u_i=g_{i\overline{j}}\cdot u^j\,\,\,\,\text{,\thinspace \thinspace
\thinspace \thinspace }^\theta \overline{\pi }_i=g_{i\overline{k}}\cdot
\,^\theta \overline{\pi }^k\text{ ,\thinspace \thinspace \thinspace
\thinspace \thinspace }^\theta \overline{s}_i=g_{i\overline{l}}\cdot
\,^\theta \overline{s}^l\text{ ,\thinspace \thinspace \thinspace }^\theta 
\overline{S}_{ij}=g_{i\overline{k}}\cdot \,^\theta \overline{S}\,^{kl}\cdot
g_{\overline{l}j}\text{ . }
\end{equation}

The symmetric energy-momentum tensor of Belinfante $_sT$ can be represented
in an analogous way by the use of the projective metrics $h^u$ and $h_u$ and
the rest mass density $\rho _T$ in the form 
\begin{equation}
_sT=(\rho _T+\frac 1e\cdot L\cdot k)\cdot u\otimes g(u)-L\cdot Kr+u\otimes
g(^T\overline{\pi })+\,^T\overline{s}\otimes g(u)+(^T\overline{S})g\text{ ,}
\label{IX.2.-51}
\end{equation}

\noindent where 
\begin{equation}  \label{IX.2.-52}
\begin{array}{c}
_sT=\,_sT_\alpha \,^\beta \cdot e_\beta \otimes e^\alpha =\,_sT_i\,^j\cdot
\partial _j\otimes dx^i \text{ ,} \\ 
_sT=\,_{sk}T-L\cdot Kr\text{ ,\thinspace \thinspace \thinspace \thinspace
\thinspace \thinspace \thinspace }_{sk}T=\,_sT+L\cdot Kr={\it T}\text{ ,}
\end{array}
\end{equation}
\begin{equation}  \label{IX.2.-53}
\varepsilon _T=\rho _T+\frac 1e\cdot L\cdot k\text{ ,}
\end{equation}
\begin{equation}  \label{IX.2.-54}
\begin{array}{c}
\rho _T=\frac 1{e^2}\cdot [g(u)](_sT)(u)=\frac 1{e^2}\cdot g_{\overline{%
\alpha }\overline{\beta }}\cdot u^\beta \cdot _sT_\gamma \,^\alpha \cdot u^{%
\overline{\gamma }}=\frac 1{e^2}\cdot _sT_\gamma \,^\alpha \cdot u_{%
\overline{\alpha }}\cdot u^{\overline{\gamma }}= \\ 
=\frac 1{e^2}\cdot g_{\overline{i}\overline{j}}\cdot u^j\cdot _sT_k\,^i\cdot
u^{\overline{k}}=\frac 1{e^2}\cdot \,_sT_k\,^i\cdot u_{\overline{i}}\cdot u^{%
\overline{k}}\text{ ,}
\end{array}
\end{equation}
\begin{equation}  \label{IX.2.-55}
^T\overline{\pi }=\frac 1e\cdot [g(u)]({\it T})h^u=\frac 1e\cdot
[g(u)](_{sk}T)h^u=\,^T\overline{\pi }^\alpha \cdot e_\alpha =\,^T\overline{%
\pi }^i\cdot \partial _i\text{ ,}
\end{equation}
\begin{equation}  \label{IX.2.-56}
^T\overline{\pi }^\alpha =\frac 1e\cdot {\it T}_\gamma \,^\beta \cdot u_{%
\overline{\beta }}\cdot h^{\overline{\gamma }\alpha }\text{ ,\thinspace
\thinspace \thinspace \thinspace }^T\overline{\pi }^i=\frac 1e\cdot {\it T}%
_l\,^k\cdot u_{\overline{k}}\cdot h^{\overline{l}i}\text{ ,}
\end{equation}
\begin{equation}  \label{IX.2.-57}
^T\overline{s}=\frac 1e\cdot h^u(g)({\it T})(u)=\,^T\overline{s}^\alpha
\cdot e_\alpha =\,^T\overline{s}^i\cdot \partial _i\text{ ,}
\end{equation}
\begin{equation}  \label{IX.2.-58}
^T\overline{s}\,^\alpha =\frac 1e\cdot h^{\alpha \beta }\cdot g_{\overline{%
\beta }\overline{\gamma }}\cdot {\it T}_\delta \,^\gamma \cdot u^{\overline{%
\delta }}\text{ ,\thinspace \thinspace \thinspace \thinspace \thinspace }^T%
\overline{s}^i=\frac 1e\cdot h^{ij}\cdot g_{\overline{j}\overline{k}}\cdot 
{\it T}_l\,^k\cdot u^{\overline{l}}\text{ ,}
\end{equation}
\begin{equation}  \label{IX.2.-59}
^T\overline{S}=h^u(g)({\it T})h^u=\,^T\overline{S}\,^{\alpha \beta }\cdot
e_\alpha \otimes e_\beta =\,^T\overline{S}\,^{ij}\cdot \partial _i\otimes
\partial _j\text{ ,}
\end{equation}
\begin{equation}  \label{IX.2.-60}
^T\overline{S}\,^{\alpha \beta }=h^{\alpha \gamma }\cdot g_{\overline{\gamma 
}\overline{\delta }}\cdot {\it T}_\kappa \,^\delta \cdot h^{\overline{\kappa 
}\beta }\text{ ,\thinspace \thinspace \thinspace \thinspace \thinspace
\thinspace \thinspace \thinspace }^T\overline{S}\,^{ij}=h^{ik}\cdot g_{%
\overline{k}\overline{l}}\cdot {\it T}_m\,^l\cdot h^{\overline{m}j}\text{ ,}
\end{equation}
\begin{equation}  \label{IX.2.-61}
\begin{array}{c}
(_sT) \overline{g}=(\rho _T+\frac 1e\cdot L\cdot k)\cdot u\otimes u-L\cdot
Kr(\overline{g})+u\otimes \,^T\overline{\pi }+\,^T\overline{s}\otimes u+\,^T%
\overline{S}= \\ 
=\,_sT^{\alpha \beta }\cdot e_\alpha \otimes e_\beta =\,_sT^{ij}\cdot
\partial _i\otimes \partial _j\text{ ,}
\end{array}
\end{equation}
\begin{equation}  \label{IX.2.-62}
_sT^{\alpha \beta }=g^{\beta \overline{\gamma }}\cdot \,_sT_\gamma \,^\alpha
=\,_sT_\gamma \,^\alpha \cdot g^{\beta \overline{\gamma }}\text{ ,\thinspace
\thinspace \thinspace \thinspace \thinspace }_sT^{ij}=\,_sT_k\,^i\cdot g^{%
\overline{k}j}\text{ ,}
\end{equation}
\begin{equation}  \label{IX.2.-63}
\begin{array}{c}
g(_sT)=(\rho _T+\frac 1e\cdot L\cdot k)\cdot g(u)\otimes g(u)-L\cdot
g(Kr)+g(u)\otimes g(^T \overline{\pi })+ \\ 
+g(^T\overline{s})\otimes g(u)+g(^T\overline{S})g\text{ ,}
\end{array}
\end{equation}
\begin{equation}  \label{IX.2.-64}
\begin{array}{c}
g(_sT)=\,_sT_{\alpha \beta }\cdot e^\alpha \otimes e^\beta =\,_sT_{ij}\cdot
dx^i\otimes dx^j \text{ ,} \\ 
_sT_{\alpha \beta }=g_{\alpha \overline{\gamma }}\cdot \,_sT_\beta \,^\gamma 
\text{ ,\thinspace \thinspace \thinspace \thinspace \thinspace \thinspace
\thinspace }_sT_{ij}=g_{i\overline{k}}\cdot \,_sT_j\text{ }^k\text{ .}
\end{array}
\end{equation}

In a co-ordinate basis $_sT$, $(_sT)\overline{g}$ and $g(_sT)$ can be
represented in the forms 
\begin{equation}  \label{IX.2.-68}
_sT_i\,^j=(\rho _T+\frac 1e\cdot L\cdot k)\cdot u_i\cdot u^j-L\cdot
g_i^j+\,^T\overline{\pi }_i\cdot u^j+u_i\cdot ^T\overline{s}^j+g_{i\overline{%
k}}\cdot \,^T\overline{S}\,^{jk}\text{ ,}
\end{equation}
\begin{equation}  \label{IX.2.-69}
_sT^{ij}=\,_sT_k\,^i\cdot g^{\overline{k}j}=(\rho _T+\frac 1e\cdot L\cdot
k)\cdot u^i\cdot u^j-L\cdot g^{\overline{i}j}+u^i\cdot \,^T\overline{\pi }%
^j+\,^T\overline{s}^i\cdot u^j+\,^T\overline{S}\,^{ij}\text{ ,}
\end{equation}
\begin{equation}  \label{IX.2.-70}
_sT_{ij}=g_{i\overline{k}}\cdot \,_sT_j\,^k=(\rho _T+\frac 1e\cdot L\cdot
k)\cdot u_i\cdot u_j-L\cdot g_{i\overline{j}}+u_i\cdot \,^T\overline{\pi }%
_j+\,^T\overline{s}_i\cdot u_j+\,g_{i\overline{k}}\cdot \,^T\overline{S}%
\,^{kl}\cdot g_{\overline{l}j}\text{ ,}
\end{equation}

\noindent where 
\[
^T\overline{\pi }_i=g_{i\overline{k}}\cdot \,^T\overline{\pi }^k\text{
,\thinspace \thinspace \thinspace \thinspace \thinspace }^T\overline{s}%
_i=g_{i\overline{l}}\cdot \,^T\overline{s}^l\text{ ,\thinspace \thinspace
\thinspace }^T\overline{S}_{ij}=g_{i\overline{k}}\cdot \,^T\overline{S}%
\,^{kl}\cdot g_{\overline{l}j}\text{ . } 
\]

The variational energy-momentum tensor of Euler-Lagrange $Q$ can be
represented in the standard manner by the use of the projective metrics $h^u$%
, $h_u$ and the rest mass density $\rho _Q$ in the form

\begin{equation}  \label{IX.2.-71}
Q=-\rho _Q\cdot u\otimes g(u)-u\otimes g(^Q\pi )-\,^Qs\otimes g(u)-(^QS)g%
\text{ ,}
\end{equation}

\noindent where 
\begin{equation}  \label{IX.2.-72}
\rho _Q=-\,\,\frac 1{e^2}\cdot [g(u)](Q)(u)\text{ , }
\end{equation}
\begin{equation}  \label{IX.2.-72a}
\begin{array}{c}
\rho _Q=-\,\,\frac 1{e^2}\cdot g_{\overline{\alpha }\overline{\beta }}\cdot
u^\beta \cdot \overline{Q}_\gamma \,^\alpha \cdot u^{\overline{\gamma }%
}=-\,\frac 1{e^2}\cdot \overline{Q}_\gamma \,^\alpha \cdot u_{\overline{%
\alpha }}\cdot u^{\overline{\gamma }}= \\ 
=-\,\frac 1{e^2}\cdot g_{\overline{i}\overline{j}}\cdot u^j\cdot \overline{Q}%
_k\,^i\cdot u^{\overline{k}}=-\,\frac 1{e^2}\cdot \overline{Q}_k\,^i\cdot u_{%
\overline{i}}\cdot u^{\overline{k}}\text{ ,}
\end{array}
\end{equation}
\begin{equation}  \label{IX.2.-73}
^Q\pi =-\frac 1e\cdot [g(u)](Q)h^u=\,^Q\pi ^\alpha \cdot e_\alpha =\,^Q\pi
^i\cdot \partial _i\text{ ,}
\end{equation}
\begin{equation}  \label{IX.2.-74}
^Q\pi ^\alpha =-\frac 1e\cdot \overline{Q}_\gamma \,^\beta \cdot u_{%
\overline{\beta }}\cdot h^{\overline{\gamma }\alpha }\text{ ,\thinspace
\thinspace \thinspace \thinspace }^Q\pi ^i=-\frac 1e\cdot \overline{Q}%
_l\,^k\cdot u_{\overline{k}}\cdot h^{\overline{l}i}\text{ ,}
\end{equation}
\begin{equation}  \label{IX.2.-75}
^Qs=-\frac 1e\cdot h^u(g)(Q)(u)=\,^Qs^\alpha \cdot e_\alpha =\,^Qs^i\cdot
\partial _i\text{ ,}
\end{equation}
\begin{equation}  \label{IX.2.-76}
^Qs^\alpha =-\frac 1e\cdot h^{\alpha \beta }\cdot g_{\overline{\beta }%
\overline{\gamma }}\cdot \overline{Q}_\delta \,^\gamma \cdot u^{\overline{%
\delta }}\text{ ,\thinspace \thinspace \thinspace \thinspace \thinspace }%
^Qs^i=-\frac 1e\cdot h^{ij}\cdot g_{\overline{j}\overline{k}}\cdot \overline{%
Q}_l\,^k\cdot u^{\overline{l}}\text{ ,}
\end{equation}
\begin{equation}  \label{IX.2.-77}
^QS=-h^u(g)(Q)h^u=\,^QS^{\alpha \beta }\cdot e_\alpha \otimes e_\beta
=\,^QS^{ij}\cdot \partial _i\otimes \partial _j\text{ ,}
\end{equation}
\begin{equation}  \label{IX.2.-78}
^QS^{\alpha \beta }=-h^{\alpha \gamma }\cdot g_{\overline{\gamma }\overline{%
\delta }}\cdot \overline{Q}_\kappa \,^\delta \cdot h^{\overline{\kappa }%
\beta }\text{ ,\thinspace \thinspace \thinspace \thinspace \thinspace
\thinspace \thinspace \thinspace }^QS^{ij}=-h^{ik}\cdot g_{\overline{k}%
\overline{l}}\cdot \overline{Q}_m\,^l\cdot h^{\overline{m}j}\text{ ,}
\end{equation}
\begin{equation}  \label{IX.2.-82}
\begin{array}{c}
(Q) \overline{g}=-\,\rho _Q\cdot u\otimes u-u\otimes \,^Q\pi -\,^Qs\otimes
u-\,^QS= \\ 
=Q^{\alpha \beta }\cdot e_\alpha \otimes e_\beta =Q^{ij}\cdot \partial
_i\otimes \partial _j\text{ ,}
\end{array}
\end{equation}
\[
Q^{\alpha \beta }=g^{\beta \overline{\gamma }}\cdot \overline{Q}_\gamma
\,^\alpha =\overline{Q}_\gamma \,^\alpha \cdot g^{\beta \overline{\gamma }}%
\text{ ,\thinspace \thinspace \thinspace \thinspace \thinspace }Q^{ij}=%
\overline{Q}_k\,^i\cdot g^{\overline{k}j}\text{ ,} 
\]
\begin{equation}  \label{IX.2.-79}
g(Q)=-\rho _Q\cdot g(u)\otimes g(u)-g(u)\otimes g(^Q\pi )-g(^Qs)\otimes
g(u)-g(^QS)g\text{ ,}
\end{equation}
\begin{equation}  \label{IX.2.-81}
g(Q)=Q_{\alpha \beta }\cdot e^\alpha \otimes e^\beta =Q_{ij}\cdot
dx^i\otimes dx^j\text{ ,\thinspace \thinspace \thinspace }Q_{\alpha \beta
}=g_{\alpha \overline{\gamma }}\cdot \overline{Q}_\beta \,^\gamma \text{
,\thinspace \thinspace \thinspace \thinspace \thinspace \thinspace
\thinspace }Q_{ij}=g_{i\overline{k}}\cdot \overline{Q}_j\text{ }^k\text{ .}
\end{equation}

In a co-ordinate basis $Q$, $(Q)\overline{g}$ and $g(Q)$ can be represented
in the forms 
\begin{equation}  \label{IX.2.-86}
\overline{Q}_i\,^j=-\rho _Q\cdot u_i\cdot u^j-\,^Q\pi _i\cdot u^j-u_i\cdot
\,^Qs^j-g_{i\overline{k}}\cdot \,^QS^{jk}\text{ ,}
\end{equation}

\begin{equation}  \label{IX.2.-87}
Q^{ij}=\overline{Q}_k\,^i\cdot g^{\overline{k}j}=-\rho _Q\cdot u^i\cdot
u^j-u^i\cdot \,^Q\pi ^j-\,^Qs^i\cdot u^j-\,^QS^{ij}\text{ ,}
\end{equation}

\begin{equation}  \label{IX.2.-88}
Q_{ij}=g_{i\overline{k}}\cdot \overline{Q}_j\,^k=-\rho _Q\cdot u_i\cdot
u_j-u_i\cdot \,^Q\pi _j-\,^Qs_i\cdot u_j-\,g_{i\overline{k}}\cdot
\,^QS^{kl}\cdot g_{\overline{l}j}\text{ ,}
\end{equation}

\noindent where 
\[
\text{\thinspace }^Q\pi _i=g_{i\overline{k}}\cdot \,^Q\pi ^k\text{
,\thinspace \thinspace \thinspace \thinspace \thinspace }^Qs_i=g_{i\overline{%
l}}\cdot \,^Qs^l\text{ ,\thinspace \thinspace \thinspace }^QS_{ij}=g_{i%
\overline{k}}\cdot \,^QS^{kl}\cdot g_{\overline{l}j}\text{ . } 
\]

The introduced abbreviations for the different projections of the
energy-momentum tensors have their analogous forms in $V_3$- and $V_4$%
-spaces, where their physical interpretations have been proposed \cite
{Landau-1}, \cite{Landau-2}, \cite{Schmutzer-1} (S.383-385), \cite
{Schmutzer-2}. The stress tensor in $V_3$-spaces has been generalized to the
energy-momentum tensor $_sT$ \thinspace \thinspace in $V_4$-spaces. The
viscosity stress tensor $_{sk}T$ appears as the tensor ${\it T}$ in the
structure of the symmetric energy-momentum tensor of Belinfante $_sT$.

On the analogy of the physical interpretation of the different projections,
the following definitions can be proposed for the quantities in the
representations of the different energy-momentum tensors:

{\it A. Generalized canonical energy-momentum tensor of the type 1} ........ 
$\theta $

(a) Generalized viscous energy-momentum tensor of the type 1 ............ $%
_k\theta $

(b) Rest mass density of the generalized canonical

energy-momentum tensor $\theta $
....................................................................$\rho
_\theta $

(c) Conductive momentum density of the generalized canonical

energy-momentum tensor $\theta $
...................................................................$^\theta
\pi $

(d) Conductive energy flux density of the generalized canonical

energy-momentum tensor $\theta $
..................................................................$e\cdot
\,^\theta s$

(e) Stress tensor of the generalized canonical

energy-momentum tensor $\theta $
....................................................................$^\theta
S$

{\it B. Symmetric energy-momentum tensor of Belinfante of the type 1} ..... $%
_sT$

(a) Symmetric viscous energy-momentum tensor of the type 1 ...............$%
...{\it T}$

(b) Rest mass density of the symmetric energy-momentum

tensor of Belinfante $_sT$
...........................................................................$%
\rho _T$

(c) Conductive momentum density of the symmetric

energy-momentum tensor of Belinfante $_sT$
..............................................$^T\pi $

(d) Conductive energy flux density of the symmetric

energy-momentum tensor of Belinfante $_sT$
............................................$e\cdot \,^Ts$

(e) Stress tensor of the symmetric energy-momentum

tensor of Belinfante $_sT$
..........................................................................$%
^TS$

{\it C. Variational (active) energy-momentum tensor of Euler-Lagrange.} .. $%
Q $

(a) Rest mass density of the variational energy-momentum

tensor of Euler-Lagrange $Q$
.................................................................... $\rho
_Q $

(b) Conductive momentum density of the variational

energy-momentum tensor of Euler-Lagrange $Q$
.........................................$^Q\pi $

(c) Conductive energy flux density of the variational

energy-momentum tensor of Euler-Lagrange $Q$
....................................... $e\cdot \,^Qs$

(d) Stress tensor of the variational energy-momentum

tensor of Euler-Lagrange $Q$
....................................................................$^QS$

The projections of the energy-momentum tensors have properties which are due
to their construction, the orthogonality of the projective metrics $h_u$ and 
$h^u$ correspondingly, and to the vector fields $u$ and $g(u)$ [$h_u(u)=0$%
,\thinspace \thinspace \thinspace \thinspace \thinspace \thinspace $%
h^u[g(u)]=0$] 
\begin{equation}
g(u,^\theta \overline{\pi })=g(^\theta \overline{\pi },u)=0\text{
,\thinspace \thinspace \thinspace \thinspace }g(u,\,^T\overline{\pi })=0%
\text{ ,\thinspace \thinspace \thinspace \thinspace }g(u,\,^Q\pi )=0\text{ ,}
\label{IX.2.-89}
\end{equation}
\begin{equation}
g(u,^\theta \overline{s})=g(^\theta \overline{s},u)=0\text{ ,\thinspace
\thinspace \thinspace \thinspace }g(u,^T\overline{s})=0\text{ ,\thinspace
\thinspace \thinspace \thinspace }g(u,^Qs)=0\text{ ,}  \label{IX.2.-90}
\end{equation}
\begin{equation}
\begin{array}{c}
g(u)(^\theta \overline{S})=0\text{, }(^\theta \overline{S})g(u)=0\text{
,\thinspace \thinspace \thinspace \thinspace }g(u)(^T\overline{S})=0\text{
,\thinspace \thinspace \thinspace \thinspace \thinspace }(^T\overline{S}%
)g(u)=0\text{ ,\thinspace \thinspace } \\ 
\text{\thinspace \thinspace }g(u)(^QS)=0\text{ , \thinspace \thinspace
\thinspace \thinspace \thinspace \thinspace }(^QS)g(u)=0\text{ .\thinspace
\thinspace }
\end{array}
\label{IX.2.-91}
\end{equation}

From the properties of the different projections, it follows that the
conductive momentum density $\pi $ (or $\overline{\pi }$) is a contravariant
vector field orthogonal to the vector field $u$. The conductive energy flux
density $e\cdot s$ (or $e\cdot \overline{s}$) is also a contravariant vector
field orthogonal to $u$. The stress tensor $S$ (or $\overline{S}$) is
orthogonal to $u$ independently of the side of the projection by means of
the vector field $u$.

The second covariant Noether identity $\theta -\,_sT\equiv Q$ can be written
by the use of the projections of the energy-momentum tensors in the form 
\begin{equation}  \label{IX.2.-92}
\begin{array}{c}
(\rho _\theta -\rho _T+\rho _Q)\cdot u\otimes g(u)+u\otimes g(^\theta 
\overline{\pi }-\,^T\overline{\pi }+\,^Q\pi )+ \\ 
+\,(^\theta \overline{s}-\,^T\overline{s}+\,^Qs)\otimes g(u)+(^\theta 
\overline{S}-\,^T\overline{S}+\,^QS)g\equiv 0\text{ .}
\end{array}
\end{equation}

After contraction of the last expression consistently with $u$ and $%
\overline{g}$ and taking into account the properties (\ref{IX.2.-89}) $\div $
(\ref{IX.2.-91}) the second covariant Noether identity disintegrates in
identities for the different projections of the energy-momentum tensors 
\begin{equation}  \label{IX.2.-94}
\rho _\theta \equiv \rho _T-\rho _Q\text{ ,\thinspace \thinspace \thinspace
\thinspace }^\theta \overline{\pi }\equiv \,^T\overline{\pi }-\,^Q\pi \text{
,\thinspace \thinspace \thinspace \thinspace }^\theta \overline{s}\equiv \,^T%
\overline{s}-\,^Qs\text{ , }^\theta \overline{S}\equiv \,^T\overline{S}-\,^QS%
\text{ .}
\end{equation}

If the covariant Euler-Lagrange equations of the type $\delta _vL/\delta
V^A\,_B=0$ are fulfilled \cite{Manoff-01} for the non-metric tensor fields
of a Lagrangian system and $_gQ=0$, then the variational energy-momentum of
Euler-Lagrange $Q=\,_vQ+\,_gQ$ is equal to zero. This fact leads to
vanishing the invariant projections of $Q$ ($\rho _Q=0$, $^Q\pi =0$, $^Qs=0$%
, $^QS=0$). The equality which follows between $\theta $ and $_sT$ has as
corollaries the identities 
\begin{equation}
\rho _\theta \equiv \rho _T\text{ ,\thinspace \thinspace \thinspace
\thinspace \thinspace \thinspace \thinspace \thinspace \thinspace \thinspace
\thinspace \thinspace \thinspace \thinspace \thinspace \thinspace }^\theta 
\overline{\pi }\equiv \,^T\overline{\pi }\text{ ,\thinspace \thinspace
\thinspace \thinspace \thinspace \thinspace \thinspace \thinspace \thinspace
\thinspace \thinspace \thinspace \thinspace \thinspace \thinspace \thinspace 
}^\theta \overline{s}\equiv \,^T\overline{s}\text{ ,\thinspace \thinspace
\thinspace \thinspace \thinspace \thinspace \thinspace \thinspace \thinspace
\thinspace \thinspace \thinspace \thinspace \thinspace \thinspace \thinspace
\thinspace \thinspace \thinspace \thinspace }^\theta \overline{S}\equiv \,^T%
\overline{S}\text{ .\thinspace }  \label{IX.2.-95}
\end{equation}

From the first identity ($\rho _\theta \equiv \rho _T$) and the identity (%
\ref{IX.2.-94}) for $\rho $ , it follows that the covariant Euler-Lagrange
equations of the type $\delta _vL/\delta V^A\,_B=0$ for non-metric fields $V$
and $_gQ=0$ appear as sufficient conditions for the unique determination of
the notion of rest mass density $\rho $ for a given Lagrangian system.

\begin{proposition}
The necessary and sufficient condition for the equality 
\begin{equation}
\rho _\theta =\rho _T  \label{IX.2.-96}
\end{equation}
\end{proposition}

\noindent {\it is the condition} $\rho _Q=0$.

Proof: It follows immediately from the first identity in (\ref{IX.2.-94}).

The condition $\rho _Q\neq 0$ leads to the violation of the unique
determination of the notion of rest mass density and to the appearance of
three different notions of rest mass density corresponding to the three
different energy-momentum tensors for a Lagrangian system. Therefore, the
violation of the covariant Euler-Lagrange equations $\delta _vL/\delta
V^A\,_B=0$ for the non-metric tensor fields or the existence of metric
tensor fields in a Lagrangian density with $_gQ\neq 0$ induce a new rest
mass density (a new rest mass respectively) for which the identity (\ref
{IX.2.-94}) is fulfilled.

The identity $\rho _\theta \equiv \rho _T-\rho _Q$ can be related to the
physical hypotheses about the inertial, passive and active gravitational
rest mass densities in models for describing the gravitational interaction.
To every energy-momentum tensor a non-null rest mass density corresponds.
The existence of the variational energy-momentum tensor of Euler-Lagrange is
connected with the existence of the gravitational interaction in a
Lagrangian system in Einstein's theory of gravitation \cite{Manoff-11} and
therefore, with the existence of a non-null active gravitational rest mass
density. When a Lagrangian system does not interact gravitationally, the
active gravitational rest mass density is equal to zero and the principle of
equivalence between the inertial and the passive rest mass density is
fulfilled \cite{Manoff-10}, \cite{Manoff-11}.

From the second covariant Noether identity of the type 1 by means of the
relations

\[
\overline{G}=g(G)\overline{g}\text{ ,\thinspace \thinspace \thinspace
\thinspace \thinspace \thinspace \thinspace \thinspace \thinspace \thinspace
\thinspace \thinspace \thinspace \thinspace \thinspace \thinspace \thinspace
\thinspace \thinspace \thinspace \thinspace \thinspace }G=\overline{g}(%
\overline{G})g\text{ ,} 
\]
one can find the second covariant Noether identity of the type 2 for the
energy-momentum tensors of the type 2 in the form 
\begin{equation}
\overline{\theta }-\,_s\overline{T}\equiv \overline{Q}\text{ ,}
\label{IX.2.-98}
\end{equation}

\noindent where 
\begin{equation}  \label{IX.2.-99}
\overline{\theta }=g(\theta )\overline{g}\text{ ,\thinspace \thinspace
\thinspace \thinspace \thinspace \thinspace \thinspace \thinspace \thinspace
\thinspace \thinspace \thinspace \thinspace }_s\overline{T}=g(_sT)\overline{g%
}\text{ ,\thinspace \thinspace \thinspace \thinspace \thinspace \thinspace
\thinspace \thinspace \thinspace \thinspace \thinspace \thinspace \thinspace
\thinspace \thinspace \thinspace \thinspace }\overline{Q}=g(Q)\overline{g}%
\text{ .}
\end{equation}

The invariant representation of the energy-momentum tensors by means of the
projective metrics $h^u$, $h_u$ and the rest mass density allows a
comparison of these tensors with the well known energy-momentum tensors from
the continuum media mechanics (for instance, with the energy-momentum tensor
of an ideal liquid in $V_4$-spaces: 
\begin{equation}
_sT_i\,^j=(\rho +\frac 1e\cdot p)\cdot u_i\cdot u^j-p\cdot g_i^j\text{
,\thinspace \thinspace \thinspace \thinspace \thinspace \thinspace
\thinspace \thinspace \thinspace \thinspace \thinspace \thinspace \thinspace 
}e=\text{const. }\neq 0\text{, \thinspace \thinspace \thinspace }k=1\text{ .}
\label{IX.2.-121}
\end{equation}

It follows from the comparison that the Lagrangian invariant $L$ can be
interpreted as the pressure $p=L$ characterizing the Lagrangian system. This
possibility for an other physical interpretation than the usual one (in the
mechanics $L$ is interpreted as the difference between the kinetic and the
potential energy) allows a description of Lagrangian systems on the basis of
phenomenological investigations determining the dependence of the pressure
on other dynamical characteristics of the system. If these relations are
given, then by the use of the method of Lagrangians with covariant
derivatives (MLCD) the corresponding covariant Euler-Lagrange equations can
be found as well as the energy-momentum tensors.

\section{Stress (tension) tensor}

Let us now consider closer the introduced stress tensors $^GS$ $(G\sim
\theta $, $_sT$, $Q)$ obtained from the different energy-momentum tensors $%
\theta $, $_sT$, and $Q$.

\subsection{Representation of the stress (tension) viscosity tensor field}

The tensor $^GS$ has the structure of a contravariant tensor of second rank
of the type 
\begin{eqnarray}
^GS &=&h^u(g)(G)h^u=h^{ik}\cdot g_{\overline{k}\overline{l}}\cdot G_{%
\overline{m}}\,^l\cdot h^{mj}\cdot \partial _i\otimes \partial _j=  \nonumber
\\
&=&\overline{g}(h_u)\overline{g}(g)(G)\overline{g}(h_u)\overline{g}=%
\overline{g}(h_u)(G)\overline{g}(h_u)\overline{g}=  \label{13.1} \\
&=&\,^GS^{ij}\cdot \partial _i\otimes \partial _j\text{\thinspace \thinspace
\thinspace \thinspace \thinspace ,}  \nonumber
\end{eqnarray}

\noindent where 
\begin{equation}
h^u=\overline{g}(h_u)\overline{g}\text{ \thinspace \thinspace \thinspace
\thinspace ,\thinspace \thinspace \thinspace \thinspace \thinspace
\thinspace \thinspace \thinspace }G=G_i\,^j\cdot \partial _j\otimes
dx^i\,\,\,\,\text{.}  \label{13.2}
\end{equation}

The corresponding covariant tensor of second rank $g(^GS)g$ will have the
forms 
\begin{eqnarray}
g(^GS)g &=&g(\overline{g})(h_u)(G)\overline{g}(h_u)\overline{g}(g)=(h_u)(G)%
\overline{g}(h_u)=  \nonumber \\
&=&h_{ik}\cdot G_m\,^{\overline{k}}\cdot g^{\overline{m}\overline{l}}\cdot
h_{lj}\cdot dx^i\otimes dx^j\text{ }  \nonumber \\
&=&h_{i\overline{k}}\cdot \,^GS^{kl}\cdot h_{\overline{l}j}\cdot dx^i\otimes
dx^j=  \nonumber \\
&=&h_u(\widetilde{B})h_u:=\widetilde{A}\text{ \thinspace \thinspace
\thinspace ,}  \label{13.3}
\end{eqnarray}

\noindent where 
\begin{equation}
\widetilde{B}=(G)\overline{g}\text{ \thinspace \thinspace \thinspace
\thinspace \thinspace \thinspace \thinspace \thinspace \thinspace \thinspace
\thinspace \thinspace ,\thinspace \thinspace \thinspace \thinspace
\thinspace \thinspace \thinspace \thinspace \thinspace \thinspace \thinspace
\thinspace \thinspace \thinspace \thinspace \thinspace \thinspace \thinspace
\thinspace \thinspace }h_u=h_u(\overline{g})h_u=g(h^u)g\text{ \thinspace
\thinspace \thinspace .}  \label{13.4}
\end{equation}

An energy-momentum tensor $G$ ($G\sim \theta $, $_sT$, $Q$) can now be
represented in the form 
\begin{eqnarray}
g(G) &=&\varepsilon _G\cdot g(u)\otimes g(u)+g(u)\otimes g(^G\pi
)+g(^Gs)\otimes g(u)+g(^GS)g=  \nonumber \\
&=&\varepsilon _G\cdot g(u)\otimes g(u)+g(u)\otimes g(^G\pi )+g(^Gs)\otimes
g(u)+h_u(\widetilde{B})h_u\text{\thinspace \thinspace \thinspace \thinspace
\thinspace .}  \label{13.5}
\end{eqnarray}

On the other side, the stress tensor $^GS$ could be expressed by its
viscosity part $^kS$ and its pressure part $L\cdot \,^{Kr}S$ 
\begin{equation}
\,^GS=\,^kS-L\cdot \,^{Kr}S\text{ \thinspace \thinspace \thinspace
\thinspace \thinspace \thinspace \thinspace \thinspace \thinspace \thinspace
\thinspace \thinspace ,\thinspace \thinspace \thinspace \thinspace
\thinspace \thinspace \thinspace \thinspace \thinspace \thinspace \thinspace
\thinspace \thinspace \thinspace \thinspace \thinspace \thinspace \thinspace
\thinspace \thinspace \thinspace }L=p\,\,\,\,\,\,\,\text{.}  \label{13.6}
\end{equation}

Since 
\begin{equation}
G=\,_kG-L\cdot Kr\text{ \thinspace \thinspace \thinspace \thinspace
\thinspace ,}  \label{13.7}
\end{equation}

\noindent we will concentrate our attention to the representation of the
viscosity tensor field $_kG$. It can be represented in analogous way as the
energy-momentum tensor $G$%
\begin{eqnarray}
_kG &=&\varepsilon _k\cdot u\otimes g(u)+u\otimes g(^k\pi )+\,^ks\otimes
g(u)+\,(^kS)g\text{ \thinspace \thinspace \thinspace \thinspace ,}
\label{13.8a} \\
g(_kG) &=&\varepsilon _k\cdot g(u)\otimes g(u)+g(u)\otimes g(^k\pi
)+\,g(^ks)\otimes g(u)+\,g(^kS)g\text{ \thinspace \thinspace \thinspace
\thinspace \thinspace \thinspace \thinspace \thinspace .}  \label{13.8b}
\end{eqnarray}

The covariant tensor of second rank $g(^kS)g$, called stress (tension)
viscosity tensor, could now be written in the forms 
\begin{eqnarray}
g(^kS)g &=&g(\overline{g})(h_u)(_kG)\overline{g}(h_u)\overline{g}%
(g)=(h_u)(_kG)\overline{g}(h_u)=  \nonumber \\
&=&h_{ik}\cdot \,_kG_m\,^{\overline{k}}\cdot g^{\overline{m}\overline{l}%
}\cdot h_{lj}\cdot dx^i\otimes dx^j=  \nonumber \\
&=&h_{i\overline{k}}\cdot \,^kS^{kl}\cdot h_{\overline{l}j}\cdot dx^i\otimes
dx^j=  \nonumber \\
&=&h_u(\overline{B})h_u:=\overline{A}\,\,\,\,\,\,\text{,}  \label{13.9}
\end{eqnarray}

\noindent where 
\begin{equation}
\overline{B}=(_kG)\overline{g}=\,_kG_m\,^i\cdot g^{\overline{m}j}\cdot
\partial _i\otimes \partial _j=\overline{B}\,^{ij}\cdot \partial _i\otimes
\partial _j\text{ \thinspace \thinspace .}  \label{13.10}
\end{equation}

The tensor $\overline{B}$ can further be expressed by its symmetric and
antisymmetric parts 
\begin{equation}
_s\overline{B}=\frac 12\cdot (\overline{B}\,^{ij}+\,\overline{B}%
\,^{ji})\cdot \partial _i.\partial _j\text{ \thinspace \thinspace \thinspace
\thinspace \thinspace ,\thinspace \thinspace \thinspace \thinspace
\thinspace \thinspace \thinspace \thinspace }_a\overline{B}=\frac 12\cdot (%
\overline{B}\,^{ij}-\,\overline{B}\,^{ji})\cdot \partial _i\wedge \partial _j%
\text{ \thinspace \thinspace \thinspace \thinspace \thinspace .\thinspace
\thinspace \thinspace \thinspace \thinspace }  \label{13.11}
\end{equation}

The {\it stress (tension) viscosity tensor} $\overline{A}$ can be
represented as a sum, containing three terms: a trace-free symmetric term,
an antisymmetric term and a trace term 
\begin{equation}
\overline{A}=\,_{ks}\overline{D}+\,_k\overline{W}+\frac 1{n-1}\cdot \,_k%
\overline{U}\cdot h_u\text{ \thinspace \thinspace \thinspace \thinspace
\thinspace \thinspace \thinspace \thinspace ,}  \label{13.12}
\end{equation}

\noindent where 
\begin{eqnarray}
_{ks}\overline{D} &=&\,_k\overline{D}-\frac 1{n-1}\cdot \overline{g}[_k%
\overline{D}]\cdot h_u=\,_k\overline{D}-\frac 1{n-1}\cdot \,_k\overline{U}%
\cdot h_u\,\text{\thinspace \thinspace \thinspace \thinspace \thinspace ,}
\label{13.13a} \\
\text{ \thinspace \thinspace \thinspace \thinspace \thinspace \thinspace
\thinspace \thinspace \thinspace \thinspace \thinspace \thinspace }\overline{%
g}[_{ks}\overline{D}] &=&g^{\overline{i}\overline{j}}\cdot \,_{ks}\overline{D%
}_{ij}=0\text{ \thinspace \thinspace \thinspace ,}  \label{13.13b}
\end{eqnarray}
\begin{eqnarray}
_k\overline{D} &=&h_u(_s\overline{B})h_u\text{ \thinspace \thinspace
\thinspace \thinspace \thinspace \thinspace \thinspace \thinspace \thinspace
,\thinspace \thinspace \thinspace \thinspace \thinspace \thinspace
\thinspace \thinspace \thinspace \thinspace \thinspace \thinspace \thinspace
\thinspace \thinspace \thinspace \thinspace \thinspace \thinspace }_k%
\overline{U}=\overline{g}[_k\overline{D}]=\overline{g}[_s\overline{A}%
]\,\,\,\,\,\,\,\,\,\,\text{,}  \label{13.14a} \\
_s\overline{A} &=&\frac 12\cdot (\overline{A}_{ij}+\overline{A}_{ji})\cdot
dx^i.dx^j\text{ \thinspace \thinspace ,}  \label{13.14b} \\
_k\overline{U} &=&g^{\overline{i}\overline{j}}\cdot \,_k\overline{D}_{ij}=g^{%
\overline{i}\overline{j}}\cdot \,_s\overline{A}_{ij}\text{ .}  \label{13.14c}
\end{eqnarray}

The trace-free symmetric tensor $_{ks}\overline{D}$ is called {\it shear
stress (tension) viscosity }tensor{\it \ (stress deviator)}, the
antisymmetric tensor $_k\overline{W}$ is called {\it rotation (vortex)
stress viscosity} tensor, the invariant $_k\overline{U}$ is called {\it %
expansion stress viscosity invariant}. The viscosity tensor field $_kG$
could now be written in the forms 
\begin{eqnarray}
g(_kG) &=&\varepsilon _k\cdot g(u)\otimes g(u)+g(u)\otimes g(^k\pi
)+\,g(^ks)\otimes g(u)+\,g(^kS)g=  \nonumber \\
&=&\varepsilon _k\cdot g(u)\otimes g(u)+g(u)\otimes g(^k\pi
)+\,g(^ks)\otimes g(u)+\overline{A}=  \nonumber \\
&=&\varepsilon _k\cdot g(u)\otimes g(u)+g(u)\otimes g(^k\pi
)+\,g(^ks)\otimes g(u)+  \nonumber \\
&&+\,_{ks}\overline{D}+\,_k\overline{W}+\frac 1{n-1}\cdot \,_k\overline{U}%
\cdot h_u\text{ \thinspace \thinspace .}  \label{13.15}
\end{eqnarray}

The energy-momentum tensor $G$ will have the form 
\begin{eqnarray}
G &=&(\rho _G+\frac 1e\cdot L\cdot k)\cdot u\otimes g(u)-L\cdot Kr+u\otimes
g(^k\pi )+\,^ks\otimes g(u)+  \nonumber \\
&&+\overline{g}(_{ks}\overline{D})+\overline{g}(_k\overline{W})+\frac
1{n-1}\cdot \,_k\overline{U}\cdot \overline{g}(h_u)\text{ ,}  \label{13.16}
\end{eqnarray}

\noindent or the form 
\begin{eqnarray}
g(G) &=&(\rho _G+\frac 1e\cdot L\cdot k)\cdot g(u)\otimes g(u)-L\cdot
g(Kr)+g(u)\otimes g(^k\pi )+g(\,^ks)\otimes g(u)+  \nonumber \\
&&+\,_{ks}\overline{D}+\,_k\overline{W}+\frac 1{n-1}\cdot \,_k\overline{U}%
\cdot h_u\,\,\,\,\,\,\,\,\,\,\,\text{.}  \label{13.17}
\end{eqnarray}

The vector field 
\begin{equation}
_kG(\xi _{(a)\perp })=g(^k\pi ,\xi _{(a)\perp })\cdot u+\overline{g}[(_{ks}%
\overline{D})(\xi _{(a)\perp })+(_k\overline{W})(\xi _{(a)\perp })]+\frac
1{n-1}\cdot \,_k\overline{U}\cdot \xi _{(a)\perp }\,\,\,\,\,\,\,
\label{13.18}
\end{equation}

\noindent is called {\it stress vector field}. It describes the stress along
the vector field $\xi _{(a)\perp }$ but is not collinear to $\xi _{(a)\perp
} $. The following relation is also fulfilled 
\begin{eqnarray}
Kr(\xi _{(a)\perp }) &=&g_j^i\cdot \xi _{(a)\perp }^{\overline{j}%
}=g_j^i\cdot g^{\overline{j}\overline{k}}\cdot h_{kl}\cdot \xi ^{\overline{l}%
}\cdot \partial _i=g^{\overline{i}\overline{k}}\cdot h_{kl}\cdot \xi ^{%
\overline{l}}\cdot \partial _i=  \nonumber \\
&=&f^i\,_m\cdot g^{m\overline{k}}\cdot h_{kl}\cdot \xi ^{\overline{l}}\cdot
\partial _i\text{\thinspace \thinspace \thinspace \thinspace \thinspace .}
\label{13.19}
\end{eqnarray}

If we write the vector field $\xi _{(a)\perp }$ in the form 
\begin{equation}
\xi _{(a)\perp }=r\cdot n_{(a)\perp }\text{ \thinspace \thinspace
,\thinspace \thinspace \thinspace \thinspace \thinspace \thinspace
\thinspace \thinspace \thinspace }g(n_{(a)\perp },n_{(a)\perp })=\varepsilon 
\text{ \thinspace \thinspace \thinspace \thinspace \thinspace ,\thinspace
\thinspace \thinspace \thinspace \thinspace \thinspace \thinspace \thinspace 
}\varepsilon =\pm 1\text{ \thinspace \thinspace ,}  \label{13.20}
\end{equation}

\noindent then 
\begin{eqnarray}
\overline{A}(\xi _{(a)\perp },\xi _{(a)\perp }) &=&r^2\cdot \overline{A}%
(n_{(a)\perp },n_{(a)\perp })=r^2\cdot \sigma _N=  \nonumber \\
&=&r^2\cdot \,_{ks}\overline{D}(n_{(a)\perp },n_{(a)\perp })\pm \,\frac
1{n-1}\cdot \,_k\overline{U}\cdot r^2\text{ \thinspace \thinspace ,}
\label{13.21}
\end{eqnarray}

\noindent where 
\begin{equation}
h_u(n_{(a)\perp },n_{(a)\perp })=\pm r^2\text{ \thinspace \thinspace
\thinspace \thinspace \thinspace ,\thinspace \thinspace \thinspace
\thinspace \thinspace \thinspace \thinspace }\sigma _N=\,_{ks}\overline{D}%
(n_{(a)\perp },n_{(a)\perp })\pm \,\frac 1{n-1}\cdot \,_k\overline{U}\text{%
\thinspace \thinspace \thinspace }\,\,\,\text{.\thinspace }  \label{13.22}
\end{equation}

The hypersurface 
\begin{equation}
\overline{A}(\xi _{(a)\perp },\xi _{(a)\perp })=\sigma _N\cdot r^2=\pm 
\overline{k}\,^2\,\,\,\,\,\,\,\,\,\text{,\thinspace \thinspace \thinspace
\thinspace \thinspace \thinspace \thinspace \thinspace \thinspace \thinspace 
}\overline{\text{\thinspace }k}=\text{const.}\,\,\,\text{,}  \label{13.23}
\end{equation}

\noindent is called {\it Cauchy stress hypersurface}. It follows, that for a
Cauchy stress hypersurface the relation 
\begin{equation}
\sigma _N=\,_{ks}\overline{D}(n_{(a)\perp },n_{(a)\perp })\pm \,\frac
1{n-1}\cdot \,_k\overline{U}=\pm \frac{\overline{k}^2}{r^2}\text{ }
\label{13.24}
\end{equation}

\noindent is fulfilled.

{\it Special case:} $_{ks}\overline{D}:=0$: 
\begin{equation}
\sigma _N=\pm \,\frac 1{n-1}\cdot \,_k\overline{U}=\pm \frac{\overline{k}^2}{%
r^2}\text{ \thinspace \thinspace \thinspace \thinspace \thinspace \thinspace
\thinspace ,\thinspace \thinspace \thinspace \thinspace \thinspace
\thinspace \thinspace \thinspace \thinspace \thinspace \thinspace \thinspace
\thinspace \thinspace \thinspace \thinspace \thinspace \thinspace \thinspace
\thinspace }\,_k\overline{U}=(n-1)\cdot \text{\thinspace }\frac{\overline{k}%
^2}{r^2}\text{ .\thinspace \thinspace \thinspace }  \label{13.25}
\end{equation}

The trace term 
\begin{equation}
\frac 1{n-1}\cdot \,_k\overline{U}\cdot h_u  \label{13.26}
\end{equation}

\noindent could be written by the use of the explicit form of the covariant
projective metric $h_u$%
\begin{eqnarray}
\frac 1{n-1}\cdot \,_k\overline{U}\cdot h_u &=&\frac 1{n-1}\cdot \,_k%
\overline{U}\cdot [g-\frac 1e\cdot g(u)\otimes g(u)]=  \nonumber \\
&=&\frac 1{n-1}\cdot [\,_k\overline{U}\cdot g-\frac 1e\cdot \,_k\overline{U}%
\cdot g(u)\otimes g(u)]=  \nonumber \\
&=&\frac 1{n-1}\cdot \,_k\overline{U}\cdot g-\frac 1{n-1}\cdot \frac 1e\cdot
\,_k\overline{U}\cdot g(u)\otimes g(u)\text{ .}  \label{13.27}
\end{eqnarray}

By means of the last expression we can represent $g(G)$ in the form 
\begin{eqnarray}
g(G) &=&(\rho _G-\frac 1{n-1}\cdot \frac 1e\cdot \,_k\overline{U}+\frac
1e\cdot L\cdot k)\cdot g(u)\otimes g(u)-L\cdot g(Kr)+\frac 1{n-1}\cdot \,_k%
\overline{U}\cdot g+  \nonumber \\
&&+g(u)\otimes g(^k\pi )+g(\,^ks)\otimes g(u)+\,_{ks}\overline{D}+\,_k%
\overline{W}  \label{13.28}
\end{eqnarray}
\begin{eqnarray}
g(G) &=&(\rho _G-\,\frac 1e\cdot \,_kE+\frac 1e\cdot L\cdot k)\cdot
g(u)\otimes g(u)-L\cdot g(Kr)+\,_kE\cdot g+  \nonumber \\
&&+g(u)\otimes g(^k\pi )+g(\,^ks)\otimes g(u)+\,_{ks}\overline{D}+\,_k%
\overline{W}\text{ \thinspace \thinspace \thinspace \thinspace ,}
\label{13.29}
\end{eqnarray}
\begin{eqnarray}
g(G) &=&[\rho _G+\frac 1e\cdot (L\cdot k-\,_kE)]\cdot g(u)\otimes
g(u)-L\cdot g(Kr)+\,_kE\cdot g+  \nonumber \\
&&+g(u)\otimes g(^k\pi )+g(\,^ks)\otimes g(u)+\,_{ks}\overline{D}+\,_k%
\overline{W}\text{ \thinspace \thinspace \thinspace \thinspace \thinspace
\thinspace ,}  \label{13.30}
\end{eqnarray}

\noindent where 
\begin{equation}
_kE=\frac 1{n-1}\cdot \,_k\overline{U}  \label{13.31}
\end{equation}

\noindent is the {\it inner energy density}. The pressure $p$ is considered
as a hydrostatic stress invariant. The tensor $p\cdot Kr+\,_kE\cdot 
\overline{g}(g)$ is called hydrostatic stress tensor. It contains a part
with the pressure $p$ and a part induced by the inner energy density $_kE$.

{\it Special case: }$(L_n,g)$-spaces:{\it \ }$S=C$. In this special case $%
f^i\,_j=g_j^i$, $k=1$, 
\begin{eqnarray}
g(Kr) &=&g(g_j^i\cdot \partial _i\otimes dx^j)=g_{k\overline{l}}\cdot
g_j^l\cdot dx^k\otimes dx^j=  \nonumber \\
&=&g_{kl}\cdot g_j^l\cdot dx^k\otimes dx^j=g_{kj}\cdot dx^k\otimes dx^j=g%
\text{ \thinspace \thinspace ,}  \label{13.32}
\end{eqnarray}
\begin{eqnarray}
g(G) &=&(\rho _G-\,\frac 1e\cdot \,_kE+\frac 1e\cdot L)\cdot g(u)\otimes
g(u)-(L-\,_kE)\cdot g+  \nonumber \\
&&+g(u)\otimes g(^k\pi )+g(\,^ks)\otimes g(u)+\,_{ks}\overline{D}+\,_k%
\overline{W}\text{ }\,\,\,\text{,}  \label{13.33}
\end{eqnarray}
\begin{eqnarray}
g(G) &=&(\rho _G+\,\frac 1e\cdot \overline{p})\cdot g(u)\otimes g(u)-%
\overline{p}\cdot g+  \nonumber \\
&&+g(u)\otimes g(^k\pi )+g(\,^ks)\otimes g(u)+\,_{ks}\overline{D}+\,_k%
\overline{W}\text{ }\,\,\,\text{,}  \label{13.34}
\end{eqnarray}

\noindent where 
\begin{equation}
\overline{p}=L-\,_kE=p-\,_kE\text{\thinspace \thinspace \thinspace .}
\label{13.35}
\end{equation}

It is obvious from the last expressions, that the structure of the invariant 
$\overline{p}$ includes not only the pressure $p$ but also the inner energy
density $_kE$, which acts in the flow on the one side as a rest mass density 
$(1/e)\cdot \,_kE$ and, on the other side, as an additional pressure $_kE$.
Usually \cite{Landau-1}, the invariant $\overline{p}$ is interpreted as the
pressure of a system in a more general sense than as the hydrostatic
pressure. The invariant $\overline{p}$ could vanish under two different
types of conditions:

(a) $\overline{p}=0$ if $p=0$ and $_kE=0$.

(b) $\overline{p}=0$ if $p=\,_kE$.

In both cases we have for $g(G)$%
\begin{equation}
g(G)=\rho _G\cdot g(u)\otimes g(u)+g(u)\otimes g(^k\pi )+g(\,^ks)\otimes
g(u)+\,_{ks}\overline{D}+\,_k\overline{W}\text{ }\,\,\,\,\text{.}
\label{13.37}
\end{equation}

For $^k\pi =0$, $\,^ks=0$, $_{ks}\overline{D}=0$, and $_k\overline{W}=0$ the
energy-momentum tensors for {\it dust matter} follow 
\begin{equation}
g(G)=\rho _G\cdot g(u)\otimes g(u)\,\,\,\,\text{.}  \label{13.38}
\end{equation}

The last result shows that dust matter could exist under different inner
conditions for the hydrostatic pressure and the inner energy.

\section{Relations between stresses (tensions) and deformations}

The considered notions of deformation velocity and deformation acceleration
are kinematic quantities of a continuous media (flow). On the other side,
the energy-momentum tensors and their corresponding viscosity tensors are
related to the dynamic characteristics of a continuous media. The assumption
that deformations generate stresses and motions in a media (flow) does not
lead to a unique (logically consistent) way for finding relations between
stresses and deformations. For any specific branch of the continuous media
mechanics (elasticity theory, plasticity theory, viscosity theory,
hydrodynamics etc.) different relations between stresses and deformations
are assumed. In general, the most of these assumptions are included in the
precondition

\begin{center}
{\sc stresses}$\,=\,stresses\,(friction\,\,velocity,\,\,\,deformations,\,\,%
\,deformation\,\,\,velocity)$.
\end{center}

The vector fields $\xi _{(a)\perp }$ [and /or their corresponding
infinitesimal vectors $\overline{\xi }_{(a)\perp }=d\lambda ^a\cdot \xi
_{(a)\perp }$ (over $a$ is not summarized)], introduced in the
preconditions, are interpreted as deformation vector fields, orthogonal to
the velocity of the continuous media in the space-time. This interpretation
of $\xi _{(a)\perp }$ and $\overline{\xi }_{(a)\perp }$ coincides with the
interpretation of $\xi _{(a)\perp }$ ($\overline{\xi }_{(a)\perp }$) as the
distance between two material points $\overline{P}$ and $P$ in a flow with
the corresponding co-ordinates $\overline{x}^i=x^i(\tau _0,\lambda
_0^a+d\lambda ^a)$ and $x^i=x^i(\tau _0,\lambda _0^a)$ after a deformation.
The point $\overline{P}$ is the point $P$ after a deformation, and the point 
$P$ is the point $P^{\prime }$ taking the place of the point $P$ after the
same deformation, i.e. after a deformation we have the changes of the points 
$P^{\prime }$ and $P$ 
\begin{eqnarray}
P^{^{\prime }} &\rightarrow &P\stackrel{\xi _{(a)\perp }}{\rightarrow }%
\overline{P}\text{ \thinspace \thinspace \thinspace \thinspace \thinspace ,}
\label{13.39} \\
\overline{x}^i &=&x^i(\tau _0,\lambda _0^a+d\lambda ^a)=x^i(\tau _0,\lambda
_0^a)+d\lambda ^a\cdot \left( \frac{dx^i}{d\lambda ^a}\right) _{(\tau
_0,\lambda _0)}=  \nonumber \\
&=&x^i+\overline{\xi }\,_{(a)\perp }^i\,\,\,\,\,\,\text{.}  \label{13.40}
\end{eqnarray}

The vector $\xi _{(a)\perp }$ is then called {\it deformation vector}. The
act of deformation itself is not considered. Only the result of the
deformation (the deformation vector $\xi _{(a)\perp }$) is taken into
account in all investigations. This means that at a given moment (time-point 
$\tau _0$) we have a picture of a continuous media after a (possible)
deformation and all future deformations, stresses and motions of this media
are further considered by the use of the tools of continuous media mechanics.

{\it Remark.} The interpretation of $\xi _{(a)\perp }$ as a deformation
vector is not necessary for the considerations of the relative velocity and
the relative acceleration with their kinematic characteristics. It is needed
if we wish to consider the different relations from elasticity theory
(Hook's law) and hydrodynamics (different types of flows) from a general
point of view.

{\it Conjecture. }The stress (tension) viscosity tensor $g(^kS)g=\overline{A}
$ could be considered in general as a function of the friction deformation
velocity tensor $R$ and the deformation velocity tensor $d$%
\begin{eqnarray}
\overline{A} &=&\overline{A}(R,d)\,\,\,\,\text{ ,\thinspace }  \nonumber \\
\overline{A} &=&\overline{A}_{ij}\cdot dx^i\otimes dx^j\,\,\,\,\text{,} 
\nonumber \\
d &=&d_{ij}\cdot dx^i\otimes dx^j\text{\thinspace \thinspace \thinspace
\thinspace ,}  \nonumber \\
R &=&R_{ij}\cdot dx^i\otimes dx^j\text{ \thinspace \thinspace \thinspace
\thinspace .}  \label{13.41}
\end{eqnarray}

On the basis of different specializations of the form of dependence of $%
\overline{A}$ on $d$ and $R$, we can find generalizations of well known
schemes from the classical continuous media mechanics in Euclidean spaces $%
E_n$ $(n=3)$.

\subsection{Linear elasticity theory}

In a linear elasticity theory the viscosity tensor $g(^kS)g=\overline{A}$
depends linearly only on the friction deformation velocity tensor $R$%
\begin{eqnarray}
\overline{A}_{ij} &=&C_{ij}\,^{km}\cdot R_{\overline{k}\overline{m}}\,\,\,%
\text{,\thinspace \thinspace \thinspace \thinspace \thinspace \thinspace
\thinspace \thinspace \thinspace \thinspace \thinspace \thinspace \thinspace
\thinspace \thinspace \thinspace }\overline{A}=C[R]\text{ ,\thinspace
\thinspace \thinspace \thinspace \thinspace \thinspace \thinspace \thinspace
\thinspace }  \nonumber \\
\text{\thinspace }C &=&C_{ij}\,^{km}\cdot dx^i\otimes dx^j\otimes \partial
_k\otimes \partial _m\in \otimes _2\,^2(M)\text{ \thinspace \thinspace
\thinspace \thinspace ,\thinspace \thinspace \thinspace }  \nonumber \\
\text{\thinspace }R &=&R_{ij}\cdot dx^i\otimes dx^j\in \otimes _2(M)\text{
\thinspace \thinspace \thinspace \thinspace \thinspace ,}  \nonumber \\
R &=&\,_\sigma R+\,_\omega R+\frac 1{n-1}\cdot \,_\theta R\cdot h_u\,\,\,\,%
\text{,}  \nonumber \\
\overline{A} &=&\,_{ks}\overline{D}+\,_k\overline{W}+\frac 1{n-1}\cdot \,_k%
\overline{U}\cdot h_u\,\,\,\,\,\,\,\,\text{,}  \nonumber \\
\overline{A}[C] &=&C[_\sigma R+\,_\omega R+\frac 1{n-1}\cdot \,_\theta
R\cdot h_u]=  \nonumber \\
&=&C[_\sigma R]+C[_\omega R]+\frac 1{n-1}\cdot \,_\theta R\cdot C[h_u]\,\,%
\text{.}  \label{13.42}
\end{eqnarray}

\noindent where $C_{ij}\,^{km}$ are components of the tensor of elasticity
coefficients. They should obey the conditions 
\begin{equation}
C_{ij}\,^{km}\cdot u^{\overline{i}}=0\,\,\,\,\,\,\,\,\,\text{,\thinspace
\thinspace \thinspace \thinspace \thinspace \thinspace \thinspace \thinspace
\thinspace \thinspace \thinspace \thinspace \thinspace }C_{ij}\,^{km}\cdot
u^{\overline{j}}=0\text{\thinspace \thinspace \thinspace \thinspace
\thinspace \thinspace \thinspace .}  \label{13.43}
\end{equation}

On the other side, we have to decompose the tensor of elasticity
coefficients $C=C_{ij}\,^{km}\cdot dx^i\otimes dx^j\otimes \partial
_k\otimes \partial _m$ with respect to the basis $dx^i\otimes dx^j$ in a
trace-free symmetric part, antisymmetric part, and trace part [in
correspondence with the structure of $\overline{A}$] 
\[
C=\,_s\overline{C}+\,_aC+\frac 1{n-1}\cdot \overline{g}[C]\cdot h_u\,\,\,\,\,%
\text{,} 
\]

\noindent where 
\begin{eqnarray}
_s\overline{C} &=&\,_sC-\frac 1{n-1}\cdot \overline{g}[C]\cdot
h_u\,\,\,\,\,\,\,\,\,\,\,\text{,\thinspace \thinspace \thinspace \thinspace
\thinspace \thinspace \thinspace \thinspace \thinspace \thinspace \thinspace
\thinspace \thinspace \thinspace \thinspace \thinspace \thinspace \thinspace
\thinspace \thinspace \thinspace \thinspace }\overline{g}[_s\overline{C}%
]=0\,\,\,\,\text{,}  \label{13.44} \\
_sC &=&\frac 12\cdot (C_{ij}\,^{km}+C_{ji}\,^{km})\cdot dx^i\otimes
dx^j\otimes \partial _k\otimes \partial _m=  \nonumber \\
&=&C_{(ij)}\,^{km}\cdot dx^i.dx^j\otimes \partial _k\otimes \partial
_m\,\,\,\,\,\text{,}  \nonumber \\
\overline{g}[C] &=&\overline{g}[_sC]=g^{\overline{i}\overline{j}}\cdot
C_{ij}\,^{km}\cdot \partial _k\otimes \partial _m\,\,\,\,\,\,\text{,} 
\nonumber \\
_aC &=&\frac 12\cdot (C_{ij}\,^{km}-C_{ji}\,^{km})\cdot dx^i\otimes
dx^j\otimes \partial _k\otimes \partial _m=  \nonumber \\
&=&C_{[ij]}\,^{km}\cdot dx^i\wedge dx^j\otimes \partial _k\otimes \partial
_m\,\text{\thinspace \thinspace \thinspace \thinspace \thinspace \thinspace .%
}  \nonumber
\end{eqnarray}

For the viscosity tensor $\overline{A}$, the following expressions are
valid: 
\begin{eqnarray}
\overline{A} &=&C[R]=(_s\overline{C}+\,_aC+\frac 1{n-1}\cdot \overline{g}[C%
]\cdot h_u)[R]=  \label{13.45} \\
&=&\,_s\overline{C}[R]+\,\,_aC[R]+\frac 1{n-1}\cdot (\overline{g}[C%
])[R]\cdot h_u=  \nonumber \\
&=&\,_{ks}\overline{D}+\,_k\overline{W}+\frac 1{n-1}\cdot \,_k\overline{U}%
\cdot h_u\,\,\,\,\,\,\,\,\text{,}  \nonumber \\
(\overline{g}[C])[R] &=&(\overline{g}[_sC])[R]=g^{\overline{i}\overline{j}%
}\cdot C_{ij}\,^{km}\cdot R_{\overline{k}\overline{m}}\,\,\,\,\,\,\text{.} 
\nonumber
\end{eqnarray}

The above relation for $\overline{A}:\overline{A}=C[R]$ is called {\it %
generalized Hook's law}. Since $\overline{A}=\,_{ks}\overline{D}+\,_k%
\overline{W}+\frac 1{n-1}\cdot \,_k\overline{U}\cdot h_u$, the generalized
Hook's law could be decomposed in its three parts 
\begin{eqnarray}
_{ks}\overline{D} &=&\,_s\overline{C}[R]=\,_s\overline{C}[_\sigma
R+\,_\omega R+\frac 1{n-1}\cdot \,_\theta R\cdot h_u]=  \nonumber \\
&=&\,_s\overline{C}[_\sigma R]+\,_s\overline{C}[_\omega R]+\frac 1{n-1}\cdot
\,_\theta R\cdot \,_s\overline{C}[h_u]\,\,\,\,\,\,\,\text{,}  \label{13.47}
\\
_k\overline{W} &=&\,_aC[R]=\,_aC[_\sigma R+\,_\omega R+\frac 1{n-1}\cdot
\,_\theta R\cdot h_u]=  \nonumber \\
&=&\,_aC[_\sigma R]+\,_aC[_\omega R]+\frac 1{n-1}\cdot \,_\theta R\cdot
\,_aC[h_u]\,\,\,\,\,\text{,}  \label{13.48} \\
_k\overline{U} &=&(\overline{g}[C])[R]=(\overline{g}[C])[_\sigma R+\,_\omega
R+\frac 1{n-1}\cdot \,_\theta R\cdot h_u]=  \nonumber \\
&=&(\overline{g}[C])[_\sigma R]+(\overline{g}[C])[_\omega R]+\frac
1{n-1}\cdot \,_\theta R\cdot (\overline{g}[C])[h_u]\,\,\,\,\,\text{.}
\label{13.49}
\end{eqnarray}

{\it Remark.} Usually, only the relation $_{ks}\overline{D}=\,_s\overline{C}[%
_\sigma R]+\frac 1{n-1}\cdot \,_\theta R\cdot \,_s\overline{C}[h_u]\,$ is
considered as generalized Hook's law in $E_3$-spaces. The other two
relations (for $_{ks}\overline{D}=\,_s\overline{C}[_\omega R]$ and for $_k%
\overline{W}=\,_aC[R]$) are not taken into account because of the assumption 
$_\omega R=0$ and consideration of the symmetric energy-momentum $\,_sT$
(for which $_k\overline{W}=0$) instead of the (generalized) canonical
energy-momentum tensor $\theta $.

\subsubsection{Isotropic elastic media}

\begin{definition}
A continuous media (flow) for which the components $C_{ij}\,^{km}$ of the
tensor of elasticity coefficients $C$ has the forms 
\begin{eqnarray}
C_{ij}\,^{km} &=&\lambda _1\cdot h_{il}\cdot h_{jn}\cdot g^{\overline{l}%
k}\cdot g^{\overline{n}m}+\lambda _2\cdot h_{ij}\cdot g^{km}\,\,\,\,\text{,}
\label{13.50} \\
C &=&\lambda _1\cdot h_u(\overline{g})\otimes h_u(\overline{g})+\lambda
_2\cdot h_u\otimes \overline{g}\text{ \thinspace \thinspace \thinspace
\thinspace }  \nonumber
\end{eqnarray}
\end{definition}

{\it is called isotropic elastic media} \cite{Mase}.

For the isotropic elastic media the following relations are fulfilled: 
\begin{eqnarray}
\overline{A}_{ij} &=&\lambda _1\cdot R_{ij}+\lambda _2\cdot \,_\theta R\cdot
h_{ij}=  \nonumber \\
&=&\lambda _1\cdot (_\sigma R_{ij}+\,_\omega R_{ij})+(\frac{\lambda _1}{n-1}%
+\lambda _2)\cdot \,_\theta R\cdot h_{ij\,}\,\,\,\,\,\,\,\text{,}
\label{13.51}
\end{eqnarray}

\noindent or 
\begin{eqnarray}
\overline{A}_{ij} &=&\lambda _1\cdot (_\sigma R_{ij}+\,_\omega R_{ij})+%
\overline{\lambda }\cdot \,_\theta R\cdot h_{ij\,}\,\,\,\,\,\,\,\,\,\,\,%
\text{,}  \label{13.52} \\
\overline{\lambda } &=&\frac{\lambda _1}{n-1}+\lambda _2\,\,\text{.} 
\nonumber
\end{eqnarray}

The constant $\lambda _1$ and $\overline{\lambda }$ are called {\it Lame
constant}.

Since the viscosity tensor $g(^kS)g=\overline{A}$ could be written in the
forms 
\begin{eqnarray}
g(^kS)g &=&\overline{A}=\,_{ks}\overline{D}+\,_k\overline{W}+\frac
1{n-1}\cdot \,_k\overline{U}\cdot h_u=  \nonumber \\
&=&\lambda _1\cdot (_\sigma R+\,_\omega R)+\overline{\lambda }\cdot
\,_\theta R\cdot h_u\,\,\text{\thinspace \thinspace \thinspace ,}
\label{13.53}
\end{eqnarray}

\noindent we can decompose the generalized Hook's law for isotropic elastic
media in its three parts 
\begin{eqnarray}
_{ks}\overline{D} &=&\lambda _1\cdot \,_\sigma R\,\,\,\,\,\text{,}
\label{13.54} \\
_k\overline{W} &=&\lambda _1\cdot \,_\omega R\,\,\,\,\,\,\text{,}
\label{13.55} \\
\,_k\overline{U} &=&\overline{\lambda }\cdot \,_\theta R\,\,\,\,\,\,\text{.}
\label{13.56}
\end{eqnarray}

The energy-momentum tensors $G\sim (\theta $, $_sT)$ will have for isotropic
elastic media the form 
\begin{eqnarray}
g(G) &=&(\rho _G+\frac 1e\cdot p\cdot k)\cdot g(u)\otimes g(u)-p\cdot
g(Kr)+g(u)\otimes g(^k\pi )+g(^ks)\otimes g(u)+  \nonumber \\
&&+\lambda _1\cdot (_\sigma R+\,_\omega R)+\overline{\lambda }\cdot
\,_\theta R\cdot h_u\text{ \thinspace \thinspace \thinspace .}  \label{13.57}
\end{eqnarray}

Let us find now the explicit form of the part $h_u(^kG)(\xi _{(a)\perp })$
of the stress vector field $^kG(\xi _{(a)\perp })$ orthogonal to the vector
field $u$ for isotropic elastic media 
\begin{eqnarray}
(h_u)(\,^kG)(\xi _{(a)\perp }) &=&[(_{ks}\overline{D})(\xi _{(a)\perp })+(_k%
\overline{W})(\xi _{(a)\perp })]+\frac 1{n-1}\cdot \,_k\overline{U}\cdot
h_u(\xi _{(a)\perp })=\overline{A}(\xi _{(a)\perp })\text{ \thinspace
\thinspace ,}  \label{13.36} \\
h_u(\,^kG)(\xi _{(a)\perp }) &=&\overline{A}(\xi _{(a)\perp })=[\lambda
_1\cdot (_\sigma R+\,_\omega R)+\overline{\lambda }\cdot \,_\theta R\cdot
h_u](\xi _{(a)\perp })=  \nonumber \\
&=&\lambda _1\cdot [_\sigma R(\xi _{(a)\perp })+_\omega R\,(\xi _{(a)\perp
})]+\overline{\lambda }\cdot \,_\theta R\cdot g(\xi _{(a)\perp })\text{
\thinspace \thinspace \thinspace .}  \label{13.58}
\end{eqnarray}

The considered relations could be implemented for description of isotropic
elastic media in $(\overline{L}_n,g)$-spaces.

\subsection{Hydrodynamics}

In the theory of fluids two major types of fluids are considered \cite{Mase}:

(a) Stokes' fluids

(b) Newton's fluids.

\begin{definition}
Stokes' fluid. A flow with $\overline{A}=\overline{A}(d)$ [$\overline{A}%
_{ij}=\overline{A}_{ij}(d_{kl})$] is called Stokes' fluid.
\end{definition}

\begin{definition}
Newton' fluid. A flow with $\overline{A}=\,_Nk[d]$ [$\overline{A}%
_{ij}=\,_Nk_{ij}\,^{km}\cdot d_{\overline{k}\overline{m}}$] and $C=\,_Nk$ is
called Newton's fluid.
\end{definition}

\subsubsection{Newton's fluids}

In the Newton fluid the viscosity tensor $g(^kS)g=\overline{A}$ depends
linearly only on the deformation velocity tensor $d$%
\begin{eqnarray}
\overline{A}_{ij} &=&\,_Nk_{ij}\,^{km}\cdot d_{\overline{k}\overline{m}%
}\,\,\,\text{,\thinspace \thinspace \thinspace \thinspace \thinspace
\thinspace \thinspace \thinspace \thinspace \thinspace \thinspace \thinspace
\thinspace \thinspace \thinspace \thinspace }\overline{A}=\,_Nk[d]\text{
,\thinspace \thinspace \thinspace \thinspace \thinspace \thinspace
\thinspace \thinspace \thinspace }  \nonumber \\
\text{\thinspace }_Nk &=&\,_Nk_{ij}\,^{km}\cdot dx^i\otimes dx^j\otimes
\partial _k\otimes \partial _m\in \otimes _2\,^2(M)\text{ \thinspace
\thinspace \thinspace \thinspace ,\thinspace \thinspace \thinspace } 
\nonumber \\
\text{\thinspace }d &=&d_{ij}\cdot dx^i\otimes dx^j\in \otimes _2(M)\text{
\thinspace \thinspace \thinspace \thinspace \thinspace ,}  \nonumber \\
d &=&\,\sigma +\,\omega +\frac 1{n-1}\cdot \,\theta \cdot h_u\,\,\,\,\text{,}
\nonumber \\
\overline{A} &=&\,_{ks}\overline{D}+\,_k\overline{W}+\frac 1{n-1}\cdot \,_k%
\overline{U}\cdot h_u\,\,\,\,\,\,\,\,\text{,}  \nonumber \\
\overline{A}[d] &=&\,_Nk[\sigma +\,\omega +\frac 1{n-1}\cdot \,\theta \cdot
h_u]=  \nonumber \\
&=&\,_Nk[\sigma ]+\,_Nk[\omega ]+\frac 1{n-1}\cdot \,\theta \cdot
\,_Nk[h_u]\,\,\text{.}  \label{13.59}
\end{eqnarray}

The components $_Nk_{ij}\,^{km}$ of the tensor $_Nk\in \otimes _2\,^2(M)$
are called {\it viscosity coefficients}. They should obey the conditions 
\begin{equation}
_Nk_{ij}\,^{km}\cdot u^{\overline{i}}=0\,\,\,\,\,\,\,\,\,\,\,\,\,\,\,\,\text{%
,\thinspace \thinspace \thinspace \thinspace \thinspace \thinspace
\thinspace \thinspace \thinspace \thinspace \thinspace \thinspace \thinspace
\thinspace \thinspace }_Nk_{ij}\,^{km}\cdot u^{\overline{j}}=0\,\,\,\,\,%
\text{.}  \label{13.60}
\end{equation}

On the other side, we have to decompose the tensor of viscosity coefficients 
$_Nk=\,_Nk_{ij}\,^{km}\cdot dx^i\otimes dx^j\otimes \partial _k\otimes
\partial _m$ with respect to the basis $dx^i\otimes dx^j$ in a trace-free
symmetric part, antisymmetric part, and trace part [in correspondence with
the structure of $\overline{A}$] 
\begin{equation}
_Nk=\,_{sN}\overline{k}+\,_{aN}k+\frac 1{n-1}\cdot \overline{g}[_Nk]\cdot
h_u\,\,\,\,\,\text{,}  \label{13.61}
\end{equation}

\noindent where 
\begin{eqnarray}
_{sN}\overline{k} &=&\,_{sN}k-\frac 1{n-1}\cdot \overline{g}[_Nk]\cdot
h_u\,\,\,\,\,\,\,\,\,\,\,\text{,\thinspace \thinspace \thinspace \thinspace
\thinspace \thinspace \thinspace \thinspace \thinspace \thinspace \thinspace
\thinspace \thinspace \thinspace \thinspace \thinspace \thinspace \thinspace
\thinspace \thinspace \thinspace \thinspace }\overline{g}[_{sN}\overline{k}%
]=0\,\,\,\,\text{,}  \label{13.63} \\
_{sN}k &=&\frac 12\cdot (_Nk_{ij}\,^{km}+\,_Nk_{ji}\,^{km})\cdot dx^i\otimes
dx^j\otimes \partial _k\otimes \partial _m=  \nonumber \\
&=&\,_Nk_{(ij)}\,^{km}\cdot dx^i.dx^j\otimes \partial _k\otimes \partial
_m\,\,\,\,\,\text{,}  \nonumber \\
\overline{g}[_Nk] &=&\overline{g}[_{sN}k]=g^{\overline{i}\overline{j}}\cdot
\,_Nk_{ij}\,^{km}\cdot \partial _k\otimes \partial _m\,\,\,\,\,\,\text{,} 
\nonumber \\
_{aN}k &=&\frac 12\cdot (_Nk_{ij}\,^{km}-\,_Nk_{ji}\,^{km})\cdot dx^i\otimes
dx^j\otimes \partial _k\otimes \partial _m=  \nonumber \\
&=&\,_Nk_{[ij]}\,^{km}\cdot dx^i\wedge dx^j\otimes \partial _k\otimes
\partial _m\,\text{\thinspace \thinspace \thinspace \thinspace \thinspace
\thinspace .}  \nonumber
\end{eqnarray}

For the viscosity tensor $\overline{A}$, the following expressions are
valid: 
\begin{eqnarray}
\overline{A} &=&\,_Nk[d]=(_{sN}\overline{k}+\,_{aN}k+\frac 1{n-1}\cdot 
\overline{g}[_Nk]\cdot h_u)[d]=  \nonumber \\
&=&\,_{sN}\overline{k}[d]+\,\,_{aN}k[d]+\frac 1{n-1}\cdot (\overline{g}[_Nk%
])[d]\cdot h_u=  \label{13.64} \\
&=&\,_{ks}\overline{D}+\,_k\overline{W}+\frac 1{n-1}\cdot \,_k\overline{U}%
\cdot h_u\,\,\,\,\,\,\,\,\text{,}  \nonumber \\
(\overline{g}[_Nk])[d] &=&(\overline{g}[_{sN}k])[d]=g^{\overline{i}\overline{%
j}}\cdot \,_Nk_{ij}\,^{km}\cdot d_{\overline{k}\overline{m}}\,\,\,\,\,\,%
\text{.}  \nonumber
\end{eqnarray}

The above relation for $\overline{A}:\overline{A}=\,_Nk[d]$ is called {\it %
generalized viscosity tensor for Newton's fluids}. Since $\overline{A}%
=\,_{ks}\overline{D}+\,_k\overline{W}+\frac 1{n-1}\cdot \,_k\overline{U}%
\cdot h_u$, the generalized Hook's law could be decomposed in its three
parts 
\begin{eqnarray}
_{ks}\overline{D} &=&\,_{sN}\overline{k}[d]=\,_{sN}\overline{k}[\sigma
+\,\omega +\frac 1{n-1}\cdot \,\theta \cdot h_u]=  \nonumber \\
&=&\,_{sN}\overline{k}[\sigma ]+\,_{sN}\overline{k}[\omega ]+\frac
1{n-1}\cdot \,\theta \cdot \,_{sN}\overline{k}[h_u]\,\,\,\,\,\,\,\text{,}
\label{13.65} \\
_k\overline{W} &=&\,_{aN}k[d]=\,_{aN}k[\sigma +\,\omega +\frac 1{n-1}\cdot
\,\theta \cdot h_u]=  \nonumber \\
&=&\,_{aN}k[\sigma ]+\,_{aN}k[\omega ]+\frac 1{n-1}\cdot \,\theta \cdot
\,_{aN}k[h_u]\,\,\,\,\,\text{,}  \label{13.66} \\
_k\overline{U} &=&(\overline{g}[_Nk])[d]=(\overline{g}[_Nk])[\sigma
+\,\omega +\frac 1{n-1}\cdot \,\theta \cdot h_u]=  \nonumber \\
&=&(\overline{g}[_Nk])[\sigma ]+(\overline{g}[_Nk])[\omega ]+\frac
1{n-1}\cdot \,\theta \cdot (\overline{g}[_Nk])[h_u]\,\,\,\,\,\text{.}
\label{13.67}
\end{eqnarray}

{\it Remark.} Usually, only the relation $_{ks}\overline{D}=\,_{sN}\overline{%
k}[\sigma ]+\frac 1{n-1}\cdot \,\theta \cdot \,_{sN}\overline{k}[h_u]\,$ is
considered as generalized viscosity tensor for Newton's fluids in $E_3$%
-spaces. The other two relations (for $_{ks}\overline{D}=\,_{sN}\overline{k}[%
\omega ]$ and for $_k\overline{W}=\,_{aN}k[d]$) are not taken into account
because of the assumption $\omega =0$ and consideration of the symmetric
energy-momentum $\,_sT$ (for which $_k\overline{W}=0$) instead of the
(generalized) canonical energy-momentum tensor $\theta $.

\subsubsection{Isotropic homogeneous Newton's fluids}

For an isotropic homogeneous Newton's fluid the components $_Nk_{ij}\,^{km}$
will have a form analogous to this for isotropic elastic media 
\begin{eqnarray}
_Nk_{ij}\,^{km} &=&k_1\cdot h_{il}\cdot h_{jn}\cdot g^{\overline{l}k}\cdot
g^{\overline{n}m}+k_2\cdot h_{ij}\cdot g^{km}\,\,\,\,\text{,}  \label{13.68}
\\
_Nk &=&k_1\cdot h_u(\overline{g})\otimes h_u(\overline{g})+k_2\cdot
h_u\otimes \overline{g}\text{ \thinspace \thinspace \thinspace \thinspace
\thinspace .}  \nonumber
\end{eqnarray}

The constant $k_1$ is called {\it first viscosity coefficient}. For the
components $\overline{A}_{ij}$ of the viscosity tensor $g(^kS)g=\overline{A}$
and for the tensor $\overline{A}$ we obtain 
\begin{eqnarray}
\overline{A}_{ij} &=&(k_1\cdot h_{il}\cdot h_{jn}\cdot g^{\overline{l}%
k}\cdot g^{\overline{n}m}+k_2\cdot h_{ij}\cdot g^{km})\cdot d_{\overline{k}%
\overline{m}}=  \nonumber \\
&=&k_1\cdot h_{il}\cdot h_{jn}\cdot g^{\overline{l}k}\cdot g^{\overline{n}%
m}\cdot d_{\overline{k}\overline{m}}+k_2\cdot h_{ij}\cdot g^{km}\cdot d_{%
\overline{k}\overline{m}}=  \nonumber \\
&=&k_1\cdot d_{ij}+k_2\cdot h_{ij}\cdot g^{km}\cdot d_{\overline{k}\overline{%
m}}=k_1\cdot d_{ij}+k_2\cdot \theta \cdot h_{ij}\,\,\,\,\,\,\text{,}
\label{13.69} \\
\overline{A} &=&k_1\cdot d+k_2\cdot \theta \cdot h_u=k_1\cdot (\sigma
+\omega +\frac 1{n-1}\cdot \theta \cdot h_u)+k_2\cdot \theta \cdot h_u= 
\nonumber \\
&=&k_1\cdot (\sigma +\omega )+(\frac{k_1}{n-1}+k_2)\cdot \theta \cdot h_u= 
\nonumber \\
&=&k_1\cdot (\sigma +\omega )+\overline{k}_2\cdot \theta \cdot h_u\text{
,\thinspace \thinspace \thinspace \thinspace \thinspace \thinspace
\thinspace \thinspace \thinspace \thinspace \thinspace \thinspace \thinspace
\thinspace \thinspace \thinspace \thinspace \thinspace \thinspace \thinspace
\thinspace \thinspace \thinspace \thinspace \thinspace \thinspace \thinspace
\thinspace \thinspace \thinspace \thinspace \thinspace \thinspace \thinspace
\thinspace \thinspace }\overline{k}_2=\frac{k_1}{n-1}+k_2\,\,\,\,\,\text{.}
\label{13.70}
\end{eqnarray}

Since the viscosity tensor $g(^kS)g=\overline{A}$ could be written in the
forms 
\begin{eqnarray}
g(^kS)g &=&\overline{A}=\,_{ks}\overline{D}+\,_k\overline{W}+\frac
1{n-1}\cdot \,_k\overline{U}\cdot h_u=  \nonumber \\
&=&k_1\cdot (\sigma +\,\omega )+\overline{k}_2\cdot \,\theta \cdot h_u\,\,%
\text{\thinspace \thinspace \thinspace ,}  \label{13.71}
\end{eqnarray}

\noindent we can decompose the generalized Hook's law for isotropic elastic
media in its three parts 
\begin{eqnarray}
_{ks}\overline{D} &=&k_1\cdot \,\sigma \,\,\,\,\,\text{,}  \label{13.72} \\
_k\overline{W} &=&k_1\cdot \,\omega \,\,\,\,\,\,\text{,}  \label{13.73} \\
\,_k\overline{U} &=&(n-1)\cdot \overline{k}_2\cdot \,\theta \,=\overline{k}%
\cdot \theta \,\,\,\,\text{.}  \label{13.74}
\end{eqnarray}

The energy-momentum tensors $G\sim (\theta $, $_sT)$ will have for isotropic
elastic media the form 
\begin{eqnarray}
g(G) &=&(\rho _G+\frac 1e\cdot p\cdot k)\cdot g(u)\otimes g(u)-p\cdot
g(Kr)+g(u)\otimes g(^k\pi )+g(^ks)\otimes g(u)+  \nonumber \\
&&+k_1\cdot (\sigma +\,\omega )+\overline{k}_2\cdot \,\theta \cdot h_u\text{
\thinspace \thinspace \thinspace .}  \label{13.75}
\end{eqnarray}

For $\overline{g}[\overline{A}]=g^{\overline{i}\overline{j}}\cdot \overline{A%
}_{ij}=g^{ij}\cdot \overline{A}_{\overline{i}\overline{j}}$, it follows the
expression 
\begin{eqnarray}
\overline{g}[\overline{A}] &=&g^{\overline{i}\overline{j}}\cdot \overline{A}%
_{ij}=(\frac{k_1}{n-1}+k_2)\cdot \theta \cdot \overline{g}[h_u%
]=[k_1+(n-1)\cdot k_2]\cdot \theta =  \label{13.76} \\
&=&\overline{k}\cdot \theta =(n-1)\cdot \overline{k}_2\cdot \theta \,\,\,\,\,%
\text{.}  \label{13.77}
\end{eqnarray}

The constant $\overline{k}$ is called {\it volume viscosity coefficient}.
The condition $\overline{k}=0$ ($n\neq 1$) is called {\it Stokes' condition}%
. It leads to $\overline{g}[\overline{A}]=0$ and to vanishing of the inner
energy (enthalpy) density 
\begin{equation}
_kE=\frac 1{n-1}\cdot \,_k\overline{U}=0\text{ \thinspace \thinspace
\thinspace \thinspace ,}  \label{13.78}
\end{equation}

\noindent because of $_k\overline{U}=\overline{g}[_s\overline{A}]=\overline{g%
}[\overline{A}]=0$.

For a fluid with $_kE=0$ we have the canonical form of the energy-momentum
tensors $G$ ($G\sim \theta $, $_sT$) 
\begin{eqnarray}
G &=&(\rho _G+\frac 1e\cdot p\cdot k)\cdot u\otimes g(u)-p\cdot Kr+u\otimes
g(^k\pi )+\,^ks\otimes g(u)+  \nonumber \\
&&+\overline{g}(_{ks}\overline{D})+\overline{g}(_k\overline{W})\text{
\thinspace \thinspace .}  \label{13.79}
\end{eqnarray}

{\it Special case:} Isotropic homogeneous Newton's fluid with $k_2=0$. For $%
k_2=0$, it follows that

\begin{eqnarray}
_Nk_{ij}\,^{km} &=&k_1\cdot h_{il}\cdot h_{jn}\cdot g^{\overline{l}k}\cdot
g^{\overline{n}m}\text{ \thinspace \thinspace \thinspace \thinspace
,\thinspace \thinspace \thinspace \thinspace \thinspace \thinspace
\thinspace \thinspace \thinspace \thinspace \thinspace \thinspace \thinspace 
}_Nk=k_1\cdot h_u(\overline{g})\otimes h_u(\overline{g})\,\,\,\,\text{,}
\label{13.80} \\
\overline{A}_{ij} &=&k_1\cdot h_{il}\cdot h_{jn}\cdot g^{\overline{l}k}\cdot
g^{\overline{n}m}\cdot d_{\overline{k}\overline{m}}=k_1\cdot
d_{ij}\,\,\,\,\,\,\,\text{,\thinspace \thinspace \thinspace \thinspace
\thinspace \thinspace \thinspace \thinspace \thinspace \thinspace \thinspace
\thinspace \thinspace \thinspace }\overline{A}=k_1\cdot d\,\,\,\,\,\text{,}
\label{13.81} \\
_k\overline{U} &=&\overline{g}[_s\overline{A}]=\overline{g}[\overline{A}%
]=k_1\cdot \overline{g}[d]=k_1\cdot \theta \,\,\,\,\,\,\,\text{,\thinspace
\thinspace \thinspace \thinspace \thinspace \thinspace \thinspace \thinspace
\thinspace \thinspace \thinspace \thinspace \thinspace \thinspace \thinspace 
}_kE=\frac{k_1}{n-1}\cdot \theta \,\,\,\,\,\,\text{.}  \label{13.82}
\end{eqnarray}

Let us now consider the trace $\overline{g}[g(G)]=g^{\overline{i}\overline{j}%
}\cdot G_{ij}$ of the energy-momentum tensors $G\sim (\theta $,\thinspace $%
_sT$). By the use of the relations 
\begin{eqnarray*}
\overline{g}[g(u)\otimes g(\xi )] &=&g^{\overline{i}\overline{j}}\cdot
g_{ik}\cdot u^{\overline{k}}\cdot g_{jl}\cdot \xi ^{\overline{l}}=g(u,\xi
)\,\,\,\,\,\text{,} \\
\overline{g}[g(G)] &=&\rho _G\cdot e+\{k-\overline{g}[g(Kr)]\}\cdot
p\,\,\,\,\,\text{,} \\
g(Kr) &=&g_{i\overline{j}}\cdot dx^i\otimes dx^j\text{ \thinspace \thinspace
\thinspace \thinspace \thinspace \thinspace \thinspace \thinspace \thinspace
,\thinspace \thinspace \thinspace \thinspace \thinspace \thinspace
\thinspace \thinspace \thinspace \thinspace \thinspace \thinspace \thinspace
\thinspace } \\
\text{\thinspace \thinspace }\overline{g}[g(Kr)] &=&g^{\overline{i}\overline{%
j}}\cdot g_{i\overline{j}}=f^k\,_j\cdot g_{ik}\cdot g^{\overline{i}\overline{%
j}}=f^k\,_j\cdot g_k^j=f^k\,_k=f\text{ \thinspace \thinspace \thinspace
\thinspace \thinspace ,}
\end{eqnarray*}

\noindent the explicit form for $\overline{g}[g(G)]$ follows as 
\begin{equation}
\overline{g}[g(G)]=\rho _G\cdot e+(k-f)\cdot p\,\,\,\,\,\,\text{.}
\label{13.83}
\end{equation}

Now we can express the pressure $p$ by means the rest mass density $\rho _G$
and the trace $\overline{g}[g(G)]$ of the energy-momentum tensor $G$%
\begin{eqnarray}
p &=&\frac 1{f-k}\cdot (\rho _G\cdot e-\overline{g}[g(G)])=  \nonumber \\
&=&\frac 1{f-k}\cdot (\rho _G\cdot g_{\overline{i}\overline{j}}\cdot
u^i\cdot u^j-g^{\overline{i}\overline{j}}\cdot G_{ij})\,\,\,\,\,\text{.}
\label{13.84}
\end{eqnarray}

If we knew the explicit form of $\rho _G$ and $G_{ij}$ (on the basis of
experimental data) we can find the corresponding Euler-Lagrange's equations
for the unknown field variables on which the pressure $p=L$ depends.

{\it Special case:} $(L_n,g)$-spaces: $S=C:f^i\,_j=g_j^i$ ,\thinspace
\thinspace \thinspace \thinspace \thinspace $f=g_k^k=n$,\thinspace
\thinspace \thinspace \thinspace $k=1$%
\begin{eqnarray}
\overline{g}[g(G)] &=&\rho _G\cdot e+(1-n)\cdot p\,\,\,\,\,\,\,\,\,\,\text{,}
\nonumber \\
p &=&\frac 1{n-1}\cdot (\rho _G\cdot e-\overline{g}[g(G)])\,\,\,\,\text{.}
\label{13.85}
\end{eqnarray}

\subsubsection{Barotropic systems}

\begin{definition}
Dynamical system for which the pressure $p$ is depending only on $\rho
_G\cdot e$ is called {\it barotropic system} \cite{Mase}.
\end{definition}

All isotropic homogeneous Newton's fluids with vanishing volume viscosity
coefficient $\overline{k}$ ($\overline{k}=0$) and vanishing trace $\overline{%
g}[g(G)]$ [$\overline{g}[g(G)]=0$] of their energy-momentum tensors $G$ are
barotropic systems fulfilling the condition 
\begin{equation}
L=p=\frac 1{f-k}\cdot \rho _G\cdot e\,\,\,\,\,\text{.}  \label{13.86}
\end{equation}

{\it Special case:} $(L_n,g)$-spaces: $S=C:f^i\,_j=g_j^i$ ,\thinspace
\thinspace \thinspace \thinspace \thinspace $f=g_k^k=n$,\thinspace
\thinspace \thinspace \thinspace $k=1$,\thinspace \thinspace \thinspace
\thinspace $\overline{g}[g(G)]=0$%
\begin{equation}
p=\frac 1{n-1}\cdot \rho _G\cdot e\,\,\,\,\,\text{.}  \label{13.87}
\end{equation}
\thinspace \thinspace

For $n=4:$%
\begin{equation}
p=\frac 13\cdot \rho _G\cdot e\,\,\,\,\,\text{.}  \label{13.88}
\end{equation}

\section{Navier-Stokes's identities. Generalized Navier-Stokes' equation}

By the use of the method of Lagrangians with covariant derivatives (MLCD) 
\cite{Manoff-01} the different energy-momentum tensors and the covariant
Noether's identities for a field theory as well as for a theory of
continuous media can be found. On the basis of the $(n-1)+1$ projective
formalism and by the use of the notion of covariant divergency of a tensor
of second rank the corresponding covariant divergencies of the
energy-momentum tensors could be found. They lead to Navier-Stokes' identity
and to the corresponding generalized Navier-Stokes' equations.

\subsection{Covariant divergency of the energy-momentum tensors and the rest
mass density}

The covariant divergency $\delta G$ of the energy-momentum tensor $\delta G$
($G\sim \,\theta $, $_sT$, $Q$) can be represented by the use of the
projective metrics $h^u$, $h_u$ of the contravariant vector field $u$ and
the rest mass density for the corresponding energy-momentum tensor. In this
case the representation of the energy-momentum tensor is in the form 
\begin{equation}
G=(\rho _G+\frac 1e\cdot L\cdot k)\cdot u\otimes g(u)-L\cdot Kr+u\otimes
g(^k\pi )+\,^ks\otimes g(u)+(^kS)g\text{ ,}  \label{14.1}
\end{equation}

\noindent where 
\begin{equation}  \label{X.3.-1}
^k\pi =\,^G\overline{\pi }\text{ ,\thinspace \thinspace \thinspace
\thinspace \thinspace \thinspace \thinspace }^ks=\,^G\overline{s}\text{
,\thinspace \thinspace \thinspace \thinspace \thinspace \thinspace
\thinspace \thinspace }^kS=\,^G\overline{S}\text{ .}
\end{equation}

By the use of the relations 
\begin{equation}
\begin{array}{c}
u(\rho _G+\frac 1e\cdot L\cdot k)=u\rho _G+L\cdot u(\frac 1e\cdot k)+\frac
1e\cdot k\cdot (uL)= \\ 
\\ 
=[\rho _{G/\alpha }+L\cdot (\frac 1e\cdot k)_{/\alpha }+\frac 1e\cdot k\cdot
L_{/\alpha }]\cdot u^\alpha = \\ 
\\ 
=[\rho _{G,j}+L\cdot (\frac 1e\cdot k)_{,j}+\frac 1e\cdot k\cdot
L_{,j}]\cdot u^j\text{ ,}
\end{array}
\label{X.3.-2}
\end{equation}
\begin{equation}
KrL=L_{/\alpha }\cdot e^\alpha =L_{,i}\cdot dx^i=\overline{\nabla }_{Kr}L%
\text{ ,}  \label{X.3.-3}
\end{equation}
\begin{equation}
\delta (L\cdot Kr)=KrL+L\cdot \delta Kr\text{ ,}  \label{X.3.-4}
\end{equation}
\begin{equation}
\begin{array}{c}
\delta (L\cdot Kr)=\frac 12\cdot [\nabla _{\overline{g}}(L\cdot
Kr)]g=(L\cdot g_\alpha ^\beta )_{/\beta }\cdot e^\alpha =(L\cdot
g_i^j)_{;j}\cdot dx^i= \\ 
=(L_{/\beta }\cdot g_\alpha ^\beta +L\cdot g_\alpha ^\beta \,_{/\beta
})\cdot e^\alpha =(L_{,i}+L\cdot g_i^j\,_{;j})\cdot dx^i\text{ ,}
\end{array}
\label{X.3.-5}
\end{equation}
\begin{equation}
\delta (u\otimes g(^G\overline{\pi }))=\delta u\cdot g(^G\overline{\pi }%
)+g(\nabla _u\,^G\overline{\pi })+(\nabla _ug)(^G\overline{\pi })\text{ ,}
\label{X.3.-6}
\end{equation}
\begin{equation}
\delta (^G\overline{s}\otimes g(u))=\delta ^G\overline{s}\cdot g(u)+g(\nabla
_{^G\overline{s}}u)+(\nabla _{^G\overline{s}}g)(u)\text{ ,}  \label{X.3.-7}
\end{equation}
\[
\delta ((^G\overline{S})g)=(g_{\alpha \overline{\gamma }}\cdot \,^G\overline{%
S}\,^{\beta \gamma })_{/\beta }\cdot e^\alpha \text{ [see (\ref{X.2.-5})],} 
\]

\noindent $\delta G$ and $\overline{g}(\delta G)$ can be found in the forms 
\begin{equation}
\begin{array}{c}
\delta G=(\rho _G+\frac 1e\cdot L\cdot k)\cdot g(a)+ \\ 
+[u(\rho _G+\frac 1e\cdot L\cdot k)+\,\,\,(\rho _G+\frac 1e\cdot L\cdot
k)\cdot \delta u+\delta ^G\overline{s}]\cdot g(u)- \\ 
-\,\,KrL-L\cdot \delta Kr+\delta u\cdot g(^G\overline{\pi })+g(\nabla _u\,^G%
\overline{\pi })+g(\nabla _{^G\overline{s}}u)+ \\ 
+\,\,(\rho _G+\frac 1e\cdot L\cdot k)\cdot (\nabla _ug)(u)+(\nabla _ug)(^G%
\overline{\pi })+(\nabla _{^G\overline{s}}g)(u)+ \\ 
+\,\,\delta ((^G\overline{S})g)\text{ ,}
\end{array}
\label{X.3.-8}
\end{equation}
\begin{equation}
\begin{array}{c}
\overline{g}(\delta G)=(\rho _G+\frac 1e\cdot L\cdot k)\cdot a+ \\ 
+[u(\rho _G+\frac 1e\cdot L\cdot k)+\,\,\,\,\,\,(\rho _G+\frac 1e\cdot
L\cdot k)\cdot \delta u+\delta ^G\overline{s}]\cdot u- \\ 
-\,\overline{g}(\,KrL)-L\cdot \overline{g}(\delta Kr)+\delta u\cdot \,^G%
\overline{\pi }+\nabla _u\,^G\overline{\pi }+\nabla _{^G\overline{s}}u+ \\ 
+\,\,(\rho _G+\frac 1e\cdot L\cdot k)\cdot \overline{g}(\nabla _ug)(u)+%
\overline{g}(\nabla _ug)(^G\overline{\pi })+\overline{g}(\nabla _{^G%
\overline{s}}g)(u)+ \\ 
+\,\,\overline{g}(\delta ((^G\overline{S})g))\text{ .}
\end{array}
\label{X.3.-9}
\end{equation}

In a co-ordinate basis $\delta G$ and $\overline{g}(\delta G)$ will have the
forms 
\begin{equation}
\begin{array}{c}
G_i\,^j\,_{;j}=(\rho _G+\frac 1e\cdot L\cdot k)\cdot a_i+ \\ 
+\,\,[(\rho _G+\frac 1e\cdot L\cdot k)_{,j}\cdot u^j+(\rho _G+\frac 1e\cdot
L\cdot k)\cdot u^j\,_{;j}+\,^G\overline{s}^j\,_{;j}]\cdot u_i- \\ 
-L_{,i}-L\cdot g_{i\,;j}^j+u^j\,_{;j}\cdot \,^G\overline{\pi }_i+g_{i%
\overline{j}}\cdot (^G\overline{\pi }^j\,_{;k}\cdot u^k+u^j\,_{;k}\cdot \,^G%
\overline{s}^k)+ \\ 
+\,\,\,g_{ij;k}\cdot [(\rho _G+\frac 1e\cdot L\cdot k)\cdot u^{\overline{j}%
}\cdot u^k+\,^G\overline{\pi }^{\overline{j}}\cdot u^k+u^{\overline{j}}\cdot
\,^G\overline{s}^k]+ \\ 
+\,(g_{i\overline{k}}\cdot \,^G\overline{S}\,^{jk})_{;j}\text{ ,}
\end{array}
\label{X.3.-12}
\end{equation}
\begin{equation}
\begin{array}{c}
\overline{g}^{i\overline{k}}\cdot G_k\,^j\,_{;j}=(\rho _G+\frac 1e\cdot
L\cdot k)\cdot a^i+ \\ 
+\,\,[(\rho _G+\frac 1e\cdot L\cdot k)_{,j}\cdot u^j+(\rho _G+\frac 1e\cdot
L\cdot k)\cdot u^j\,_{;j}+\,^G\overline{s}^j\,_{;j}]\cdot u^i- \\ 
-L_{,j}\cdot g^{i\overline{j}}-L\cdot g^{i\overline{k}}\cdot
g_{k\,;j}^j+u^j\,_{;j}\cdot \,^G\overline{\pi }^i+\,^G\overline{\pi }%
^i\,_{;j}\cdot u^j+u^i\,_{;j}\cdot \,^G\overline{s}^j+ \\ 
+\,\,\,g^{i\overline{l}}\cdot g_{lj;k}\cdot [(\rho _G+\frac 1e\cdot L\cdot
k)\cdot u^{\overline{j}}\cdot u^k+\,^G\overline{\pi }^{\overline{j}}\cdot
u^k+u^{\overline{j}}\cdot \,^G\overline{s}^k]+ \\ 
+\,g^{i\overline{l}}\cdot (g_{l\overline{k}}\cdot \,^G\overline{S}%
\,^{jk})_{;j}\text{ .}
\end{array}
\label{X.3.-13}
\end{equation}

\subsection{Covariant divergency of the energy-momentum tensors for linear
elastic theory}

For linear elastic theory we have to specialize the structure of the term $%
\delta ((^G\overline{S})g)$ in the expression for the divergency of the
energy-momentum tensor $G\sim (\theta $,\thinspace $_sT)$. Since $(^G%
\overline{S})g=(^kS)g=\overline{g}(g(^kS)g)=\overline{g}(\overline{A})$, we
have 
\begin{equation}
\delta ((^G\overline{S})g)=\delta (\overline{g}(\overline{A}))\,\,\,\text{.}
\label{14.19}
\end{equation}

\subsubsection{Representation of $\delta (\overline{g}(\overline{A}))\,$}

For $\overline{g}(\overline{A})=\overline{g}(\,_{ks}\overline{D})+\,%
\overline{g}(_k\overline{W})+\frac 1{n-1}\cdot \,_k\overline{U}\cdot 
\overline{g}(h_u)$ we obtain by the use of the general formulae for
covariant divergency of a tensor field 
\begin{equation}
\delta (\overline{g}(\overline{A}))=\delta (\overline{g}(\,_{ks}\overline{D}%
))+\,\delta (\overline{g}(_k\overline{W}))+\frac 1{n-1}\cdot \,\delta (_k%
\overline{U}\cdot \overline{g}(h_u))\,\,\,\,\,\,\,\text{,}  \label{14.20}
\end{equation}

\noindent where 
\begin{eqnarray}
\delta (\overline{g}(\overline{A})) &=&(g^{i\overline{j}}\overline{A}%
_{jk})_{;i}\cdot dx^k=(g^{ij}\,_{;i}\cdot \overline{A}_{\overline{j}k}+g^{i%
\overline{j}}\cdot \overline{A}_{jk;i})\cdot dx^k=  \nonumber \\
&=&g^{ij}\,_{;i}\cdot (_{ks}\overline{D}_{\overline{j}k}+\,_k\overline{W}_{%
\overline{j}k}+\frac 1{n-1}\cdot \,_k\overline{U}\cdot h_{\overline{j}%
k})\cdot dx^k+  \nonumber \\
&&+g^{i\overline{j}}\cdot [_{ks}\overline{D}_{\overline{j}k;i}+\,_k\overline{%
W}_{\overline{j}k;i}+\frac 1{n-1}\cdot (_k\overline{U}_{,i}\cdot h_{jk}+\,_k%
\overline{U}\cdot h_{jk;i})]\cdot dx^k\,\,\,\,\text{,}  \label{14.21}
\end{eqnarray}
\begin{eqnarray}
\delta (\overline{g}(\,_{ks}\overline{D})) &=&\delta (\overline{g}(_s%
\overline{C}[_\sigma R]))+\delta (\overline{g}(_s\overline{C}[_\omega R%
]))+\frac 1{n-1}\cdot \delta (_\theta R\cdot \overline{g}(_s\overline{C}[h_u%
]))\,\,\,\,\,\text{,}  \nonumber \\
\delta (\overline{g}(\,_k\overline{W})) &=&\delta (\overline{g}(_aC[_\sigma
R]))+\delta (\overline{g}(_aC[_\omega R]))+\frac 1{n-1}\cdot \delta (_\theta
R\cdot \overline{g}(_aC[h_u]))\,\,\,\text{,}  \nonumber \\
\delta (_k\overline{U}\cdot \overline{g}(h_u)) &=&\delta ((\overline{g}[C%
])[R]\cdot \overline{g}(h_u))\,\,\,\,\,\text{,}  \nonumber \\
\overline{g}(\,_{ks}\overline{D}) &=&g^{i\overline{k}}\cdot \,_{ks}\overline{%
D}_{kj}\cdot \partial _i\otimes dx^j\,\,\,\,\,\text{,}  \nonumber \\
\delta (\overline{g}(\,_{ks}\overline{D})) &=&(g^{i\overline{k}}\cdot \,_{ks}%
\overline{D}_{kj})_{;i}\cdot dx^j=(g^{ij}\,_{;i}\cdot \,_{ks}\overline{D}_{%
\overline{j}k}+g^{i\overline{j}}\cdot \,_{ks}\overline{D}_{\overline{j}%
k;i})\cdot dx^k\,\text{\thinspace ,}  \label{14.22} \\
\delta (\overline{g}(\,_k\overline{W})) &=&(g^{i\overline{j}}\cdot \,_k%
\overline{W}_{jk})_{;i}\cdot dx^k=(g^{ij}\,_{;i}\cdot \,_k\overline{W}_{%
\overline{j}k}+g^{i\overline{j}}\cdot \,_k\overline{W}_{\overline{j}%
k;i})\cdot dx^k\,\,\,\,\text{,}  \nonumber \\
\delta (_k\overline{U}\cdot \overline{g}(h_u)) &=&\delta (_k\overline{U}%
\cdot g^{i\overline{j}}\cdot h_{jk}\cdot \partial _i\otimes dx^k)=  \nonumber
\\
&=&[_k\overline{U}_{,i}\cdot g^{i\overline{j}}\cdot h_{jk}+\,_k\overline{U}%
\cdot (g^{ij}\,_{;i}\cdot h_{\overline{j}k}+g^{i\overline{j}}\cdot
h_{jk;i})]\cdot dx^k\text{ ,}  \nonumber \\
g^{i\overline{j}} &=&f^j\,_l\cdot g^{il}=f^j\,_l\cdot g^{li}=g^{\overline{j}%
i\,}\,\,\,\,\,\,\,\text{,}  \nonumber
\end{eqnarray}
\begin{eqnarray}
\overline{g}(_s\overline{C}[_\sigma R]) &=&g^{i\overline{j}}\cdot \,_s%
\overline{C}_{jk}\,^{mn}\cdot \,_\sigma R_{\overline{m}\overline{n}}\cdot
\partial _i\otimes dx^k\,\,\,\text{,}  \nonumber \\
\delta (\overline{g}(_s\overline{C}[_\sigma R])) &=&(g^{i\overline{j}}\cdot
\,_s\overline{C}_{jk}\,^{mn}\cdot \,_\sigma R_{\overline{m}\overline{n}%
})_{;i}\cdot dx^k\,\,\,\,\,\,\text{,}  \nonumber \\
(g^{i\overline{j}}\cdot \,_s\overline{C}_{jk}\,^{mn}\cdot \,_\sigma R_{%
\overline{m}\overline{n}})_{;i} &=&g^{ij}\,_{;i}\cdot _s\overline{C}%
_{jk}\,^{mn}\cdot \,_\sigma R_{\overline{m}\overline{n}}+  \nonumber \\
&&+g^{i\overline{j}}\cdot (_s\overline{C}_{jk}\,^{mn}\,_{;i}\cdot \,_\sigma
R_{\overline{m}\overline{n}}+_s\overline{C}_{jk}\,^{mn}\cdot \,_\sigma R_{%
\overline{m}\overline{n};i})\text{ \thinspace \thinspace ,}  \label{14.23} \\
\delta ((\overline{g}[C])[R]\cdot \overline{g}(h_u)) &=&\delta (g^{\overline{%
r}\overline{l}}\cdot C_{rl}\,^{mn}\cdot R_{\overline{m}\overline{n}}\cdot
g^{i\overline{j}}\cdot h_{jk})\cdot \partial _i\otimes dx^k\text{ \thinspace
\thinspace ,}  \label{14.25} \\
\delta (g^{\overline{r}\overline{l}}\cdot C_{rl}\,^{mn}\cdot R_{\overline{m}%
\overline{n}}\cdot g^{i\overline{j}}\cdot h_{jk}) &=&g^{rl}\,_{;i}\cdot
C_{rl}\,^{mn}\cdot R_{\overline{m}\overline{n}}\cdot g^{i\overline{j}}\cdot
h_{jk}+  \nonumber
\end{eqnarray}
\begin{eqnarray}
&&+g^{\overline{r}\overline{l}}\cdot (C_{rl}\,^{mn}\,_{;i}\cdot R_{\overline{%
m}\overline{n}}+C_{rl}\,^{\overline{m}\overline{n}}\cdot R_{mn;i})\cdot g^{i%
\overline{j}}\cdot h_{jk}+  \nonumber \\
&&+g^{\overline{r}\overline{l}}\cdot C_{rl}\,^{mn}\cdot R_{\overline{m}%
\overline{n}}\cdot (g^{ij}\,_{;i}\cdot h_{\overline{j}k}+g^{i\overline{j}%
}\cdot h_{jk;i})\text{ \thinspace \thinspace \thinspace \thinspace
\thinspace \thinspace .}  \label{14.26}
\end{eqnarray}

\subsubsection{Isotropic elastic media}

For isotropic elastic media $\overline{A}=\lambda _1\cdot (_\sigma
R+\,_\omega R)+\overline{\lambda }\cdot \,_\theta R\cdot h_u$. Then, it
follows for $(^kS)g=\overline{g}(\overline{A})$%
\begin{eqnarray}
\overline{g}(\overline{A}) &=&\lambda _1\cdot \overline{g}(_\sigma
R+\,_\omega R)+\overline{\lambda }\cdot \,_\theta R\cdot \overline{g}(h_u)= 
\nonumber \\
&=&\lambda _1\cdot \overline{g}(_\sigma R)+\,\overline{g}(_\omega R)+%
\overline{\lambda }\cdot \,_\theta R\cdot \overline{g}(h_u)\text{ \thinspace
\thinspace \thinspace \thinspace ,\thinspace \thinspace \thinspace
\thinspace \thinspace \thinspace \thinspace }\lambda _1\text{,\thinspace
\thinspace \thinspace }\overline{\lambda }=\text{const.\thinspace \thinspace
\thinspace \thinspace ,}  \nonumber \\
\delta (\overline{g}(\overline{A})) &=&\lambda _1\cdot \delta (\overline{g}%
(_\sigma R))+\lambda _1\cdot \,\delta (\overline{g}(_\omega R))+\overline{%
\lambda }\cdot \delta (\,_\theta R\cdot \overline{g}(h_u))\text{ \thinspace
\thinspace \thinspace \thinspace ,}  \label{14.27}
\end{eqnarray}

\noindent where 
\begin{eqnarray}
\delta (\overline{g}(_\sigma R)) &=&\delta (\overline{g}(_\sigma R))=(g^{i%
\overline{j}}\cdot \,_\sigma R_{jk})_{;i}\cdot dx^k=(g^{ij}\,_{;i}\cdot
\,_\sigma R_{\overline{j}k}+g^{i\overline{j}}\cdot \,_\sigma R_{\overline{j}%
k;i})\cdot dx^k\,\,\,\,\text{,}  \nonumber \\
\delta (\overline{g}(_\omega R)) &=&\delta (\overline{g}(_\omega R))=(g^{i%
\overline{j}}\cdot \,_\omega R_{jk})_{;i}\cdot dx^k=(g^{ij}\,_{;i}\cdot
\,_\omega R_{\overline{j}k}+g^{i\overline{j}}\cdot \,_\omega R_{\overline{j}%
k;i})\cdot dx^k\,\,\,\,\text{,}  \nonumber \\
\delta (\,_\theta R\cdot \overline{g}(h_u)) &=&\delta (_\theta R\cdot g^{i%
\overline{j}}\cdot h_{jk}\cdot \partial _i\otimes dx^k)=  \nonumber \\
&=&[_\theta R_{,i}\cdot g^{i\overline{j}}\cdot h_{jk}+\,_\theta R\cdot
(g^{ij}\,_{;i}\cdot h_{\overline{j}k}+g^{i\overline{j}}\cdot h_{jk;i})]\cdot
dx^k\,\,\,\,\,\text{.}  \label{14.28}
\end{eqnarray}

For the covariant divergency $\delta (\overline{g}(\overline{A}))$ of the
viscosity tensor $(^kS)g=\overline{g}(\overline{A})$ we obtain in a
co-ordinate (or in a non-co-ordinate basis if $dx^k\rightarrow e^k$)

\begin{eqnarray}
\delta ((^kS)g) &=&\delta (\overline{g}(\overline{A}))=  \nonumber \\
&=&\{\lambda _1\cdot (g^{jk}\,_{;j}\cdot \,_\sigma R_{\overline{k}i}+g^{j%
\overline{k}}\cdot \,_\sigma R_{\overline{k}i;j}+g^{jk}\,_{;j}\cdot
\,_\omega R_{\overline{k}i}+g^{j\overline{k}}\cdot \,_\omega R_{ki;j})+ 
\nonumber \\
&&+\overline{\lambda }\cdot [_\theta R_{,j}\cdot g^{j\overline{k}}\cdot
h_{ki}+\,_\theta R\cdot (g^{jk}\,_{;j}\cdot h_{\overline{k}i}+g^{j\overline{k%
}}\cdot h_{ki;j})]\}\cdot dx^i\,\,\,\,\,\,\,\text{.}  \label{14.29}
\end{eqnarray}

The energy-momentum tensors $G\sim (\theta $,\thinspace $_sT)$ could be
found in the forms 
\[
\begin{array}{c}
\delta G=(\rho _G+\frac 1e\cdot L\cdot k)\cdot g(a)+ \\ 
+[u(\rho _G+\frac 1e\cdot L\cdot k)+\,\,\,(\rho _G+\frac 1e\cdot L\cdot
k)\cdot \delta u+\delta ^G\overline{s}]\cdot g(u)- \\ 
-\,\,KrL-L\cdot \delta Kr+\delta u\cdot g(^G\overline{\pi })+g(\nabla _u\,^G%
\overline{\pi })+g(\nabla _{^G\overline{s}}u)+ \\ 
+\,\,(\rho _G+\frac 1e\cdot L\cdot k)\cdot (\nabla _ug)(u)+(\nabla _ug)(^G%
\overline{\pi })+(\nabla _{^G\overline{s}}g)(u)+ \\ 
+\,\text{ }\lambda _1\cdot \delta (\overline{g}(_\sigma R))+\lambda _1\cdot
\,\delta (\overline{g}(_\omega R))+\overline{\lambda }\cdot \delta
(\,_\theta R\cdot \overline{g}(h_u))\text{,}
\end{array}
\]

\[
\begin{array}{c}
G_i\,^j\,_{;j}=(\rho _G+\frac 1e\cdot L\cdot k)\cdot a_i+ \\ 
+\,\,[(\rho _G+\frac 1e\cdot L\cdot k)_{,j}\cdot u^j+(\rho _G+\frac 1e\cdot
L\cdot k)\cdot u^j\,_{;j}+\,^G\overline{s}^j\,_{;j}]\cdot u_i- \\ 
-L_{,i}-L\cdot g_{i\,;j}^j+u^j\,_{;j}\cdot \,^G\overline{\pi }_i+g_{i%
\overline{j}}\cdot (^G\overline{\pi }^j\,_{;k}\cdot u^k+u^j\,_{;k}\cdot \,^G%
\overline{s}^k)+ \\ 
+\,\,\,g_{ij;k}\cdot [(\rho _G+\frac 1e\cdot L\cdot k)\cdot u^{\overline{j}%
}\cdot u^k+\,^G\overline{\pi }^{\overline{j}}\cdot u^k+u^{\overline{j}}\cdot
\,^G\overline{s}^k]+ \\ 
+\lambda _1\cdot (g^{jk}\,_{;j}\cdot \,_\sigma R_{\overline{k}i}+g^{j%
\overline{k}}\cdot \,_\sigma R_{\overline{k}i;j}+g^{jk}\,_{;j}\cdot
\,_\omega R_{\overline{k}i}+g^{j\overline{k}}\cdot \,_\omega R_{ki;j}) \\ 
+\overline{\lambda }\cdot [_\theta R_{,j}\cdot g^{j\overline{k}}\cdot
h_{ki}+\,_\theta R\cdot (g^{jk}\,_{;j}\cdot h_{\overline{k}i}+g^{j\overline{k%
}}\cdot h_{ki;j})]\,\,\,\,\,\,\,\,\text{.}
\end{array}
\]

\subsection{Covariant divergency of the energy-momentum tensors for Newton's
fluids}

Analogous considerations as in the case of linear elasticity theory could be
made for Newton's fluids. A comparison of $\overline{A}=\,_Nk[d]$
,\thinspace \thinspace \thinspace \thinspace $\overline{A}%
_{ij}=\,\,_Nk_{ij}{}^{km}\cdot d_{\overline{k}\overline{m}}$ with $\overline{%
A}=\,C[R]$ ,\thinspace \thinspace \thinspace \thinspace $\overline{A}%
_{ij}=\,C_{ij}{}^{km}\cdot R_{\overline{k}\overline{m}}$ shows that the
results for the Newton fluids could be formally found from the results for
linear elasticity theory by the use of the following substitutions 
\begin{eqnarray}
C &\rightarrow &\,_Nk\text{ ,\thinspace \thinspace \thinspace \thinspace
\thinspace \thinspace \thinspace \thinspace \thinspace \thinspace }%
R\rightarrow d\,\,\,\,\text{,\thinspace \thinspace \thinspace \thinspace
\thinspace \thinspace \thinspace \thinspace \thinspace \thinspace \thinspace 
}_\sigma R\rightarrow \sigma \text{ \thinspace \thinspace \thinspace
\thinspace \thinspace \thinspace ,\thinspace \thinspace \thinspace
\thinspace \thinspace \thinspace \thinspace \thinspace \thinspace \thinspace
\thinspace }_\omega R\rightarrow \omega \,\,\,\,\,\,\text{,\thinspace
\thinspace \thinspace \thinspace \thinspace \thinspace \thinspace \thinspace
\thinspace \thinspace }_\theta R\rightarrow \theta \text{\thinspace
\thinspace \thinspace \thinspace \thinspace ,}  \nonumber \\
\lambda _1 &\rightarrow &k_1\,\,\,\,\,\text{,\thinspace \thinspace
\thinspace \thinspace \thinspace \thinspace \thinspace \thinspace \thinspace
\thinspace }\overline{\lambda }\rightarrow \overline{k}_2  \label{14.30}
\end{eqnarray}

For $\overline{g}(\overline{A})=\overline{g}(\,_{ks}\overline{D})+\,%
\overline{g}(_k\overline{W})+\frac 1{n-1}\cdot \,_k\overline{U}\cdot 
\overline{g}(h_u)$ we obtain by the use of the general formulae for
covariant divergency of a tensor field 
\begin{equation}
\delta (\overline{g}(\overline{A}))=\delta (\overline{g}(\,_{ks}\overline{D}%
))+\,\delta (\overline{g}(_k\overline{W}))+\frac 1{n-1}\cdot \,\delta (_k%
\overline{U}\cdot \overline{g}(h_u))\,\,\,\,\,\,\,\text{,}  \label{14.31}
\end{equation}

\noindent where 
\begin{eqnarray}
\delta (\overline{g}(\overline{A})) &=&(g^{i\overline{j}}\overline{A}%
_{jk})_{;i}\cdot dx^k=(g^{ij}\,_{;i}\cdot \overline{A}_{\overline{j}k}+g^{i%
\overline{j}}\cdot \overline{A}_{jk;i})\cdot dx^k=  \nonumber \\
&=&g^{ij}\,_{;i}\cdot (_{ks}\overline{D}_{\overline{j}k}+\,_k\overline{W}_{%
\overline{j}k}+\frac 1{n-1}\cdot \,_k\overline{U}\cdot h_{\overline{j}%
k})\cdot dx^k+  \nonumber \\
&&+g^{i\overline{j}}\cdot [_{ks}\overline{D}_{\overline{j}k;i}+\,_k\overline{%
W}_{\overline{j}k;i}+\frac 1{n-1}\cdot (_k\overline{U}_{,i}\cdot h_{jk}+\,_k%
\overline{U}\cdot h_{jk;i})]\cdot dx^k\,\,\,\,\text{,}  \label{14.32}
\end{eqnarray}
\begin{eqnarray}
\delta (\overline{g}(\,_{ks}\overline{D})) &=&\delta (\overline{g}(_{sN}%
\overline{k}[\sigma ]))+\delta (\overline{g}(_{sN}\overline{k}[\omega %
]))+\frac 1{n-1}\cdot \delta (\theta \cdot \overline{g}(_{sN}\overline{k}[h_u%
]))\,\,\,\,\,\text{,}  \nonumber \\
\delta (\overline{g}(\,_k\overline{W})) &=&\delta (\overline{g}%
(_{aN}k[\sigma ]))+\delta (\overline{g}(_{aN}k[\omega ]))+\frac 1{n-1}\cdot
\delta (\theta \cdot \overline{g}(_{aN}k[h_u]))\,\,\,\text{,}  \nonumber \\
\delta (_k\overline{U}\cdot \overline{g}(h_u)) &=&\delta ((\overline{g}[_Nk%
])[d]\cdot \overline{g}(h_u))\,\,\,\,\,\text{,}  \nonumber \\
\overline{g}(\,_{ks}\overline{D}) &=&g^{i\overline{k}}\cdot \,_{ks}\overline{%
D}_{kj}\cdot \partial _i\otimes dx^j\,\,\,\,\,\text{,}  \nonumber \\
\delta (\overline{g}(\,_{ks}\overline{D})) &=&(g^{i\overline{k}}\cdot \,_{ks}%
\overline{D}_{kj})_{;i}\cdot dx^j=(g^{ij}\,_{;i}\cdot \,_{ks}\overline{D}_{%
\overline{j}k}+g^{i\overline{j}}\cdot \,_{ks}\overline{D}_{\overline{j}%
k;i})\cdot dx^k\,\text{\thinspace ,}  \label{14.33} \\
\delta (\overline{g}(\,_k\overline{W})) &=&(g^{i\overline{j}}\cdot \,_k%
\overline{W}_{jk})_{;i}\cdot dx^k=(g^{ij}\,_{;i}\cdot \,_k\overline{W}_{%
\overline{j}k}+g^{i\overline{j}}\cdot \,_k\overline{W}_{\overline{j}%
k;i})\cdot dx^k\,\,\,\,\text{,}  \nonumber \\
\delta (_k\overline{U}\cdot \overline{g}(h_u)) &=&\delta (_k\overline{U}%
\cdot g^{i\overline{j}}\cdot h_{jk}\cdot \partial _i\otimes dx^k)=  \nonumber
\\
&=&[_k\overline{U}_{,i}\cdot g^{i\overline{j}}\cdot h_{jk}+\,_k\overline{U}%
\cdot (g^{ij}\,_{;i}\cdot h_{\overline{j}k}+g^{i\overline{j}}\cdot
h_{jk;i})]\cdot dx^k\text{ ,}  \nonumber \\
g^{i\overline{j}} &=&f^j\,_l\cdot g^{il}=f^j\,_l\cdot g^{li}=g^{\overline{j}%
i\,}\,\,\,\,\,\,\,\text{,}  \nonumber
\end{eqnarray}
\begin{eqnarray}
\overline{g}(_{sN}\overline{k}[\sigma ]) &=&g^{i\overline{j}}\cdot \,_{sN}%
\overline{k}_{jk}\,^{mn}\cdot \,\sigma _{\overline{m}\overline{n}}\cdot
\partial _i\otimes dx^k\,\,\,\text{,}  \nonumber \\
\delta (\overline{g}(_{sN}\overline{k}[\sigma ])) &=&(g^{i\overline{j}}\cdot
\,_{sN}\overline{k}_{jk}\,^{mn}\cdot \,\sigma _{\overline{m}\overline{n}%
})_{;i}\cdot dx^k\,\,\,\,\,\,\text{,}  \nonumber \\
(g^{i\overline{j}}\cdot \,_{sN}\overline{k}_{jk}\,^{mn}\cdot \,\sigma _{%
\overline{m}\overline{n}})_{;i} &=&g^{ij}\,_{;i}\cdot \,_{sN}\overline{k}%
_{jk}\,^{mn}\cdot \,\sigma _{\overline{m}\overline{n}}+  \nonumber \\
&&+g^{i\overline{j}}\cdot (_{sN}\overline{k}_{jk}\,^{mn}\,_{;i}\cdot
\,\sigma _{\overline{m}\overline{n}}+\,_{sN}\overline{k}_{jk}\,^{mn}\cdot
\,\sigma _{\overline{m}\overline{n};i})\text{ \thinspace \thinspace ,}
\label{14.34} \\
\delta ((\overline{g}[_Nk])[d]\cdot \overline{g}(h_u)) &=&\delta (g^{%
\overline{r}\overline{l}}\cdot \,_Nk_{rl}\,^{mn}\cdot d_{\overline{m}%
\overline{n}}\cdot g^{i\overline{j}}\cdot h_{jk})\cdot \partial _i\otimes
dx^k\text{ \thinspace \thinspace ,}  \nonumber \\
\delta (g^{\overline{r}\overline{l}}\cdot \,_Nk_{rl}\,^{mn}\cdot d_{%
\overline{m}\overline{n}}\cdot g^{i\overline{j}}\cdot h_{jk})
&=&g^{rl}\,_{;i}\cdot \,_Nk_{rl}\,^{mn}\cdot d_{\overline{m}\overline{n}%
}\cdot g^{i\overline{j}}\cdot h_{jk}+  \nonumber
\end{eqnarray}
\begin{eqnarray}
&&+g^{\overline{r}\overline{l}}\cdot (_Nk_{rl}\,^{mn}\,_{;i}\cdot d_{%
\overline{m}\overline{n}}+\,_Nk_{rl}\,^{\overline{m}\overline{n}}\cdot
d_{mn;i})\cdot g^{i\overline{j}}\cdot h_{jk}+  \nonumber \\
&&+g^{\overline{r}\overline{l}}\cdot \,_Nk_{rl}\,^{mn}\cdot d_{\overline{m}%
\overline{n}}\cdot (g^{ij}\,_{;i}\cdot h_{\overline{j}k}+g^{i\overline{j}%
}\cdot h_{jk;i})\text{ \thinspace \thinspace \thinspace \thinspace
\thinspace \thinspace .}  \label{14.36}
\end{eqnarray}

\subsubsection{Isotropic homogeneous fluid}

For isotropic homogeneous fluid $\overline{A}=k_1\cdot (\sigma +\,\omega )+%
\overline{k}_2\cdot \,\theta \cdot h_u$. Then, it follows for $(^kS)g=%
\overline{g}(\overline{A})$%
\begin{eqnarray}
\overline{g}(\overline{A}) &=&k_1\cdot \overline{g}(\sigma +\,\omega )+%
\overline{k}_2\cdot \,\theta \cdot \overline{g}(h_u)=  \nonumber \\
&=&k_1\cdot \overline{g}(\sigma )+\,\overline{g}(\omega )+\overline{k}%
_2\cdot \,\theta \cdot \overline{g}(h_u)\text{ \thinspace \thinspace
\thinspace \thinspace ,\thinspace \thinspace \thinspace \thinspace
\thinspace \thinspace \thinspace \thinspace \thinspace \thinspace \thinspace
\thinspace \thinspace \thinspace }k_1\text{,\thinspace \thinspace \thinspace 
}\overline{k}_2=\text{const.\thinspace \thinspace \thinspace \thinspace ,}
\label{14.37} \\
\delta (\overline{g}(\overline{A})) &=&k_1\cdot \delta (\overline{g}(\sigma
))+k_1\cdot \,\delta (\overline{g}(\omega ))+\overline{k}_2\cdot \delta
(\,\theta \cdot \overline{g}(h_u))\text{ \thinspace \thinspace \thinspace
\thinspace ,}  \nonumber
\end{eqnarray}

\noindent where 
\begin{eqnarray}
\delta (\overline{g}(\sigma )) &=&\delta (\overline{g}(\sigma ))=(g^{i%
\overline{j}}\cdot \,\sigma _{jk})_{;i}\cdot dx^k=(g^{ij}\,_{;i}\cdot
\,\sigma _{\overline{j}k}+g^{i\overline{j}}\cdot \,\sigma _{\overline{j}%
k;i})\cdot dx^k\,\,\,\,\text{,}  \label{14.38} \\
\delta (\overline{g}(\omega )) &=&\delta (\overline{g}(\omega ))=(g^{i%
\overline{j}}\cdot \,\omega _{jk})_{;i}\cdot dx^k=(g^{ij}\,_{;i}\cdot
\,\omega _{\overline{j}k}+g^{i\overline{j}}\cdot \,\omega _{\overline{j}%
k;i})\cdot dx^k\,\,\,\,\text{,}  \label{14.39} \\
\delta (\,\theta \cdot \overline{g}(h_u)) &=&\delta (\theta \cdot g^{i%
\overline{j}}\cdot h_{jk}\cdot \partial _i\otimes dx^k)=  \nonumber \\
&=&[\theta _{,i}\cdot g^{i\overline{j}}\cdot h_{jk}+\,\theta \cdot
(g^{ij}\,_{;i}\cdot h_{\overline{j}k}+g^{i\overline{j}}\cdot h_{jk;i})]\cdot
dx^k\,\,\,\,\,\text{.}  \label{14.40}
\end{eqnarray}

For the covariant divergency $\delta (\overline{g}(\overline{A}))$ of the
viscosity tensor $(^kS)g=\overline{g}(\overline{A})$ we obtain in a
co-ordinate (or in a non-co-ordinate basis if $dx^k\rightarrow e^k$)

\begin{eqnarray}
\delta ((^kS)g) &=&\delta (\overline{g}(\overline{A}))=  \nonumber \\
&=&\{k_1\cdot (g^{jk}\,_{;j}\cdot \,\sigma _{\overline{k}i}+g^{j\overline{k}%
}\cdot \,\sigma _{\overline{k}i;j}+g^{jk}\,_{;j}\cdot \,\omega _{\overline{k}%
i}+g^{j\overline{k}}\cdot \,\omega _{ki;j})+  \nonumber \\
&&+\overline{k}_2\cdot [\theta _{,j}\cdot g^{j\overline{k}}\cdot
h_{ki}+\,\theta \cdot (g^{jk}\,_{;j}\cdot h_{\overline{k}i}+g^{j\overline{k}%
}\cdot h_{ki;j})]\}\cdot dx^i\,\,\,\,\,\,\,\text{.}  \label{14.41}
\end{eqnarray}

The energy-momentum tensors $G\sim (\theta $,\thinspace $_sT)$ could be
found in the forms 
\begin{equation}
\begin{array}{c}
\delta G=(\rho _G+\frac 1e\cdot L\cdot k)\cdot g(a)+ \\ 
+[u(\rho _G+\frac 1e\cdot L\cdot k)+\,\,\,(\rho _G+\frac 1e\cdot L\cdot
k)\cdot \delta u+\delta ^G\overline{s}]\cdot g(u)- \\ 
-\,\,KrL-L\cdot \delta Kr+\delta u\cdot g(^G\overline{\pi })+g(\nabla _u\,^G%
\overline{\pi })+g(\nabla _{^G\overline{s}}u)+ \\ 
+\,\,(\rho _G+\frac 1e\cdot L\cdot k)\cdot (\nabla _ug)(u)+(\nabla _ug)(^G%
\overline{\pi })+(\nabla _{^G\overline{s}}g)(u)+ \\ 
+\,\text{ }k_1\cdot \delta (\overline{g}(\sigma ))+k_1\cdot \,\delta (%
\overline{g}(\omega ))+\overline{k}_2\cdot \delta (\theta \cdot \overline{g}%
(h_u))\text{,}
\end{array}
\label{14.42}
\end{equation}

\begin{equation}
\begin{array}{c}
G_i\,^j\,_{;j}=(\rho _G+\frac 1e\cdot L\cdot k)\cdot a_i+ \\ 
+\,\,[(\rho _G+\frac 1e\cdot L\cdot k)_{,j}\cdot u^j+(\rho _G+\frac 1e\cdot
L\cdot k)\cdot u^j\,_{;j}+\,^G\overline{s}^j\,_{;j}]\cdot u_i- \\ 
-L_{,i}-L\cdot g_{i\,;j}^j+u^j\,_{;j}\cdot \,^G\overline{\pi }_i+g_{i%
\overline{j}}\cdot (^G\overline{\pi }^j\,_{;k}\cdot u^k+u^j\,_{;k}\cdot \,^G%
\overline{s}^k)+ \\ 
+\,\,\,g_{ij;k}\cdot [(\rho _G+\frac 1e\cdot L\cdot k)\cdot u^{\overline{j}%
}\cdot u^k+\,^G\overline{\pi }^{\overline{j}}\cdot u^k+u^{\overline{j}}\cdot
\,^G\overline{s}^k]+ \\ 
+k_1\cdot (g^{jk}\,_{;j}\cdot \,\sigma _{\overline{k}i}+g^{j\overline{k}%
}\cdot \,\sigma _{\overline{k}i;j}+g^{jk}\,_{;j}\cdot \,\omega _{\overline{k}%
i}+g^{j\overline{k}}\cdot \,\omega _{ki;j}) \\ 
+\overline{k}_2\cdot [\theta _{,j}\cdot g^{j\overline{k}}\cdot
h_{ki}+\,\theta \cdot (g^{jk}\,_{;j}\cdot h_{\overline{k}i}+g^{j\overline{k}%
}\cdot h_{ki;j})]\,\,\,\,\,\,\,\,\text{.}
\end{array}
\label{14.43}
\end{equation}

\subsection{Explicit form of the energy-momentum tensors $\theta $, $_sT$
and $Q$}

On the grounds of the representations of $\delta G$ and $\overline{g}(\delta
G)$ the representation of the different energy-momentum tensors $\theta $, $%
_sT$ and $Q$ can be found.

The covariant divergency $\delta \theta $ of the generalized canonical
energy-momentum tensor (GC-EMT) $\theta $ can be written in the form 
\begin{equation}
\begin{array}{c}
\delta \theta =(\rho _\theta +\frac 1e\cdot L\cdot k)\cdot g(a)+ \\ 
+[u(\rho _\theta +\frac 1e\cdot L\cdot k)+(\rho _\theta +\frac 1e\cdot
L\cdot k)\cdot \delta u+\delta ^\theta \overline{s}]\cdot g(u)- \\ 
-\,\,KrL-L\cdot \delta Kr+\delta u\cdot g(^\theta \overline{\pi })+g(\nabla
_u\,^\theta \overline{\pi })+g(\nabla _{^\theta \overline{s}}u)+ \\ 
+\,\,(\rho _\theta +\frac 1e\cdot L\cdot k)\cdot (\nabla _ug)(u)+(\nabla
_ug)(^\theta \overline{\pi })+(\nabla _{^\theta \overline{s}}g)(u)+ \\ 
+\,\,\delta ((^\theta \overline{S})g)\text{ ,}
\end{array}
\label{X.3.-14}
\end{equation}

\noindent or in the form 
\begin{equation}
\begin{array}{c}
\overline{g}(\delta \theta )=(\rho _\theta +\frac 1e\cdot L\cdot k)\cdot a+
\\ 
+[u(\rho _\theta +\frac 1e\cdot L\cdot k)+(\rho _\theta +\frac 1e\cdot
L\cdot k)\cdot \delta u+\delta ^\theta \overline{s}]\cdot u- \\ 
-\,\overline{g}(\,KrL)-L\cdot \overline{g}(\delta Kr)+\delta u\cdot
\,^\theta \overline{\pi }+\nabla _u\,^\theta \overline{\pi }+\nabla
_{^\theta \overline{s}}u+ \\ 
+\,\,(\rho _\theta +\frac 1e\cdot L\cdot k)\cdot \overline{g}(\nabla _ug)(u)+%
\overline{g}(\nabla _ug)(^\theta \overline{\pi })+\overline{g}(\nabla
_{^\theta \overline{s}}g)(u)+ \\ 
+\,\,\overline{g}(\delta ((^\theta \overline{S})g))\text{ .}
\end{array}
\label{X.3.-15}
\end{equation}

In a co-ordinate basis $\delta \theta $ and $\overline{g}(\delta \theta )$
will have the forms 
\begin{equation}  \label{X.3.-18}
\begin{array}{c}
\overline{\theta }_i\,^j\,_{;j}=(\rho _\theta +\frac 1e\cdot L\cdot k)\cdot
a_i+ \\ 
+\,\,[(\rho _\theta +\frac 1e\cdot L\cdot k)_{,j}\cdot u^j+\,\,\,(\rho
_\theta +\frac 1e\cdot L\cdot k)\cdot u^j\,_{;j}+\,^\theta \overline{s}%
^j\,_{;j}]\cdot u_i- \\ 
-L_{,i}-L\cdot g_{i\,;j}^j+u^j\,_{;j}\cdot \,^\theta \overline{\pi }_i+g_{i%
\overline{j}}\cdot (^\theta \overline{\pi }^j\,_{;k}\cdot
u^k+u^j\,_{;k}\cdot \,^\theta \overline{s}^k)+ \\ 
+\,\,\,g_{ij;k}\cdot [(\rho _\theta +\frac 1e\cdot L\cdot k)\cdot u^{%
\overline{j}}\cdot u^k+\,^\theta \overline{\pi }^{\overline{j}}\cdot u^k+u^{%
\overline{j}}\cdot \,^\theta \overline{s}^k]+ \\ 
+\,(g_{i\overline{k}}\cdot \,^\theta \overline{S}\,\,^{jk})_{;j}\text{ ,}
\end{array}
\end{equation}
\begin{equation}  \label{X.3.-19}
\begin{array}{c}
\overline{g}^{i\overline{k}}\cdot \overline{\theta }_k\,^j\,_{;j}=(\rho
_\theta +\frac 1e\cdot L\cdot k)\cdot a^i+ \\ 
+\,\,[(\rho _\theta +\frac 1e\cdot L\cdot k)_{,j}\cdot u^j+(\rho _\theta
+\frac 1e\cdot L\cdot k)\cdot u^j\,_{;j}+\,^\theta \overline{s}%
^j\,_{;j}]\cdot u^i- \\ 
-L_{,j}\cdot g^{i \overline{j}}-L\cdot g^{i\overline{k}}\cdot
g_{k\,;j}^j+u^j\,_{;j}\cdot \,^\theta \overline{\pi }^i+\,^\theta \overline{%
\pi }^i\,_{;j}\cdot u^j+u^i\,_{;j}\cdot \,^\theta \overline{s}^j+ \\ 
+\,\,\,g^{i \overline{l}}\cdot g_{lj;k}\cdot [(\rho _\theta +\frac 1e\cdot
L\cdot k)\cdot u^{\overline{j}}\cdot u^k+\,^\theta \overline{\pi }^{%
\overline{j}}\cdot u^k+u^{\overline{j}}\cdot \,^\theta \overline{s}^k]+ \\ 
+\,g^{i\overline{l}}\cdot (g_{l\overline{k}}\cdot \,^\theta \overline{S}%
\,^{jk})_{;j}\text{ .}
\end{array}
\end{equation}

The covariant divergency $\delta _sT$ of the symmetric energy-momentum
tensor of Belinfante (S-EMT-B) $_sT$, represented in the form 
\[
_sT=(\rho _T+\frac 1e\cdot L\cdot k)\cdot u\otimes g(u)-L\cdot Kr+u\otimes
g(^T\overline{\pi })+\,^T\overline{s}\otimes g(u)+(^T\overline{S})g\text{ ,} 
\]

\noindent can be found in the form 
\begin{equation}  \label{X.3.-20}
\begin{array}{c}
\delta _sT=(\rho _T+\frac 1e\cdot L\cdot k).g(a)+ \\ 
+[u(\rho _T+\frac 1e\cdot L\cdot k)+(\rho _T+\frac 1e\cdot L\cdot k)\cdot
\delta u+\delta ^T \overline{s}]\cdot g(u)- \\ 
-\,\,KrL-L\cdot \delta Kr+\delta u\cdot g(^T \overline{\pi })+g(\nabla _u\,^T%
\overline{\pi })+g(\nabla _{^T\overline{s}}u)+ \\ 
+\,\,(\rho _T+\frac 1e\cdot L\cdot k)\cdot (\nabla _ug)(u)+(\nabla _ug)(^T 
\overline{\pi })+(\nabla _{^T\overline{s}}g)(u)+ \\ 
+\,\,\delta ((^T\overline{S})g)\text{ ,}
\end{array}
\end{equation}

\noindent or in the form 
\begin{equation}  \label{X.3.-21}
\begin{array}{c}
\overline{g}(\delta _sT)=(\rho _T+\frac 1e\cdot L\cdot k)\cdot a+ \\ 
+[u(\rho _T+\frac 1e\cdot L\cdot k)+\,\,\,\,\,\,(\rho _T+\frac 1e\cdot
L\cdot k)\cdot \delta u+\delta ^T \overline{s}]\cdot u- \\ 
-\, \overline{g}(\,KrL)-L\cdot \overline{g}(\delta Kr)+\delta u\cdot \,^T%
\overline{\pi }+\nabla _u\,^T\overline{\pi }+\nabla _{^T\overline{s}}u+ \\ 
+\,\,(\rho _T+\frac 1e\cdot L\cdot k)\cdot \overline{g}(\nabla _ug)(u)+%
\overline{g}(\nabla _ug)(^T\overline{\pi })+\overline{g}(\nabla _{^T%
\overline{s}}g)(u)+ \\ 
+\,\,\overline{g}(\delta ((^T\overline{S})g))\text{ .}
\end{array}
\end{equation}

In a co-ordinate basis $\delta _sT$ and $\overline{g}(\delta _sT)$ will have
the forms 
\begin{equation}  \label{X.3.-24}
\begin{array}{c}
_sT_i\,^j\,_{;j}=(\rho _T+\frac 1e\cdot L\cdot k)\cdot a_i+ \\ 
+\,\,[(\rho _T+\frac 1e\cdot L\cdot k)_{,j}\cdot u^j+\,\,\,\,\,\,(\rho
_T+\frac 1e\cdot L\cdot k)\cdot u^j\,_{;j}+\,^T \overline{s}^j\,_{;j}]\cdot
u_i- \\ 
-L_{,i}-L\cdot g_{i\,;j}^j+u^j\,_{;j}\cdot \,^T \overline{\pi }_i+g_{i%
\overline{j}}\cdot (^T\overline{\pi }^j\,_{;k}\cdot u^k+u^j\,_{;k}\cdot \,^T%
\overline{s}^k)+ \\ 
+\,\,\,g_{ij;k}\cdot [(\rho _T+\frac 1e\cdot L\cdot k)\cdot u^{\overline{j}%
}\cdot u^k+\,^T\overline{\pi }^{\overline{j}}\cdot u^k+u^{\overline{j}}\cdot
\,^T\overline{s}^k]+ \\ 
+\,(g_{i\overline{k}}\cdot \,^T\overline{S}\,^{jk})_{;j}\text{ ,}
\end{array}
\end{equation}
\begin{equation}  \label{X.3.-25}
\begin{array}{c}
\overline{g}^{i\overline{k}}\cdot \,_sT_k\,^j\,_{;j}=(\rho _T+\frac 1e\cdot
L\cdot k)\cdot a^i+ \\ 
+\,\,[(\rho _T+\frac 1e\cdot L\cdot k)_{;j}\cdot u^j+(\rho _T+\frac 1e\cdot
L\cdot k)\cdot u^j\,_{;j}+\,^T \overline{s}^j\,_{;j}]\cdot u^i- \\ 
-L_{,j}\cdot g^{i \overline{j}}-L\cdot g^{i\overline{k}}\cdot
g_{k\,;j}^j+u^j\,_{;j}\cdot \,^T\overline{\pi }^i+\,^T\overline{\pi }%
^i\,_{;j}\cdot u^j+u^i\,_{;j}\cdot \,^T\overline{s}^j+ \\ 
+\,\,\,g^{i \overline{l}}\cdot g_{lj;k}\cdot [(\rho _T+\frac 1e\cdot L\cdot
k)\cdot u^{\overline{j}}\cdot u^k+\,^T\overline{\pi }^{\overline{j}}\cdot
u^k+u^{\overline{j}}\cdot \,^T\overline{s}^k]+ \\ 
+\,g^{i\overline{l}}\cdot (g_{l\overline{k}}\cdot \,^T\overline{S}%
\,^{jk})_{;j}\text{ .}
\end{array}
\end{equation}

The covariant divergency $\delta Q$ of the variational energy-momentum
tensor of Euler-Lagrange (V-EMT-EL) $Q$, represented in the form 
\[
Q=-\,\rho _Q\cdot u\otimes g(u)-u\otimes g(^Q\pi )-\,^Qs\otimes g(u)-(^QS)g%
\text{ ,} 
\]

\noindent follows in the form 
\begin{equation}  \label{X.3.-26}
\begin{array}{c}
\delta Q=-\,\rho _Q\cdot g(a)-(u\rho _Q+\rho _Q\cdot \delta u+\delta
^Qs)\cdot g(u)- \\ 
-\,\,\delta u\cdot g(^Q\pi )-g(\nabla _u\,^Q\pi )-g(\nabla _{^Qs}u)- \\ 
-\,\,\rho _Q\cdot (\nabla _ug)(u)-(\nabla _ug)(^Q\pi )-(\nabla
_{^Qs}g)(u)-\,\delta ((^QS)g)\text{ ,}
\end{array}
\end{equation}

\noindent or in the form 
\begin{equation}  \label{X.3.-27}
\begin{array}{c}
\overline{g}(\delta Q)=-\,\rho _Q\cdot a-(u\rho _Q+\rho _Q\cdot \delta
u+\delta ^Qs)\cdot u- \\ 
-\,\,\delta u\cdot \,^Q\pi -\nabla _u\,^Q\pi -\nabla _{^Qs}u- \\ 
-\,\,\rho _Q\cdot \overline{g}(\nabla _ug)(u)-\overline{g}(\nabla _ug)(^Q\pi
)-\overline{g}(\nabla _{^Qs}g)(u)-\,\overline{g}(\delta ((^QS)g))\text{ .}
\end{array}
\end{equation}

In a co-ordinate basis $\delta Q$ and $\overline{g}(\delta Q)$ will have the
forms 
\begin{equation}
\begin{array}{c}
\overline{Q}_i\,^j\,_{;j}=-\,\rho _Q\cdot a_i-(\rho _Q{}_{,j}\cdot u^j+\rho
_Q\cdot u^j\,_{;j}+\,^Qs^j\,_{;j})\cdot u_i- \\ 
-\,u^j\,_{;j}\cdot \,^Q\pi _i-g_{i\overline{j}}\cdot (^Q\pi ^j\,_{;k}\cdot
u^k+u^j\,_{;k}\cdot \,^Qs^k)- \\ 
-\,\,\,g_{ij;k}\cdot (\rho _Q\cdot u^{\overline{j}}\cdot u^k+\,^Q\pi ^{%
\overline{j}}\cdot u^k+u^{\overline{j}}\cdot \,^Qs^k)-(g_{i\overline{k}%
}\cdot \,^QS^{jk})_{;j}\text{ ,}
\end{array}
\label{X.3.-30}
\end{equation}
\begin{equation}
\begin{array}{c}
\overline{g}^{i\overline{k}}\cdot \overline{Q}_k\,^j\,_{;j}=-\,\rho _Q\cdot
a^i-(\rho _{Q,j}\cdot u^j+\rho _Q\cdot u^j\,_{;j}+\,^Qs^j\,_{;j})\cdot u^i-
\\ 
-\,\,u^j\,_{;j}\cdot \,^Q\pi ^i-\,^Q\pi ^i\,_{;j}\cdot u^j-u^i\,_{;j}\cdot
\,^Qs^j- \\ 
-\,\,\,g^{i\overline{l}}\cdot g_{lj;k}\cdot (\rho _Q\cdot u^{\overline{j}%
}\cdot u^k+\,^Q\pi ^{\overline{j}}\cdot u^k+u^{\overline{j}}\cdot
\,^Qs^k]-g^{i\overline{l}}\cdot (g_{l\overline{k}}\cdot \,^QS^{jk})_{;j}%
\text{ .}
\end{array}
\label{X.3.-31}
\end{equation}

\subsection{Covariant Noether's identities and relations between their
structures}

The covariant Noether identities 
\[
\overline{F}_\alpha +\overline{\theta }_\alpha \,^\beta \,_{/\beta }\equiv 0%
\text{ , \thinspace \thinspace \thinspace \thinspace \thinspace \thinspace
\thinspace \thinspace \thinspace \thinspace \thinspace }\overline{F}_i+%
\overline{\theta }_i\,^j\,_{;j}\equiv 0\text{\thinspace ,\thinspace
\thinspace } 
\]
\[
\overline{\theta }_\alpha \,^\beta -\,_sT_\alpha \,^\beta \equiv \overline{Q}%
_\alpha \,^\beta \text{ ,\thinspace \thinspace \thinspace \thinspace
\thinspace \thinspace \thinspace \thinspace \thinspace \thinspace \thinspace
\thinspace \thinspace \thinspace \thinspace \thinspace \thinspace \thinspace 
}\overline{\theta }_i\,^j-\,_sT_i\,^j\equiv \overline{Q}_i\,^j\text{
,\thinspace \thinspace } 
\]

\noindent for the mixed tensor fields of second rank of the type 1 could be
written in index-free form. $\theta $, $_sT$ and $Q$ can be written in
index-free form by the use of the covariant divergency as 
\[
F+\delta \theta \equiv
0,\,\,\,\,\,\,\,\,\,\,\,\,\,\,\,\,\,\,\,\,\,\,\,\,\,\theta -\,_sT\equiv Q, 
\]
\begin{equation}
\overline{g}(F)+\overline{g}(\delta \theta )\equiv
0,\,\,\,\,\,\,\,\,\,\,\,\,\,\,\,\,\,\,\,\,\,\,\,\,\,\,\,\,(\theta )\overline{%
g}-(\,_sT)\overline{g}\equiv (Q)\overline{g},  \label{X.6.-3}
\end{equation}

\noindent where 
\begin{equation}  \label{X.6.-4}
F=\,_vF+\,_gF\text{ ,}
\end{equation}
\begin{equation}  \label{X.6.-8}
\begin{array}{c}
_vF=\,_{va}F+\,_vW \text{ , \thinspace \thinspace \thinspace \thinspace
\thinspace }_gF=\,_{ga}F+\,_gW\text{ ,\thinspace \thinspace \thinspace
\thinspace \thinspace \thinspace \thinspace }_aF=\;_{va}F+\,_{ga}F\text{
,\thinspace \thinspace } \\ 
W=\,_vW+\,_gW\text{ ,\thinspace }_vF=\,_v\overline{F}_\alpha \cdot e^\alpha
=\,_v\overline{F}_i\cdot dx^i\text{ ,\thinspace \thinspace \thinspace
\thinspace }_vW=\,_v\overline{W}_\alpha \cdot e^\alpha \text{ ,}
\end{array}
\end{equation}
\begin{equation}  \label{X.6.-9}
_{ga}F=\frac{\delta _gL}{\delta g_{\beta \gamma }}\cdot g_{\beta \gamma
/\alpha }\cdot e^\alpha \text{ ,\thinspace \thinspace \thinspace \thinspace
\thinspace \thinspace \thinspace \thinspace \thinspace \thinspace \thinspace 
}_gW=\,_g\overline{W}_\alpha \cdot e^\alpha \text{ .}
\end{equation}

From the second Noether identity ($\theta -\,_sT\equiv Q$) the relation
between the covariant divergencies of the energy-momentum tensors $\theta $, 
$_sT$ and $Q$ follows $\delta \theta \equiv \delta _sT+\delta Q$,\thinspace $%
\delta _sT\equiv \delta \theta -\delta Q$.

\begin{definition}
{\it Local covariant conserved quantity }$G${\it \ of the type of an
energy-momentum tensor of the type 1. Mixed tensor field }$G${\it \ of the
type 1 with vanishing covariant divergency, i. e. }$\delta G=0$, $G_\alpha
\,^\beta \,_{/\beta }=0$, $G_i\,^j\,_{;j}=0$.
\end{definition}

If a given energy-momentum tensor has to fulfil conditions for a local
covariant conserved quantity, then relations follow from the covariant
Noether identities (CNIs) between the covariant divergencies of the other
energy-momentum tensors and the covariant vector field $F$

\begin{center}
$
\begin{array}{cccc}
\text{No.} & \text{Condition for }\delta G & \text{Condition for }F & \text{%
Corollaries from CNIs} \\ 
1. & \delta \theta =0 & F=0 & \delta _sT=-\,\,\delta Q \\ 
2. & \delta _sT=0 & F\neq 0 & \delta \theta =\delta Q=-\,\,F \\ 
&  & F=0 & \delta \theta =\delta Q \\ 
3. & \delta Q=0 & F\neq 0 & \delta \theta =\delta _sT=-\,\,F \\ 
&  & F=0 & \delta \theta =\delta _sT=0
\end{array}
$
\end{center}

{\it Special case}: 
\[
\frac{\delta _vL}{\delta V^A\,_B}=0,\,\,\,\,\,\,\,\,\,\,\,\,\,\,\,\,\,\,\,\,%
\,\,\,\,\,\,\,\,\,\,\,\frac{\delta _gL}{\delta g_{\alpha \beta }}=0\,\,\,%
\text{.} 
\]
\begin{equation}  \label{X.6.-11}
_{va}F=0\text{ ,\thinspace \thinspace \thinspace \thinspace \thinspace
\thinspace \thinspace \thinspace \thinspace \thinspace \thinspace \thinspace 
}_{ga}F=0\text{ ,\thinspace \thinspace \thinspace \thinspace \thinspace
\thinspace \thinspace \thinspace \thinspace \thinspace \thinspace }Q=0\text{
,\thinspace \thinspace \thinspace \thinspace \thinspace \thinspace
\thinspace \thinspace \thinspace \thinspace \thinspace \thinspace }_vF=\,_vW%
\text{ ,\thinspace \thinspace \thinspace \thinspace \thinspace \thinspace
\thinspace \thinspace \thinspace \thinspace \thinspace \thinspace \thinspace
\thinspace \thinspace \thinspace \thinspace \thinspace \thinspace }_gF=\,_gW%
\text{ ,}
\end{equation}
\begin{equation}  \label{X.6.-12}
_aF=\,_{va}F+\,_{ga}F=0\text{ ,}
\end{equation}
\begin{eqnarray}
F &=&W:W+\delta \theta =0\text{ ,\thinspace \thinspace \thinspace }\theta
=\,_sT\text{ ,\thinspace \thinspace \thinspace \thinspace }\delta \theta
=\delta _sT=-\,\,W\text{ .}  \label{X.6.-14} \\
\text{For\thinspace \thinspace \thinspace \thinspace }W &=&0:\delta \theta
=0,\delta _sT=0\text{ .}  \nonumber
\end{eqnarray}

The finding out the covariant Noether identities for a given Lagrangian
density ${\bf L}=\sqrt{-d_g}\cdot L$ along with the energy-momentum tensors $%
\theta $, $_sT$ and $Q$ allow the construction of a rough scheme of the
structures of a Lagrangian theory over a differentiable manifold with
contravariant and covariant affine connections and a metric:%
\newline%

\begin{center}
$
\begin{array}{ccccccccc}
& \leftarrow &  &  & \leftarrow \fbox{L}\rightarrow &  &  & \rightarrow & 
\\ 
\downarrow &  &  &  &  &  &  &  & \downarrow \\ 
&  &  &  &  &  &  &  &  \\ 
&  &  &  & \downarrow &  &  &  & \downarrow \\ 
&  &  &  & \fbox{$_s$T} &  &  &  & \fbox{$\delta $L/$\delta $K$^A$\thinspace 
$_B$} \\ 
\downarrow &  &  &  &  &  &  &  & \downarrow \\ 
\fbox{$\theta $} &  &  &  &  &  &  &  & \fbox{Q} \\ 
\downarrow &  &  &  &  &  &  &  & \downarrow \\ 
&  &  &  & \downarrow &  &  &  &  \\ 
\downarrow & \rightarrow &  & \rightarrow & \fbox{$\theta \,$-\thinspace $_s$%
T$\,\equiv \,$Q} & \leftarrow &  & \leftarrow & \downarrow \\ 
&  &  &  &  &  &  &  & \fbox{F} \\ 
\downarrow &  &  &  & \downarrow &  &  &  & \downarrow \\ 
& \rightarrow &  & \rightarrow & \fbox{F\thinspace +$\,\delta \theta \equiv
\,$0} & \leftarrow &  & \leftarrow & 
\end{array}
$

\vspace{1mm}%
$\_\_\_\_\_\_\_\_\_\_\_\_\_\_\_\_\_\_\_\_\_\_\_\_\_\_\_\_\_\_\_\_\_\_\_\_\_%
\_\_\_\_\_\_\_\_\_\_\_\_\_\_$

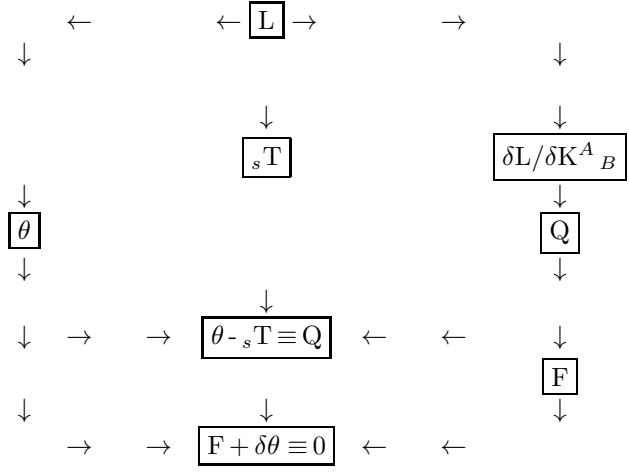
{\bf Fig. 1}. {\it Scheme of the main structure of a Lagrangian theory}
\end{center}

The symmetric energy-momentum tensor of Hilbert $_{gsh}T$ appears as a
construction related to the functional variation of the metric field
variables $g_{\alpha \beta }$ [as a part of the variables $K^A\,_B\sim
(V^A\,_B$, $g_C)$] and interpreted as a symmetric energy-momentum tensor of
a Lagrangian system. This tensor does not exist as a relevant element of the
scheme for obtaining Lagrangian structures by the method of Lagrangians with
covariant derivatives (MLCD). It takes in the scheme a separate place and
has different relations than the usual for the other elements.

\begin{center}
$
\begin{array}{cccccccccc}
& \leftarrow &  &  & \leftarrow \fbox{L}\rightarrow &  &  & \rightarrow & 
\rightarrow &  \\ 
\downarrow &  &  &  & \downarrow &  &  &  & \downarrow & \downarrow \\ 
&  &  &  &  &  &  &  &  &  \\ 
&  &  &  & \downarrow &  &  &  & \downarrow &  \\ 
&  &  &  & \fbox{$_s$T} &  &  &  & \fbox{$\delta $L/$\delta $K$^A$\thinspace 
$_B$} & \downarrow \\ 
\downarrow &  &  &  &  &  &  &  & \downarrow & \fbox{$_{gsh}$T} \\ 
\fbox{$\theta $} &  &  &  &  &  &  &  & \fbox{Q} &  \\ 
&  &  &  &  &  &  &  &  &  \\ 
\downarrow &  &  &  & \downarrow &  &  &  &  &  \\ 
& \rightarrow &  & \rightarrow & \fbox{$\theta \,$-\thinspace $_s$T$\,\equiv
\,$Q} & \leftarrow &  & \leftarrow & \downarrow &  \\ 
&  &  &  &  &  &  &  & \fbox{F} &  \\ 
\downarrow &  &  &  & \downarrow &  &  &  & \downarrow &  \\ 
& \rightarrow &  & \rightarrow & \fbox{F\thinspace +$\,\delta \theta \equiv
\,$0} & \leftarrow &  & \leftarrow &  & 
\end{array}
$

\vspace{1mm}%
$\_\_\_\_\_\_\_\_\_\_\_\_\_\_\_\_\_\_\_\_\_\_\_\_\_\_\_\_\_\_\_\_\_\_\_\_\_%
\_\_\_\_\_\_\_\_\_\_\_\_$

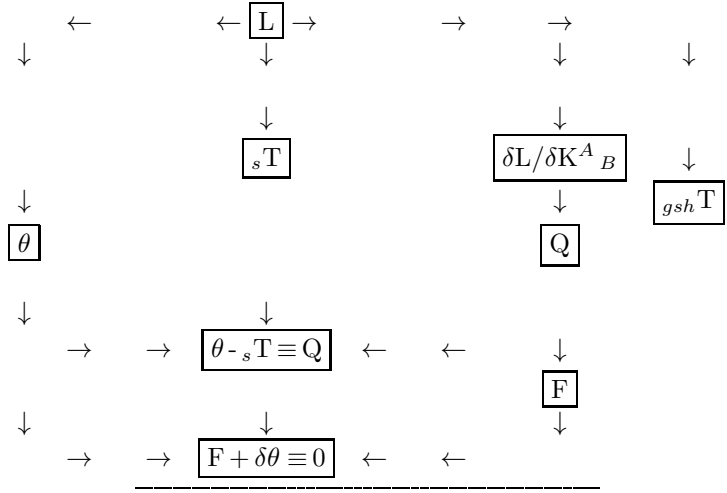
{\bf Fig. 2}. {\it The main structures of a Lagrangian theory }

{\it and the energy-momentum tensor of Hilbert}
\end{center}

Classical field theories involve relations between the different structures
of Lagrangian systems. For the most part of Lagrangian systems equations of
the type of the Euler-Lagrange equations have been imposed and the symmetric
energy-momentum tensor of Hilbert has been used.

\subsection{Navier-Stokes' identities}

If we consider the projections of the first Noether's identity along a
non-null (non-isotropic) vector field $u$ and its corresponding
contravariant and covariant projective metrics $h^u$ and $h_u$ we will find
the first and second Navier-Stokes' identities.

From the Noether's identities in the form 
\begin{eqnarray*}
\overline{g}(F)+\overline{g}(\delta \theta ) &\equiv &0\text{,\thinspace
\thinspace \thinspace \thinspace \thinspace \thinspace \thinspace \thinspace
\thinspace \thinspace \thinspace \thinspace \thinspace \thinspace \thinspace
\thinspace \thinspace }\,\,\,\,\text{(first Noether's identity)}\,\text{,}%
\,\, \\
\,\,\,\,\,\,\,\,\,\,\,\,\,\,\,\,(\theta )\overline{g}-(\,_sT)\overline{g}
&\equiv &(Q)\overline{g}\text{,\thinspace \thinspace \thinspace \thinspace
\thinspace \thinspace \thinspace \thinspace \thinspace \thinspace \thinspace
(second Noether's identity) ,}
\end{eqnarray*}

\noindent we can find the projections of the first Noether's identity along
a contravariant non-null vector field $u=u^i\cdot \partial _i$ and
orthogonal to $u$.

Since 
\begin{eqnarray}
g(\overline{g}(F),u) &=&g_{ik}\cdot g^{\overline{k}\overline{l}}\cdot
F_l\cdot u^{\overline{i}}=g_i^l\cdot F_l\cdot u^{\overline{i}}=F_i\cdot u^{%
\overline{i}}=F(u)\,\,\,\,\,\,\,\text{, \thinspace \thinspace \thinspace
\thinspace }F=F_k\cdot dx^k\text{\thinspace ,}  \label{14.2a} \\
g(\overline{g}(\delta \theta ),u) &=&(\delta \theta )(u)  \label{14.2b}
\end{eqnarray}

\noindent we obtain the{\it \ first Navier-Stokes' identity} in the form 
\begin{equation}
F(u)+(\delta \theta )(u)\equiv 0\text{ \thinspace \thinspace .}  \label{14.3}
\end{equation}

By the use of the relation 
\begin{eqnarray}
\overline{g}[h_u(\overline{g})(F) &=&\overline{g}(h_u[\overline{g}%
(F)])=h^u(F)\,\text{\thinspace \thinspace \thinspace \thinspace \thinspace
,\thinspace \thinspace \thinspace \thinspace \thinspace \thinspace
\thinspace \thinspace \thinspace \thinspace }\overline{g}(h_u)\overline{g}%
=h^u\,\,\,\,\text{,}  \label{14.4a} \\
\overline{g}[h_u(\overline{g})(\delta \theta )] &=&\overline{g}(h_u[%
\overline{g}(\delta \theta )])=h^u(\delta \theta )\text{ ,}  \label{14.4b}
\end{eqnarray}

\noindent the first Noether's identity could be written in the forms 
\begin{eqnarray}
h_u[\overline{g}(F)]+h_u[\overline{g}(\delta \theta )] &\equiv &0\text{
\thinspace \thinspace \thinspace ,}  \label{14.5a} \\
h^u(F)+h^u(\delta \theta ) &\equiv &0\,\,\,\,\,\text{.}  \label{14.5b}
\end{eqnarray}

The last two forms of the first Nother's identity represent the {\it second
Navier-Stokes' identity}.

If the projection $h^u(F)$, orthogonal to $u$, of the volume force\thinspace 
$F$ is equal to zero, we obtain the{\it \ generalized Navier-Stokes' equation%
} in the form 
\begin{equation}
h^u(\delta \theta )=0\text{ \thinspace ,}  \label{14.6}
\end{equation}

\noindent or in the form 
\begin{equation}
h_u[\overline{g}(\delta \theta )]=0\text{ \thinspace \thinspace .}
\label{14.7}
\end{equation}

Let us now find the explicit form of the first and second Navier-Stokes'
identities and the explicit form of the generalized Navier-Stokes' equation.
For this purpose we can use the explicit form of the covariant divergency $%
\delta \theta $ of the generalized canonical energy-momentum tensor $\theta $%
.

(a) The first Navier-Stokes' identity follows in the form 
\begin{eqnarray*}
&&F(u)+(\rho _\theta +\frac 1e\cdot L\cdot k)\cdot g(a,u)+ \\
&&+e\cdot [u(\rho _\theta +\frac 1e\cdot L\cdot k)+(\rho _\theta +\frac
1e\cdot L\cdot k)\cdot \delta u+\delta ^\theta \overline{s}]- \\
&&-(KrL)(u)-L\cdot (\delta Kr)(u)+g(\nabla _u\,^\theta \overline{\pi }%
,u)+g(\nabla _{^\theta \overline{s}}u,u)+ \\
&&+(\rho _\theta +\frac 1e\cdot L\cdot k)\cdot (\nabla _ug)(u,u)+(\nabla
_ug)(^\theta \overline{\pi },u)+(\nabla _{^\theta \overline{s}}g)(u,u)+
\end{eqnarray*}
\begin{equation}
+[\delta ((^\theta \overline{S})g)](u)\equiv 0\,\,\,\,\,\,\,\text{.}
\label{14.8}
\end{equation}

(b) The second Navier-Stokes' identity can be found in the form 
\[
h_u[\overline{g}(F)]+h_u[\overline{g}(\delta \theta )]\equiv 
\]
\begin{eqnarray}
&\equiv &(\rho _\theta +\frac 1e\cdot L\cdot k)\cdot h_u(a)-  \nonumber \\
&&-h_u[\overline{g}(KrL)]-L\cdot h_u[\overline{g}(\delta Kr)]+\delta u\cdot
h_u(^\theta \overline{\pi })+  \nonumber \\
&&+h_u(\nabla _u\,^\theta \overline{\pi })+h_u(\nabla _{^\theta \overline{s}%
}u)+  \nonumber \\
&&+(\rho _\theta +\frac 1e\cdot L\cdot k)\cdot h_u[\overline{g}(\nabla
_ug)(u)]+h_u[\overline{g}(\nabla _ug)(^\theta \overline{\pi })]+  \nonumber
\\
&&+h_u[\overline{g}(\nabla _{^\theta \overline{s}}g)(u)]+h_u[\overline{g}%
(\delta ((^\theta \overline{S})g))]+  \nonumber \\
+h_u[\overline{g}(F)] &\equiv &0\text{ .}  \label{14.9}
\end{eqnarray}

(c) The generalized Navier-Stokes' equation $h_u[\overline{g}(\delta \theta
)]=0$ follows from the second Navier-Stokes' identity under the condition $%
h_u[\overline{g}(F)]=0$ or under the condition $F=0$%
\begin{eqnarray*}
&&(\rho _\theta +\frac 1e\cdot L\cdot k)\cdot h_u(a)- \\
&&-h_u[\overline{g}(KrL)]-L\cdot h_u[\overline{g}(\delta Kr)]+\delta u\cdot
h_u(^\theta \overline{\pi })+ \\
&&+h_u(\nabla _u\,^\theta \overline{\pi })+h_u(\nabla _{^\theta \overline{s}%
}u)+ \\
&&+(\rho _\theta +\frac 1e\cdot L\cdot k)\cdot h_u[\overline{g}(\nabla
_ug)(u)]+h_u[\overline{g}(\nabla _ug)(^\theta \overline{\pi })]+ \\
&&+h_u[\overline{g}(\nabla _{^\theta \overline{s}}g)(u)]+h_u[\overline{g}%
(\delta ((^\theta \overline{S})g))]
\end{eqnarray*}
\begin{eqnarray}
&=&0\,\,\,\,\,\,\,\text{,}  \label{14.10} \\
h_u(a) &=&g(a)-\frac 1e\cdot g(u,a)\cdot g(u)\text{ \thinspace \thinspace .}
\label{14.11}
\end{eqnarray}

{\it Special case:} $(L_n,g)$-spaces: $S=C$, $f^i\,_j=g_j^i$, $g(u,u)=e=$
const. $\neq 0$, $k=1$. 
\begin{equation}
\delta Kr=0\text{ ,}  \label{14.12}
\end{equation}

(a) First Navier-Stokes' identity

\begin{eqnarray*}
&&F(u)+(\rho _\theta +\frac 1e\cdot L)\cdot g(a,u)+ \\
&&+e\cdot [u(\rho _\theta +\frac 1e\cdot L)+(\rho _\theta +\frac 1e\cdot
L)\cdot \delta u+\delta ^\theta \overline{s}]- \\
&&-(KrL)(u)+g(\nabla _u\,^\theta \overline{\pi },u)+g(\nabla _{^\theta 
\overline{s}}u,u)+ \\
&&+(\rho _\theta +\frac 1e\cdot L)\cdot (\nabla _ug)(u,u)+(\nabla
_ug)(^\theta \overline{\pi },u)+(\nabla _{^\theta \overline{s}}g)(u,u)+
\end{eqnarray*}
\begin{equation}
+[\delta ((^\theta \overline{S})g)](u)\equiv 0\,\,\,\,\,\text{.}
\label{14.13}
\end{equation}

(b) Second Navier-Stokes' identity

\begin{eqnarray*}
&&(\rho _\theta +\frac 1e\cdot L)\cdot h_u(a)- \\
&&-h_u[\overline{g}(KrL)]+\delta u\cdot h_u(^\theta \overline{\pi })+ \\
&&+h_u(\nabla _u\,^\theta \overline{\pi })+h_u(\nabla _{^\theta \overline{s}%
}u)+ \\
&&+(\rho _\theta +\frac 1e\cdot L)\cdot h_u[\overline{g}(\nabla _ug)(u)]+h_u[%
\overline{g}(\nabla _ug)(^\theta \overline{\pi })]+ \\
&&+h_u[\overline{g}(\nabla _{^\theta \overline{s}}g)(u)]+h_u[\overline{g}%
(\delta ((^\theta \overline{S})g))]
\end{eqnarray*}
\begin{equation}
+h_u[\overline{g}(F)]\equiv 0\text{ .}  \label{14.14}
\end{equation}

(c) Generalized Navier-Stokes' equation $h_u[\overline{g}(\delta \theta )]=0$

\begin{eqnarray*}
&&(\rho _\theta +\frac 1e\cdot L)\cdot h_u(a)- \\
&&-h_u[\overline{g}(KrL)]+\delta u\cdot h_u(^\theta \overline{\pi })+ \\
&&+h_u(\nabla _u\,^\theta \overline{\pi })+h_u(\nabla _{^\theta \overline{s}%
}u)+ \\
&&+(\rho _\theta +\frac 1e\cdot L)\cdot h_u[\overline{g}(\nabla _ug)(u)]+h_u[%
\overline{g}(\nabla _ug)(^\theta \overline{\pi })]+ \\
&&+h_u[\overline{g}(\nabla _{^\theta \overline{s}}g)(u)]+h_u[\overline{g}%
(\delta ((^\theta \overline{S})g))]
\end{eqnarray*}
\begin{equation}
=0\,\,\,\,\,\,\,\text{.}  \label{14.15}
\end{equation}

{\it Special case: }$V_n$-spaces: $S=C$, $f^i\,_j=g_j^i$, $\nabla _\xi g=0$
for $\forall \xi \in T(M)$, $g(u,u)=e=$ const. $\neq 0$, $k=1$, $g(a,u)=0$.

(a) First Navier-Stokes' identity

\begin{eqnarray*}
&&F(u)+ \\
&&+e\cdot [u(\rho _\theta +\frac 1e\cdot L)+(\rho _\theta +\frac 1e\cdot
L)\cdot \delta u+\delta ^\theta \overline{s}]-(KrL)(u)+
\end{eqnarray*}
\begin{equation}
+[\delta ((^\theta \overline{S})g)](u)\equiv 0\,\,\,\,\,\text{.}
\label{14.16}
\end{equation}

(b) Second Navier-Stokes' identity

\begin{equation}
(\rho _\theta +\frac 1e\cdot L)\cdot h_u(a)-h_u[\overline{g}(KrL)]+\delta
u\cdot h_u(^\theta \overline{\pi })+h_u[\overline{g}(\delta ((^\theta 
\overline{S})g))]+h_u[\overline{g}(F)]\equiv 0\text{ .}  \label{14.17}
\end{equation}

Generalized Navier-Stokes' equation $h_u[\overline{g}(\delta \theta )]=0$

\begin{equation}
(\rho _\theta +\frac 1e\cdot L)\cdot h_u(a)-h_u[\overline{g}(KrL)]+\delta
u\cdot h_u(^\theta \overline{\pi })+h_u[\overline{g}(\delta ((^\theta 
\overline{S})g))]=0\text{ .}  \label{14.18}
\end{equation}

If we express the stress (tension) tensor $(^\theta \overline{S})g$ by the
use of the shear stress tensor $_{ks}\overline{D}$, rotation (vortex) stress
tensor $_k\overline{W}$, and the expansion stress invariant $_k\overline{U}$
then the covariant divergency of the corresponding tensors could be found
and at the end we will have the explicit form of the Navier-Stokes'
identities and the generalized Navier-Stokes' equation including all
necessary tensors for further applications.

\subsection{Navier-Stokes' identities in linear elasticity theory}

If we express the stress (tension) tensor $(^\theta \overline{S})g$ by its
explicit form for linear elasticity theory we can found the Navier-Stokes'
identities in linear elasticity theory

(a) First Navier-Stokes' identity

\begin{eqnarray*}
&&F(u)+(\rho _\theta +\frac 1e\cdot L)\cdot g(a,u)+ \\
&&+e\cdot [u(\rho _\theta +\frac 1e\cdot L)+(\rho _\theta +\frac 1e\cdot
L)\cdot \delta u+\delta ^\theta \overline{s}]- \\
&&-(KrL)(u)+g(\nabla _u\,^\theta \overline{\pi },u)+g(\nabla _{^\theta 
\overline{s}}u,u)+ \\
&&+(\rho _\theta +\frac 1e\cdot L)\cdot (\nabla _ug)(u,u)+(\nabla
_ug)(^\theta \overline{\pi },u)+(\nabla _{^\theta \overline{s}}g)(u,u)+
\end{eqnarray*}

\begin{equation}
+[\delta (\overline{g}(\,_{ks}\overline{D}))+\,\delta (\overline{g}(_k%
\overline{W}))+\frac 1{n-1}\cdot \,\delta (_k\overline{U}\cdot \overline{g}%
(h_u))](u)\equiv 0\,\,\,\,\,\text{.}  \label{14.44}
\end{equation}

(b) The second Navier-Stokes' identity can be found in the form 
\[
h_u[\overline{g}(F)]+h_u[\overline{g}(\delta \theta )]\equiv 
\]
\begin{eqnarray}
&\equiv &(\rho _\theta +\frac 1e\cdot L\cdot k)\cdot h_u(a)-  \nonumber \\
&&-h_u[\overline{g}(KrL)]-L\cdot h_u[\overline{g}(\delta Kr)]+\delta u\cdot
h_u(^\theta \overline{\pi })+  \nonumber \\
&&+h_u(\nabla _u\,^\theta \overline{\pi })+h_u(\nabla _{^\theta \overline{s}%
}u)+  \nonumber \\
&&+(\rho _\theta +\frac 1e\cdot L\cdot k)\cdot h_u[\overline{g}(\nabla
_ug)(u)]+h_u[\overline{g}(\nabla _ug)(^\theta \overline{\pi })]+  \nonumber
\\
&&+h_u[\overline{g}(\nabla _{^\theta \overline{s}}g)(u)]+  \nonumber \\
&&+h_u[\overline{g}(\delta (\overline{g}(\,_{ks}\overline{D})))+\,\overline{g%
}(\delta (\overline{g}(_k\overline{W})))+\frac 1{n-1}\cdot \,\overline{g}%
(\delta (_k\overline{U}\cdot \overline{g}(h_u)))]+  \nonumber \\
+h_u[\overline{g}(F)] &\equiv &0\text{ .}  \label{14.45}
\end{eqnarray}

(c) The generalized Navier-Stokes' equation $h_u[\overline{g}(\delta \theta
)]=0$ follows from the second Navier-Stokes' identity under the condition $%
h_u[\overline{g}(F)]=0$ or under the condition $F=0$%
\begin{eqnarray*}
&&(\rho _\theta +\frac 1e\cdot L\cdot k)\cdot h_u(a)- \\
&&-h_u[\overline{g}(KrL)]-L\cdot h_u[\overline{g}(\delta Kr)]+\delta u\cdot
h_u(^\theta \overline{\pi })+ \\
&&+h_u(\nabla _u\,^\theta \overline{\pi })+h_u(\nabla _{^\theta \overline{s}%
}u)+ \\
&&+(\rho _\theta +\frac 1e\cdot L\cdot k)\cdot h_u[\overline{g}(\nabla
_ug)(u)]+h_u[\overline{g}(\nabla _ug)(^\theta \overline{\pi })]+ \\
&&+h_u[\overline{g}(\nabla _{^\theta \overline{s}}g)(u)]+
\end{eqnarray*}
\begin{equation}
+h_u[\overline{g}(\delta (\overline{g}(\,_{ks}\overline{D})))+\,\overline{g}%
(\delta (\overline{g}(_k\overline{W})))+\frac 1{n-1}\cdot \,\overline{g}%
(\delta (_k\overline{U}\cdot \overline{g}(h_u)))]=0\,\,\,\,\,\,\,\text{,}
\label{14.46}
\end{equation}
\begin{equation}
h_u(a)=g(a)-\frac 1e\cdot g(u,a)\cdot g(u)\text{ \thinspace \thinspace .}
\label{14.47}
\end{equation}

From the first Noether identity $\overline{g}(F)+\overline{g}(\delta \theta
)\equiv 0$ for the case of isotropic elastic media with $\overline{g}[F]=0$,
the equation $\overline{g}(\delta \theta )=0$ follows in the form 
\begin{equation}
\begin{array}{c}
\overline{g}^{i\overline{k}}\cdot \overline{\theta }_k\,^j\,_{;j}=(\rho
_\theta +\frac 1e\cdot L\cdot k)\cdot a^i+ \\ 
+\,\,[(\rho _\theta +\frac 1e\cdot L\cdot k)_{,j}\cdot u^j+(\rho _\theta
+\frac 1e\cdot L\cdot k)\cdot u^j\,_{;j}+\,^\theta \overline{s}%
^j\,_{;j}]\cdot u^i- \\ 
-L_{,j}\cdot g^{i\overline{j}}-L\cdot g^{i\overline{k}}\cdot
g_{k\,;j}^j+u^j\,_{;j}\cdot \,^\theta \overline{\pi }^i+\,^\theta \overline{%
\pi }^i\,_{;j}\cdot u^j+u^i\,_{;j}\cdot \,^\theta \overline{s}^j+ \\ 
+\,\,\,g^{i\overline{l}}\cdot g_{lj;k}\cdot [(\rho _\theta +\frac 1e\cdot
L\cdot k)\cdot u^{\overline{j}}\cdot u^k+\,^\theta \overline{\pi }^{%
\overline{j}}\cdot u^k+u^{\overline{j}}\cdot \,^\theta \overline{s}^k]+ \\ 
+\lambda _1\cdot g^{i\overline{l}}\cdot [g^{mk}\,_{;m}\cdot (_\sigma R_{%
\overline{k}l}+\,_\omega R_{\overline{k}l})+g^{m\overline{k}}\cdot (_\sigma
R_{kl;m}+\,_\omega R_{kl;m})]+ \\ 
+\overline{\lambda }\cdot g^{i\overline{l}}\cdot [_\theta R\cdot
(g^{mk}\,_{;m}\cdot h_{\overline{k}l}+g^{m\overline{k}}\cdot
h_{kl;m})+\,_\theta R_{,m}\cdot g^{m\overline{k}}\cdot h_{kl}]=0\text{
\thinspace \thinspace \thinspace .}
\end{array}
\label{14.48}
\end{equation}

If the vector field $u$ is a given vector field [$e=g(u,u)\neq 0$] and $%
_\sigma R$, $_\omega R$, and $_\theta R$ are given in their explicitly form
as function of the components $\xi _{\perp }^i$ and their covariant
derivatives $\xi _{\perp ;k}^i$ of the deformation vector $\xi _{\perp }$
then the last equation with respect to $\xi _{\perp }$ could be called
generalized Navier-Chauchy equation (or generalized Lame equation) \cite
{Mase}.

\subsection{Navier-Stokes' identities for Newton's fluids}

In analogous way as in linear elasticity theory the Navier-Stokes identities
could be found for Newton's fluids.

If we express the stress (tension) tensor $(^\theta \overline{S})g$ by its
explicit form for Newton's fluids we can found the Navier-Stokes' identities
in hydromechanics for the Newton fluids.

(a) First Navier-Stokes' identity

\begin{eqnarray*}
&&F(u)+(\rho _\theta +\frac 1e\cdot L)\cdot g(a,u)+ \\
&&+e\cdot [u(\rho _\theta +\frac 1e\cdot L)+(\rho _\theta +\frac 1e\cdot
L)\cdot \delta u+\delta ^\theta \overline{s}]- \\
&&-(KrL)(u)+g(\nabla _u\,^\theta \overline{\pi },u)+g(\nabla _{^\theta 
\overline{s}}u,u)+ \\
&&+(\rho _\theta +\frac 1e\cdot L)\cdot (\nabla _ug)(u,u)+(\nabla
_ug)(^\theta \overline{\pi },u)+(\nabla _{^\theta \overline{s}}g)(u,u)+
\end{eqnarray*}

\begin{equation}
+[\delta (\overline{g}(\,_{ks}\overline{D}))+\,\delta (\overline{g}(_k%
\overline{W}))+\frac 1{n-1}\cdot \,\delta (_k\overline{U}\cdot \overline{g}%
(h_u))](u)\equiv 0\,\,\,\,\,\text{.}  \label{14.49}
\end{equation}

(b) The second Navier-Stokes' identity can be found in the form 
\[
h_u[\overline{g}(F)]+h_u[\overline{g}(\delta \theta )]\equiv 
\]
\begin{eqnarray}
&\equiv &(\rho _\theta +\frac 1e\cdot L\cdot k)\cdot h_u(a)-  \nonumber \\
&&-h_u[\overline{g}(KrL)]-L\cdot h_u[\overline{g}(\delta Kr)]+\delta u\cdot
h_u(^\theta \overline{\pi })+  \nonumber \\
&&+h_u(\nabla _u\,^\theta \overline{\pi })+h_u(\nabla _{^\theta \overline{s}%
}u)+  \nonumber \\
&&+(\rho _\theta +\frac 1e\cdot L\cdot k)\cdot h_u[\overline{g}(\nabla
_ug)(u)]+h_u[\overline{g}(\nabla _ug)(^\theta \overline{\pi })]+  \nonumber
\\
&&+h_u[\overline{g}(\nabla _{^\theta \overline{s}}g)(u)]+  \nonumber \\
&&+h_u[\overline{g}(\delta (\overline{g}(\,_{ks}\overline{D})))+\,\overline{g%
}(\delta (\overline{g}(_k\overline{W})))+\frac 1{n-1}\cdot \,\overline{g}%
(\delta (_k\overline{U}\cdot \overline{g}(h_u)))]+  \nonumber \\
+h_u[\overline{g}(F)] &\equiv &0\text{ .}  \label{14.50}
\end{eqnarray}

(c) The generalized Navier-Stokes' equation $h_u[\overline{g}(\delta \theta
)]=0$ follows from the second Navier-Stokes' identity under the condition $%
h_u[\overline{g}(F)]=0$ or under the condition $F=0$%
\begin{eqnarray*}
&&(\rho _\theta +\frac 1e\cdot L\cdot k)\cdot h_u(a)- \\
&&-h_u[\overline{g}(KrL)]-L\cdot h_u[\overline{g}(\delta Kr)]+\delta u\cdot
h_u(^\theta \overline{\pi })+ \\
&&+h_u(\nabla _u\,^\theta \overline{\pi })+h_u(\nabla _{^\theta \overline{s}%
}u)+ \\
&&+(\rho _\theta +\frac 1e\cdot L\cdot k)\cdot h_u[\overline{g}(\nabla
_ug)(u)]+h_u[\overline{g}(\nabla _ug)(^\theta \overline{\pi })]+ \\
&&+h_u[\overline{g}(\nabla _{^\theta \overline{s}}g)(u)]+
\end{eqnarray*}
\begin{equation}
+h_u[\overline{g}(\delta (\overline{g}(\,_{ks}\overline{D})))+\,\overline{g}%
(\delta (\overline{g}(_k\overline{W})))+\frac 1{n-1}\cdot \,\overline{g}%
(\delta (_k\overline{U}\cdot \overline{g}(h_u)))]=0\,\,\,\,\,\,\,\text{,}
\label{14.51}
\end{equation}
\[
h_u(a)=g(a)-\frac 1e\cdot g(u,a)\cdot g(u)\text{ \thinspace \thinspace .} 
\]

From the first Noether identity $\overline{g}(F)+\overline{g}(\delta \theta
)\equiv 0$ for the case of Newton's fluids with $\overline{g}[F]=0$, the
equation $\overline{g}(\delta \theta )=0$ follows in the form 
\begin{equation}
\begin{array}{c}
\overline{g}^{i\overline{k}}\cdot \overline{\theta }_k\,^j\,_{;j}=(\rho
_\theta +\frac 1e\cdot p\cdot k)\cdot a^i+ \\ 
+\,\,[(\rho _\theta +\frac 1e\cdot p\cdot k)_{,j}\cdot u^j+(\rho _\theta
+\frac 1e\cdot p\cdot k)\cdot u^j\,_{;j}+\,^\theta \overline{s}%
^j\,_{;j}]\cdot u^i- \\ 
-p_{,j}\cdot g^{i\overline{j}}-p\cdot g^{i\overline{k}}\cdot
g_{k\,;j}^j+u^j\,_{;j}\cdot \,^\theta \overline{\pi }^i+\,^\theta \overline{%
\pi }^i\,_{;j}\cdot u^j+u^i\,_{;j}\cdot \,^\theta \overline{s}^j+ \\ 
+\,\,\,g^{i\overline{l}}\cdot g_{lj;k}\cdot [(\rho _\theta +\frac 1e\cdot
p\cdot k)\cdot u^{\overline{j}}\cdot u^k+\,^\theta \overline{\pi }^{%
\overline{j}}\cdot u^k+u^{\overline{j}}\cdot \,^\theta \overline{s}^k]+ \\ 
+k_1\cdot g^{i\overline{l}}\cdot [g^{mk}\,_{;m}\cdot (\sigma _{\overline{k}%
l}+\omega _{\overline{k}l})+g^{m\overline{k}}\cdot (\sigma _{kl;m}+\,\omega
_{kl;m})]+ \\ 
+\overline{k}_2\cdot g^{i\overline{l}}\cdot [\theta \cdot
(g^{mk}\,_{;m}\cdot h_{\overline{k}l}+g^{m\overline{k}}\cdot
h_{kl;m})+\,\theta _{,m}\cdot g^{m\overline{k}}\cdot h_{kl}]=0\text{
\thinspace \thinspace \thinspace .}
\end{array}
\label{14.52}
\end{equation}

If the vector field $u$ is a given vector field [$e=g(u,u)\neq 0$] and $%
\sigma $, $\omega $, and $\theta $ are given in their explicitly form as
function of the components $u^i$ and their covariant derivatives $%
u_{\,\,;k}^i$ of the velocity vector $u$ then the last equation with respect
to $u$ is called generalized Navier-Stokes equation.

On the basis of the considered results many problems of the continuous media
mechanics could be further specialized and solved.

\section{Conclusion}

In the present paper continuous media mechanics is consider over $(\overline{%
L}_n,g)$-spaces with respect to its basic notions from the point of view of
the differential-geometric structures related to

(a) deformations,

(b) stresses (tensions),

(c) relation between stressed and deformations on the basis of the method of
Lagrangians with covariant derivatives.

The presented results were specialized for elasticity theory and
hydrodynamics in $(\overline{L}_n,g)$-spaces. These results could also be
used in all spaces considered as special cases of the $(\overline{L}_n,g)$%
-spaces [Euclidean $E_n$-spaces, (pseudo) Euclidean $M_n$-spaces, (pseudo)
Riemannian $V_n$- and $U_n$-spaces, Weyl's spaces $W_n$ and $Y_n$ etc.] as
well as in general relativity as an extension in some aspects of the
relativistic continuous media mechanics. Many new features of the continuous
media mechanics are found at a local level [e.g. the existence of local
rotations velocity tensors for isotropic elastic media and Newton's fluids,
existence of stresses generated by rotation velocity tensor etc.].

The existence of classical field theories for describing dynamical systems
in different models of space or space-time allows a closer comparison with
the structure of a theory of continuous media. All basic notions of
continuous mechanics could be used in a classical field theory and vice
versa - basic notions of a classical field theory could be applied in a
theory of continuous media.

The covariant and contravariant metrics introduced over differentiable
manifolds with contravariant and covariant affine connections allow
applications for mathematical models of dynamic systems described over $(%
\overline{L}_n,g)$-spaces. On the other side, different type of geometries
can be considered by imposing certain additional conditions of the type of
metric transports on the metric. Additional conditions determined by
different ''draggings along'' of the metric can have physical interpretation
connected with changes of the length of a vector field and with changes of
the angle between two vector fields.

The introduction of contravariant and covariant projective metrics
corresponding to a non-isotropic (non-null) contravariant vector field
allows the evolution of tensor analysis over sub manifolds of a manifold
with contravariant and covariant connections and metrics and its
applications for descriptions of the evolution of physical systems over $(%
\overline{L}_n,g)$-spaces.

The kinematic characteristics, connected with the introduced notions of
relative velocity and relative acceleration can be used for description of
different dynamic systems by means of mathematical models, using
differentiable manifold $M$ with contravariant and covariant affine
connections and metrics as a model of space-time ($\dim M=4$) [ETG in $V_n$%
-spaces, Einstein-Cartan theory in $U_n$-spaces, metric-affine theories in $%
(L_n,g)$-spaces], or as a model for the consideration of dynamic
characteristics of some physical systems [theories of the type of
Kaluza-Klein in $V_n$-spaces ($n>4$), relativistic hydrodynamics etc.]. All
basic notions related to the relative velocity and relative acceleration
could be related to deformations and stresses (tensions) in the continuous
media mechanics in $(\overline{L}_n,g)$-spaces. At the same time the
kinematic characteristics can be used for a more correct formulation of
problems, connected with the experimental check-up of modern gravitational
theories.

In the case of general relativity theory one of the propositions can be used
for describing the characteristics of gravitational detectors: If test
particles are considered to move in an external gravitational field ($%
R_{ij}=0$), then their relative acceleration will be caused only by the
curvature shear acceleration. Therefore, gravitational wave detectors have
to be able to detect accelerations of the type of shear acceleration (and
not of the type of expansion acceleration), if the energy-momentum tensor of
the detector is neglected as a source of a gravitational field.

The conjecture connecting the stresses and deformations in continuous media
mechanics on the basis of the kinematic characteristics and the
energy-momentum tensors could lead to a more exact and comprehensive
description of dynamical systems in continuous media mechanics and in the
classical (non-quantized) field theories in $V_n$-, $U_n$-, $(L_n,g)$-, and $%
(\overline{L}_n,g)$-spaces as well as in their special cases.

\end{document}